\documentclass[12pt,preprint,dvips,deluxetable]{aastex}

\def \msun{$\mathrm{M}_\odot$}

\def \kms{km~s$^{-1}$}
\def \1pap{Paper I}
\def \pap2{Paper II}
\def \deg{$^\circ$}

\makeatletter
\newcommand\cutinheadb[1]{%
 \noalign{\vskip .8ex}%
 \@ptabularcr
 \noalign{\vskip -4ex}%
 \multicolumn{\pt@ncol}{c}{#1}%
 \@ptabularcr
 \noalign{\vskip .8ex}%
 \hline
 \@ptabularcr
 \noalign{\vskip -1.5ex}%
}%
\def\cutinheadb@ppt#1{%
 \noalign{\vskip .8ex}%
 \@ptabularcr
 \noalign{\vskip -1.5ex}% Style Note: in apj, it is -1.5ex
 \multicolumn{\pt@ncol}{c}{#1}%
 \@ptabularcr
 \noalign{\vskip .8ex}%
 \hline
 \@ptabularcr
 \noalign{\vskip -1.5ex}%
}%
\makeatother

\makeatletter
\newcommand\cutinheadc[1]{%
 \hline
 \noalign{\vskip 1.5ex}%
 \@ptabularcr
 \noalign{\vskip -4ex}%
 \multicolumn{\pt@ncol}{c}{#1}%
 \@ptabularcr
 \noalign{\vskip .8ex}%
 \hline
 \@ptabularcr
 \noalign{\vskip -1.5ex}%
}%
\def\cutinheadc@ppt#1{%
 \noalign{\vskip .8ex}%
 \@ptabularcr
 \noalign{\vskip -1.5ex}% Style Note: in apj, it is -1.5ex
 \multicolumn{\pt@ncol}{c}{#1}%
 \@ptabularcr
 \noalign{\vskip .8ex}%
 \hline
 \@ptabularcr
 \noalign{\vskip -1.5ex}%
}%
\makeatother

\author{Henry~A.~Kobulnicky\altaffilmark{1},
Daniel~C.~Kiminki\altaffilmark{2},
Michael J. Lundquist\altaffilmark{1},
Jamison Burke\altaffilmark{1, 3},
James Chapman\altaffilmark{1, 4},
Erica Keller\altaffilmark{1, 5},
Kathryn Lester\altaffilmark{1, 6},
Emily K. Rolen\altaffilmark{1, 7},
Eric Topel\altaffilmark{1, 8},
Anirban Bhattacharjee\altaffilmark{1},
Rachel A. Smullen\altaffilmark{1},
Carlos A. Vargas A\'lvarez\altaffilmark{1},
Jessie C. Runnoe\altaffilmark{1, 9},
Daniel A. Dale\altaffilmark{1},
Michael M. Brotherton\altaffilmark{1}}

\altaffiltext{1}{Dept. of Physics \& Astronomy, University 
of Wyoming, Laramie, WY 82070, USA \\ chipk@uwyo.edu}
\altaffiltext{2}{Dept. of Astronomy, University of Arizona, Tucson, AZ 85721, USA} 
\altaffiltext{3}{Department of Physics and Astronomy, Swarthmore College, Swarthmore, PA 19081, USA \\ jburke2@swarthmore.edu } 
\altaffiltext{4}{Massachusetts College of Liberal Arts, 375 Church St., North Adams, MA 01247, USA \\ jc6380@mcla.edu } 
\altaffiltext{5}{Department of Astronomy, Mt. Holyoke College, 50 College Street, South Hadley, MA 01075, USA \\ kelle22e@mtholyoke.edu}
\altaffiltext{6}{Department of Physics, Lehigh University, 16 Memorial Drive East, Bethlehem, PA 18015, USA \\ kvl214@lehigh.edu}
\altaffiltext{7}{Department of Physics and Astronomy, Vanderbilt University, Nashville, TN 37235, USA \\ emily.k.rolen@vanderbilt.edu}
\altaffiltext{8}{Department of Physics, 1520 St. Olaf Avenue, Northfield, MN 55057, USA \\ topel@stolaf.edu}
\altaffiltext{9}{Department of Physics, Department of Astronomy \& Astrophysics, The Pennsylvania State University, 525 Davey Lab, University Park, PA 16802, USA}
\begin{document}
\slugcomment{Accepted for Publication in ApJ}

\title{TOWARD COMPLETE STATISTICS OF MASSIVE BINARY STARS: 
PENULTIMATE RESULTS FROM THE CYGNUS OB2 RADIAL VELOCITY SURVEY}

\begin{abstract}  

We analyze orbital solutions for 48 massive multiple-star systems
in the Cygnus~OB2 Association, 23 of which are newly
presented here, to find that the observed distribution of
orbital periods is approximately uniform in $\log~P$ for
$P<$45 d,  but it is not scale-free.
Inflections in the cumulative distribution near 6 d, 14, d,
and 45 d, suggest key physical scales of $\simeq$0.2,
$\simeq$0.4, and $\simeq$1 A.U. where yet-to-be-identified
phenomena create distinct features.  No single power law 
provides a statistically compelling prescription, but if
features are ignored, a power law with exponent
$\beta\simeq-0.22$ provides a crude approximation over
$P$=1.4 -- 2000 d, as does a piece-wise linear function with
a break near 45 d.  The cumulative period distribution
flattens at $P>$45 d, even after correction for completeness,
indicating either a lower binary fraction or a shift toward
low-mass companions.   A high degree of similarity (91\%
likelihood) between the Cyg~OB2 period distribution and that
of other surveys  suggests that the  binary
properties at $P\lesssim$25 d are determined  by local physics of disk/clump
fragmentation and are relatively insensitive to
environmental and evolutionary factors.   Fully 30\% of the
unbiased parent sample is a binary with period $P<$45 d.
Completeness corrections imply a binary fraction near 55\%
for $P<$5000 d.  The observed distribution of mass ratios
$0.2<q<1$ is consistent with uniform,  while the  observed
distribution of eccentricities $0.1<e<0.6$ is consistent
with uniform plus an excess of $e\simeq0$ systems.   We
identify six stars, all supergiants, that exhibit aperiodic
velocity variations of $\sim$30 \kms\ attributed to
atmospheric fluctuations.  

\end{abstract}


\section{Introduction}

Massive stars (M$>$8\msun) earlier than about B2.5V 
dominate the cosmic production of ionizing photons and 
stellar wind momentum before terminating in nature's most 
energetic explosive events, leaving behind  neutron stars
and black holes.   Formation scenarios for these explosions
and subsequent compact stellar remnants require the
existence of a close stellar companion.   In  X-ray binaries
the massive star becomes a neutron star accreting from an
evolved star.  Mergers of compact objects (neutron
star-neutron star or black hole-neutron star) produce
gravitational waves (GW) that should be detected by the
immanent generation of GW experiments \citep{Dominik2013}.  
``Runaway'' or ``high-velocity stars also owe their extreme
kinematics to hard interactions with  massive binary systems
\citep{blaauw, giesbolton, Hoogerwerf2001}. In particular,
stars in the range  8--25 \msun\ that have close companions
may become the progenitors of Type Ibc supernovae when the H
envelope of the more massive star is stripped during a phase
of common envelope evolution \citep{Nomoto1995, Smartt2009,
Eldridge2013, Smith2014}. Close binaries may even produce
all type Ibc supernova if single H-poor Wolf-Rayet stars
collapse to become black holes without producing a supernova
\citep{Fryer2007}. Furthermore, supernovae of type Ic and
$\gamma$-ray bursts appear to happen simultaneously,
suggesting a connection between the two types of events that
may have their origins in massive binary progenitors
\citep{woosley2006}.   Close massive binaries also produce
the population of low-mass X-ray binary systems through
evolution of the more massive star to a supernova,
production of a neutron star or black hole, and then the
subsequent evolution of the lower-mass companion that
becomes the mass donor \citep{vanden83}.  

The Cygnus OB2 Radial Velocity Survey \citep{Kiminki07}  is
an optical spectroscopic survey of 128 photometrically
selected O and early-B stars comprising an unbiased (with
respect to binarity) sample  within the core region of the
nearby Cygnus OB2 Association ($\sim$1.4 kpc;
\citealt{Hanson03}).  This sample contains 45 stars of
spectral type O and 83 stars of spectral type B. Luminosity
class V (i.e., unevolved) stars account for 91 of the 128.  
The Survey is designed to statistically measure the binary
properties of massive stars within a common formation
environment at a similar age.   Although massive star
formation in Cyg~OB2 may not be coeval,  studies suggest an
age of 3--4 Myr,  young enough that the most massive stars
are still present and the majority of these stars are still
on the main sequence \citep{Hanson03}.   Papers I--VI in
this series (\citealt{Kiminki07}, \citealt{Kiminki08},
\citealt{Kiminki09}, \citealt{Kiminki2012a},
\citealt{Kiminki2012b}, \citealt{Kobulnicky2012})  describe
the Survey and document orbital parameters for 25 massive
binary systems known from  the inception of the Survey in
1999  through the 2011 observing season.  In particular,
Paper V \citep{Kiminki2012b} provides a high-level overview
of the Survey's constraints on statistics of massive binary
stars by use of a Monte Carlo analysis. Stars down to
spectral type $\sim$B2.5V are included in the Survey because
they are expected to dominate the population of supernova
progenitors, given typical initial mass functions.  Other
large studies of binarity among massive stars include the
spectroscopic  surveys of \citet{Garmany1980},
\citet{Sana2012}, and \citet{Chini2012}  as well as the
imaging surveys of \citet{Kouwenhoven2007} and
\citet{Mason2009}.    

In this contribution we report the observations (Section 2)
and orbital solutions for 22 new single-lined spectroscopic 
binaries and one new double-lined binary (Section 3).  These
data contribute to the growing census of massive binary
statistics in a  complete sample of massive stars defined in
Paper I. This increasingly complete compilation of orbital
periods, eccentricities, and mass ratios provides a rich
dataset  to constrain our understanding of massive star
formation, evolution through binary channels, and massive
star end states.  Section 3 also contains a list of 16 stars
that  show minimal velocity variations after extensive
observation and six stars---all supergiants---that exhibit
irregular variations presumed to originate from atmospheric
pulsations.   An appendix provides an update on orbital
solutions for six stars from Papers II and III for
consistency with the most recent Cyg~OB2 analyses.  We
conclude by tabulating  orbital parameters for the 48 known
binary (and several triple) systems and conducting a
high-level analysis of the  distribution of orbital
parameters (Section 4).  Nomenclature of objects  discussed
herein follows the MT91~\#\#\# notation of \citet{MT91},
with ``S \#'' indicating the numeration of \citet{Schulte58}
and ``CPR2002-A\#\#'' or ``B\#\#'' for that of
\citet{comeron02}.

\section{Spectroscopic Observations, Reductions, and Radial Velocity 
Measurement}

Observational methodologies and data reduction procedures
closely follow those described in  \citet{Kobulnicky2012}
and earlier papers.  Paper I describes instrumental setups
and dates of observation for observing runs using Keck+HIRES
spectrograph  (1999--2001), Lick+Hamilton echelle
spectrograph (1999--2000), and WIYN+Hydra spectrograph
(2001--2008).  Results presented here principally use 
spectra obtained at the Wyoming Infrared Observatory (WIRO)
2.3 meter telescope+Longslit optical spectrograph using a
2000 l mm$^{-1}$ grating in first order during the
2011--2013 observing seasons, with a small number of
observations  obtained in 2014 May.   Resolutions of
R$\approx$4500 were achieved using a 1\farcs2 slit. 
Multiple 600 s spectra of each target were obtained over the
wavelength range 5400--6700 \AA.    Data from WIRO in 2010
and earlier used either a 600 l mm$^{-1}$ grating in second
order or an 1800 l mm$^{-1}$ grating in first order to
achieve spectral resolutions of R$\approx$2500 and 4000,
respectively.  These observations are more fully described
in earlier papers in this series.  Routine CuAr lamp
exposures provided wavelength solutions having a typical rms
of 0.012--0.030  \AA.  Spectra were combined and transformed
into the  Heliocentric reference frame, yielding a final
signal-to-noise ratio (SNR) of 60--200 in the vicinity of
the \ion{He}{1} 5876 \AA\ line used to measure radial
velocities.   The ensemble of spectra spanning hours to
years were shifted by small amounts (2 -- 8) \kms\ so that
the interstellar \ion{Na}{2} $\lambda\lambda$ 5889/96 lines
align  at the mean velocity across all observations.  This
corrects for small changes in the wavelength solution zero
point across epochs.   Data from Keck and WIYN did not cover
the  \ion{Na}{2} lines so these spectra potentially have
systematic velocity zero point differences compared to the
WIRO data.  However, our analysis of constant-velocity stars
in the Survey shows good agreement, at the level of 3--6
\kms, between the velocities obtained from Keck, WIYN, and
WIRO spectra. Nevertheless, solutions are obtained solely
from WIRO data, where possible.  When necessary, data from
Keck and/or WIYN are included in the solutions. Data from
Lick were generally of lower quality and are not used.

Radial velocities were measured by fitting Gaussian profiles
to the  \ion{He}{1} 5876 \AA\ photospheric line (\ion{He}{1}
4471 \AA\ for the Keck and WIYN spectra) adopting a rest
wavelength of 5875.69 \AA\ (4471.55 \AA\ for Keck and WIYN
spectra).  Constant-velocity stars showed good agreement
(3--6 \kms) between velocities from the various telescopes
and instruments, lending confidence to the solutions that
involve all three sources of data.   Our fitting
code\footnote{We use the robust curve-fitting algorithm
MPFIT as implemented in IDL \citep{mpfit}.}  fixes the
Gaussian width and depth to be the mean determined from all
the spectra, after rejecting outliers, and it solves for the
best-fitting line center and its uncertainty.  Since the
primary goal of the Cygnus~OB2 radial velocity survey is to
obtain orbital parameters for massive binaries, a goal that
requires good \emph{relative} radial velocities, we did not
 observe radial velocity standard stars, and consequently the
absolute space velocities reported are likely to be 
accurate to $\sim$3--5 \kms.  Nevertheless,
there is good agreement between the mean velocities of
our Cyg~OB2 OB star sample ($V_{ave}$=-15.6
\kms; $\sigma_V$=8.2 \kms) and results from 
other workers ($V_{ave}\simeq$-18 \kms; N. Wright, private communication).  

\begin{figure}
\epsscale{1.0}
\centering
\plotone{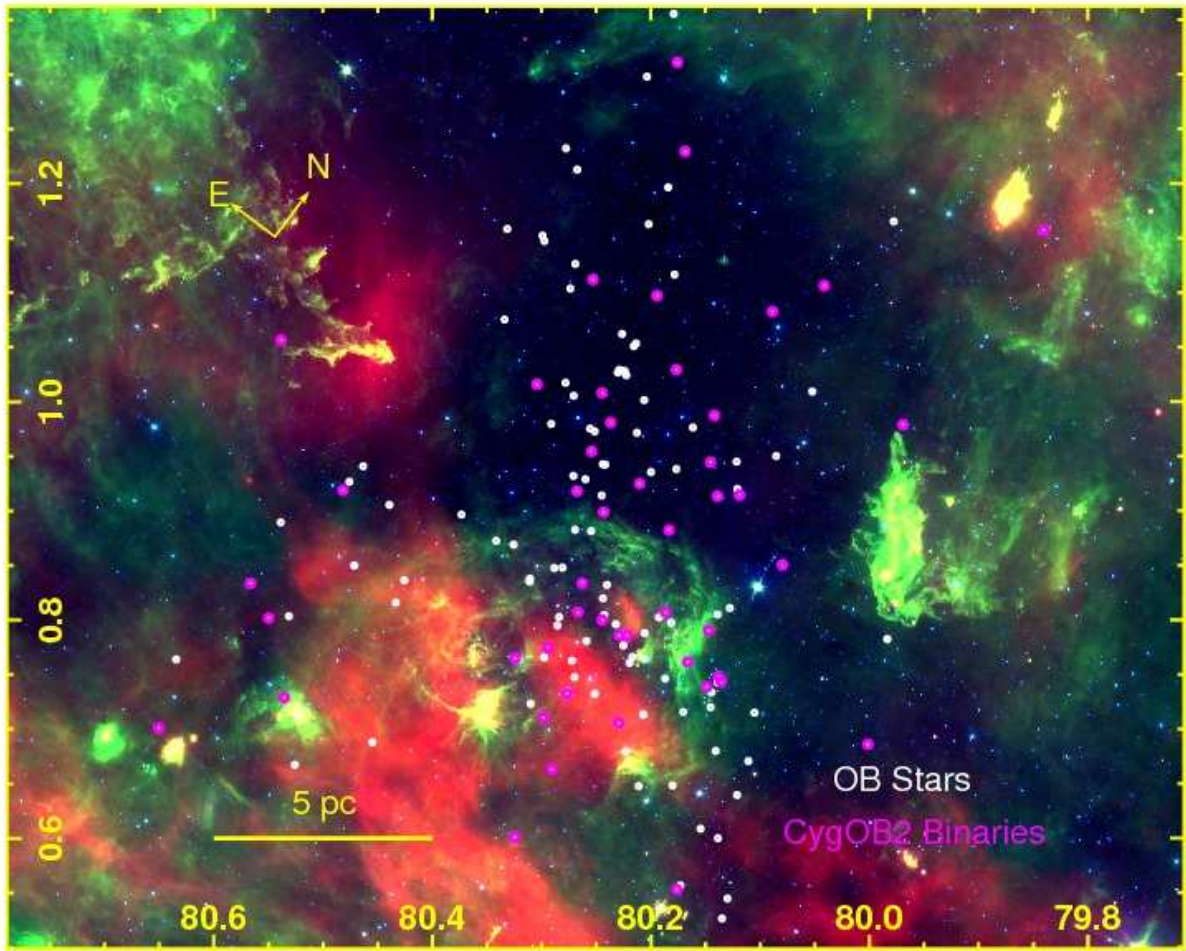}
\caption{Three-color image of the Cygnus~OB2 region with the
{\it Spitzer} 4.5 $\mu$m, 8.0 $\mu$m, and 24 $\mu$m images
in blue/green/red.  
White symbols denote O- and early B massive stars while
larger magenta symbols mark the 48 known binary/multiple systems.  
The early-B SB2 system CPR2002~A45
lies just off the field of view to the upper right.  The
bar at lower left shows a linear scale of 5 pc at a distance of 1.4 kpc.
\label{color}}
\end{figure}

Table~\ref{vel.tab} records the heliocentric Julian Date,
orbital phase, the heliocentric radial velocity, the
velocity uncertainty, and the observed-minus-computed (O-C) 
velocity for each measurement based on the orbital solutions
that follow.  All systems, with the exception of MT91~646,
are single-lined spectroscopic binaries so that only one
velocity is reported for each epoch.  MT91~268 is a probable
triple system exhibiting two periodicities, 
so both velocity components appear in
Table~\ref{vel.tab}. Figure~\ref{color}  shows a three-color
view of the Cygnus~OB2 region with 4.5,
8.0, and 24 $\mu$m mosaic images from the {\it Spitzer
Cygnus-X Legacy Survey} \citep{Hora} in blue, green, and
red, respectively. White points depict massive stars in the
Cygnus~OB2 Radial Velocity Survey parent sample, while magenta
points mark known binary or higher-order systems.  The 5
pc bar at lower left marks the linear scale at the
adopted distance of 1.4 kpc.  Binaries are apparently distributed
across the face of the Association without preference
for radial distance from center.

\section{Orbital Solutions}
 
We analyzed the radial velocity power spectrum for each
object to select likely periods and then examined the folded
velocity curve for periods corresponding to the strongest
peaks.   In most cases the strongest peak yielded a clear,
unambiguous period and a convincing  phased velocity curve. 
Secondary peaks and possible aliases could be eliminated by
visual inspection owing to the much larger dispersion in the
data at any given phase.  We used the binary orbital
solution package ``BINARY'' by D.
Gudehus\footnote{http://www.chara.gsu.edu/$\sim$gudehus/binary.html}
with these initial period estimates and the radial velocity
data to solve for the full suite of orbital parameters and
associated uncertainties.  Tables~\ref{solutions1.tab}
through \ref{solutions5.tab} compile these best-fitting
parameters and uncertainties for each object.  Listed within
the table are the period in days ($P$), eccentricity of the
orbit ($e$), longitude of periastron in degrees ($\omega$),
systemic radial velocity ($\gamma$), epoch of periastron
($T_0$), primary velocity  semi-amplitude  ($K_1$) and (if
applicable) secondary velocity semi-amplitude ($K_2$), 
spectral classifications from this survey (S.C.$_1$ \&\
S.C.$_2$, if available), estimates of the inclination ($i$),
the adopted primary  stellar mass ($M_1$) and (if
applicable) secondary mass ($M_2$), mass ratio ($q$),
semi-major axis ($a$), and reduced chi squared values of the
best fitting solution.   Tables~\ref{solutions1.tab} through
\ref{solutions5.tab} list orbital parameters for 22 new
systems reported in this work.   Masses for OB stars
are taken from \citet{Martins2005} for O stars
and \citet{Hunter2008} for early B stars.  Upper and lower limits
on the inclinations are obtained, in most cases, 
by adopting 90\degr\ and the lowest inclination compatible with the 
absence of secondary spectral features (i.e., where the mass of the
secondary approaches that of the primary).

\subsection{New Binary Systems}

\begin{figure}
\epsscale{1.0}
\centering
\plotone{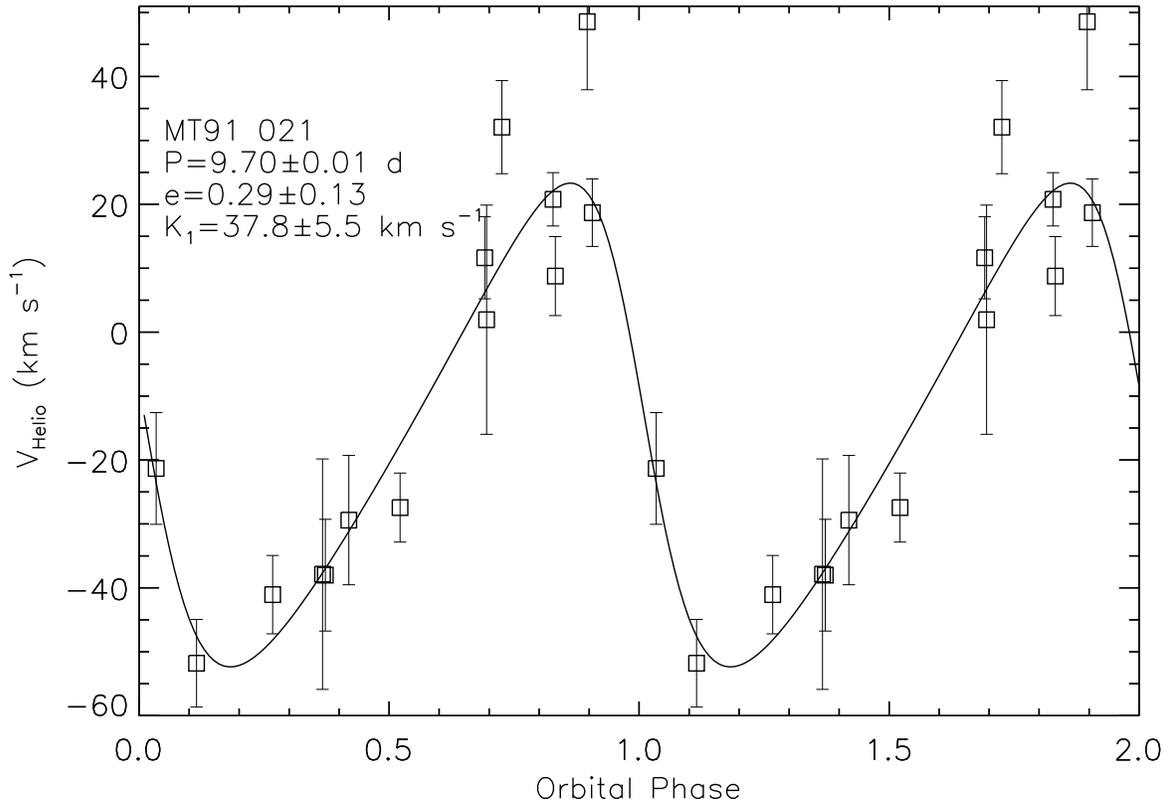}
\caption{Folded radial velocity curve and 
best-fitting solution for MT91~021.   The large
reduced $\chi^2$ of 3.2 suggests the presence of an additional
velocity component in this single-lined, potentially triple
system.
\label{sol021}}
\end{figure}

{\it MT91~021---}\citet{Kiminki07} classified the MT91~021
primary  (V=13.74; \citealt[][]{MT91}) as a probable
single-lined binary system on the basis of four data
points.  Using a combination of three WIYN spectra we
revise  the spectral type to B1.5V by comparison in the
\citet{WF90} spectral atlas.   Table~\ref{vel.tab} lists the
14 radial velocity measurements  from WIRO spanning 2010
August -- 2013 September. Figure~\ref{sol021} shows the
best-fitting orbital solution  (solid curve) and folded
velocity  data (points with error bars).  
Table~\ref{solutions1.tab} summarizes the full suite of
orbital parameters.  If we adopt a mass of 12 \msun\ for the
primary (MT91~021a), an inclination of $i$=85\degr\ implies
a secondary (MT91~021b) mass of 2.2 \msun, while $i$=15\degr\ 
requires the secondary (MT91~021b) mass to approach that of
the primary.  This system is interesting for having such a
high eccentricity despite  its short period. However, the
eccentricity is  strongly dependent on just a few data
points, and a solution with  fixed eccentricity of zero
still yields a similar $\chi^2$ value.

Finally, we note that the reduced $\chi^2$ of the fit (3.2)
is large.  The O-C residuals appear to show evidence for
periodic variation at  $\sim$4.3 days with an amplitude near
15 \kms, suggesting the possibility  of this being a triple
system.  This would make the tertiary, MT91~021c, a probable
solar mass star with a minimum mass of 0.5 \msun.  
Additional measurements will be needed to confirm
the periodicity and amplitude of a second velocity component
in this single-lined system.   Inclusion of older (but less
precise) WIYN (2 spectra) and Keck (2 spectra)  measurements
strengthen the evidence for more than one velocity
component in MT91~021.

\begin{figure}
\epsscale{1.0}
\centering
\plotone{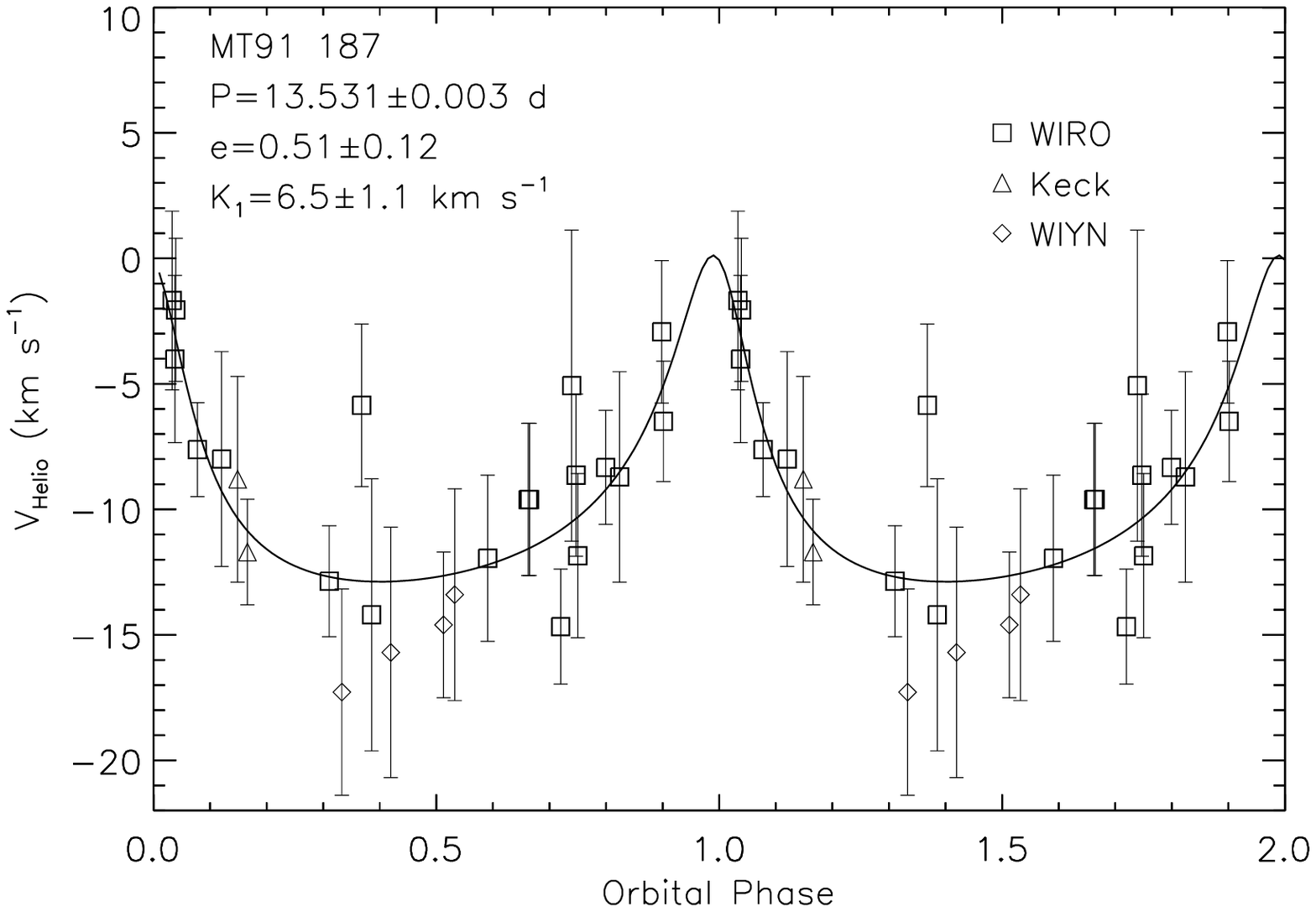}
\caption{Folded radial velocity data and 
best-fitting solution for MT91~187.   
\label{sol187}}
\end{figure}

{\it MT91~187---}The B1~V primary of this V=13.24 system was
not initially detected as a probable binary on the basis of
five measurements \citep{Kiminki07}.  The present dataset
consisting of two Keck spectra, four WIYN spectra, and  19
WIRO spectra spanning 1999--2013  (Table~\ref{vel.tab})
reveals this systems to be a $P$=13.531$\pm$0.002 d binary
with $K_1$=6.5$\pm$1.1 \kms.   Figure~\ref{sol187} shows the
best-fitting orbital solution  (solid line) and folded
velocity curve data (points with error bars).  This is one
of the lowest amplitude systems yet detected in the Survey.
The velocity variability is consistent with a binary
(reduced $\chi^2$=0.82 for 12 degrees of freedom) but not a
constant-velocity system (reduced $\chi^2$=1.8 for 19
degrees of freedom, yielding a probability of $<$2\%). The
eccentricity is not well constrained at $e$=0.51$\pm$0.12. 
An alternative period of $P\simeq$6.1 d is also possible,
but less probable based on the power spectrum and the higher
eccentricity required in the best fit.   There are no known
eclipses from this system, so the inclination  is
essentially unconstrained.   If we adopt a mass of 14 \msun\
for the B1V primary, an inclination of $i$=85\degr\ implies
a secondary (MT91~187b) mass of 0.43 \msun, while
$i$=3\degr\ allows a secondary mass approaching that of the
primary.  Given the probable low-mass secondary and extreme
mass ratio, this system is a candidate for being a
progenitor of a low-mass  X-ray binary system if the system
remains bound once the primary reaches its end-state as a
neutron star.  

\begin{figure}
\epsscale{1.0}
\centering
\plotone{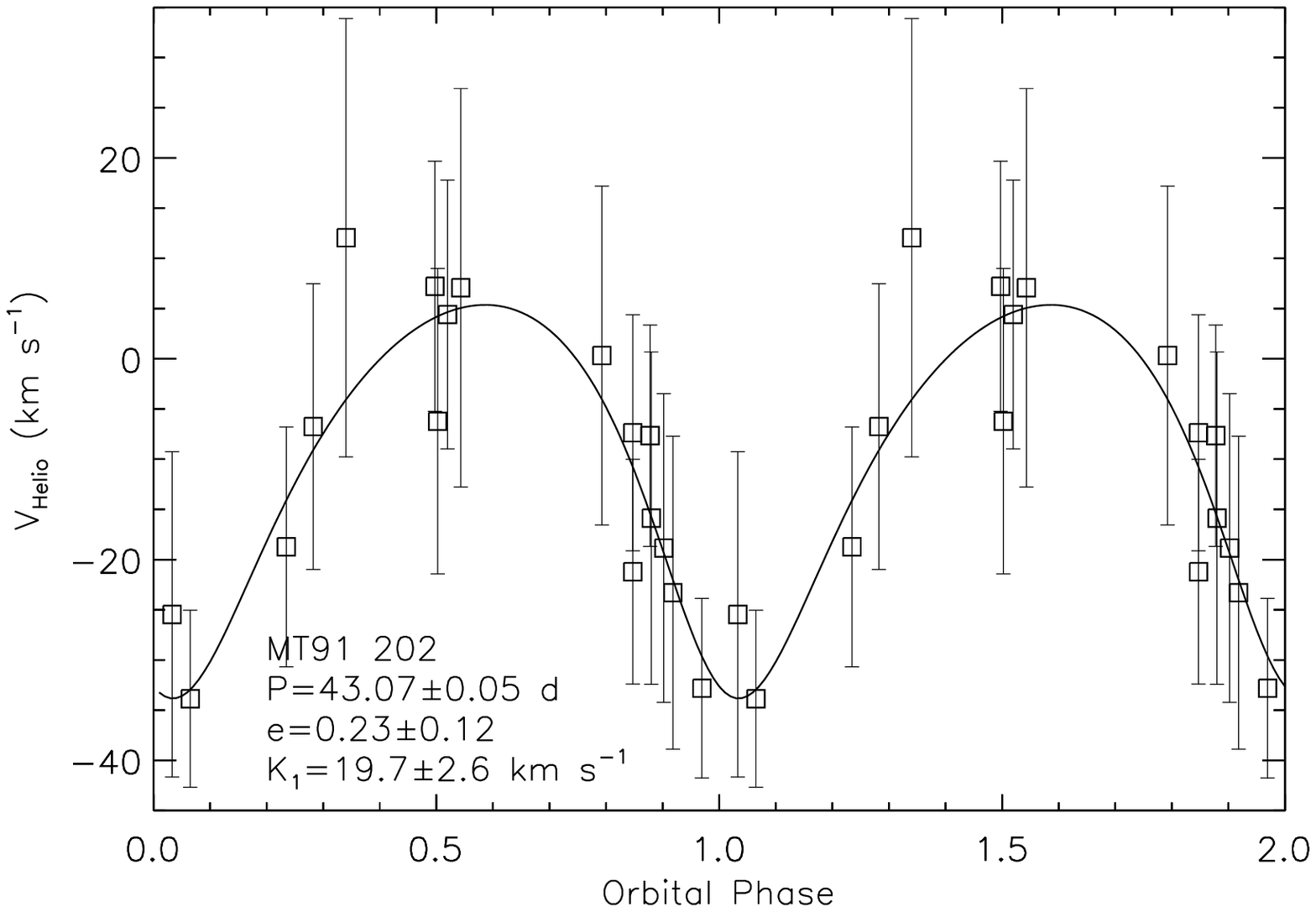}
\caption{Folded radial velocity data and 
best-fitting solution for MT91~202.   
\label{sol202}}
\end{figure}

{\it MT91~202---}This $V$=14.40 binary is listed as an SB1
by \citet{Kiminki07}  on the basis of five data points, and
they type the primary as B2V.  We find an orbital period of
$P$=43.07$\pm$0.05 d with no credible aliases and an
amplitude of $K_1$=19.7$\pm$2.6 km s$^{-1}$ using 17
measurements from WIRO.   The eccentricity is 
$e$=0.23$\pm$0.11. Adopting 11 \msun\
for the primary yields $M_2$=2.0 \msun\ if $i$=90\degr. The
lack of known eclipses does not provide useful constraints
on the inclination given the relatively long period.  The
mass of the secondary (MT91~202b) approaches  that of the
primary if $i$=15\degr.  These rough mass constraints mean
that the secondary spectral type probably lies in the range
A to mid-B, assuming a main-sequence star.  Although there
are hints of line width variations that may suggest a
double-lined system, the low SNR of our spectra on this
faint star coupled with the broad linewidth and small
velocity amplitude means that no attempt was made to
separate components.  Figure~\ref{sol202} shows the
best-fitting orbital solution and folded
velocity data.

{\it MT91~234---}This V=13.25 system is dominated by its
B1.5V (revised from B2V in Paper I)
primary star that exhibits long-term radial velocity
variations.  The solution required a combination of data
spanning the whole range of  the Cygnus OB2 Radial Velocity
Survey from 1999 October using Keck+HIRES
 (two measurements) through 2001 August--2008
June using WIYN+Hydra (10 measurements),
and 2011--2013 October using WIRO (14 measurements).  

\begin{figure}
\epsscale{1.0}
\centering
\plotone{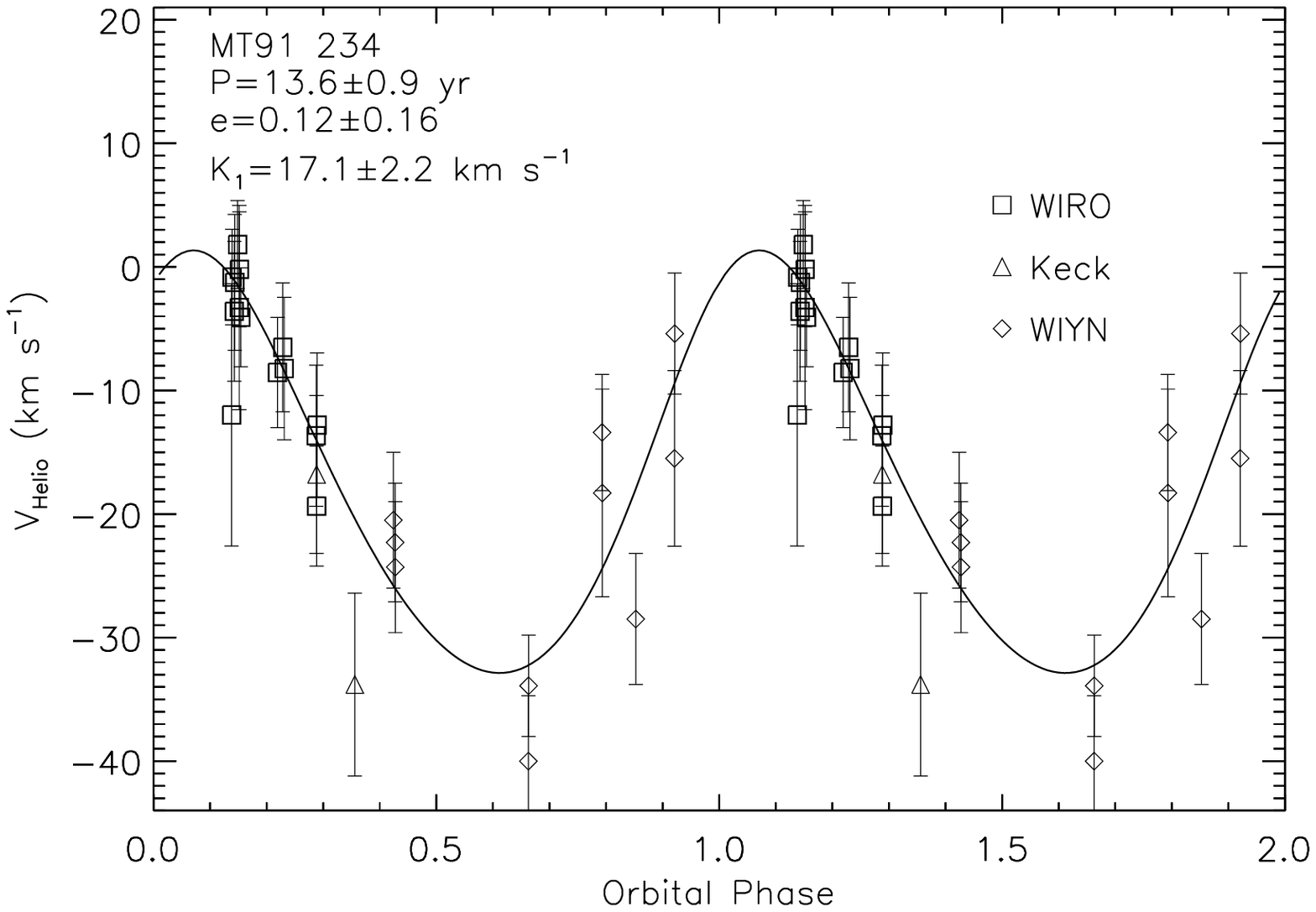}
\caption{Folded radial velocity data and best-fitting solution for MT91~234.   
\label{sol234}}
\end{figure}

Figure~\ref{sol234} shows the best-fitting orbital solution 
and folded velocity data.  Symbol types designate data from 
Keck (triangles), WIYN (diamonds), and WIRO (squares).   The
period of 13.6$\pm$0.9 years makes this the longest period
system yet uncovered in Cyg~OB2.  The eccentricity is small
(0.12$\pm$0.16) and  consistent with zero.  A velocity
amplitude of $K_1$=17.1$\pm$2.2 \kms\ is large for such a
long-period system and implies a fairly massive companion.  
Adopting a primary mass of 12 \msun\ yields a minimum
secondary (MT91~234b) mass of 11 \msun\ for $i=90$\degr.
This suggests that the system is seen nearly edge-on and
that the secondary is also a B star.  

\begin{figure}
\epsscale{1.0}
\centering
\plotone{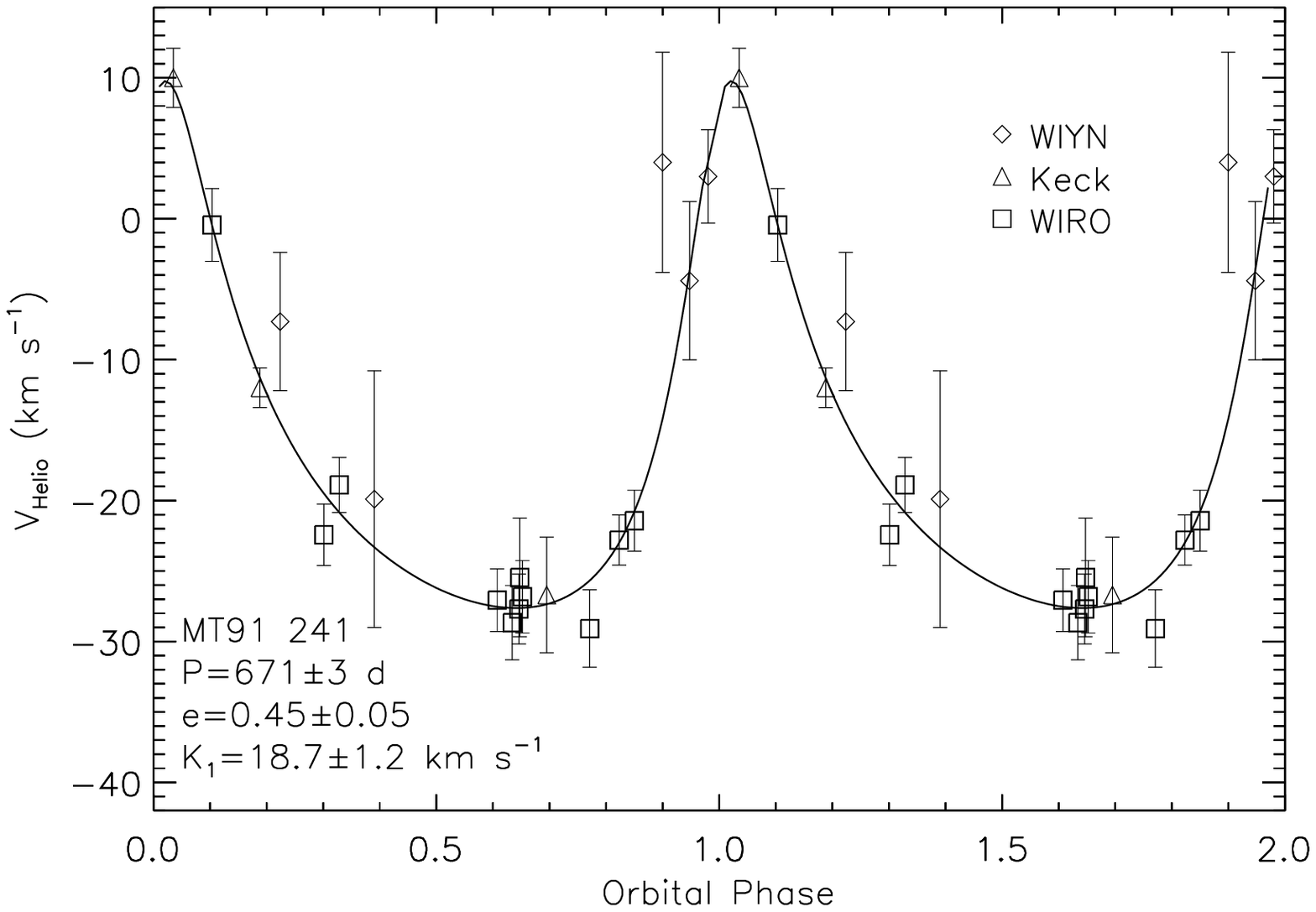}
\caption{Folded radial velocity data and best-fitting solution for MT91~241.   
\label{sol241}}
\end{figure}

{\it MT91~241---}A B1.5V (revised here from B2V in Paper I) 
primary star, the 11 WIRO data
combined with three Keck and five WIYN measurements show
that MT91~241a has a period of 671$\pm$2 d (1.83 yr) and
semi-amplitude 18.7$\pm$1.2 \kms\ with eccentricity of
0.45$\pm$0.06.  Figure~\ref{sol241} shows the best-fitting
orbital solution and folded velocity  data.  Adopting a
primary mass of 12 \msun, the implied secondary mass
(MT91~241b) is 5.2 \msun\ for $i$=90\degr.  The very narrow
\ion{He}{1} lines show no sign of a second component, so it seems
likely that the secondary star mass is closer to this lower
limit than to the primary mass, implying a  mass ratio in
the range 0.42$<q\lesssim$0.8.  

\begin{figure}
\epsscale{1.0}
\centering
\plotone{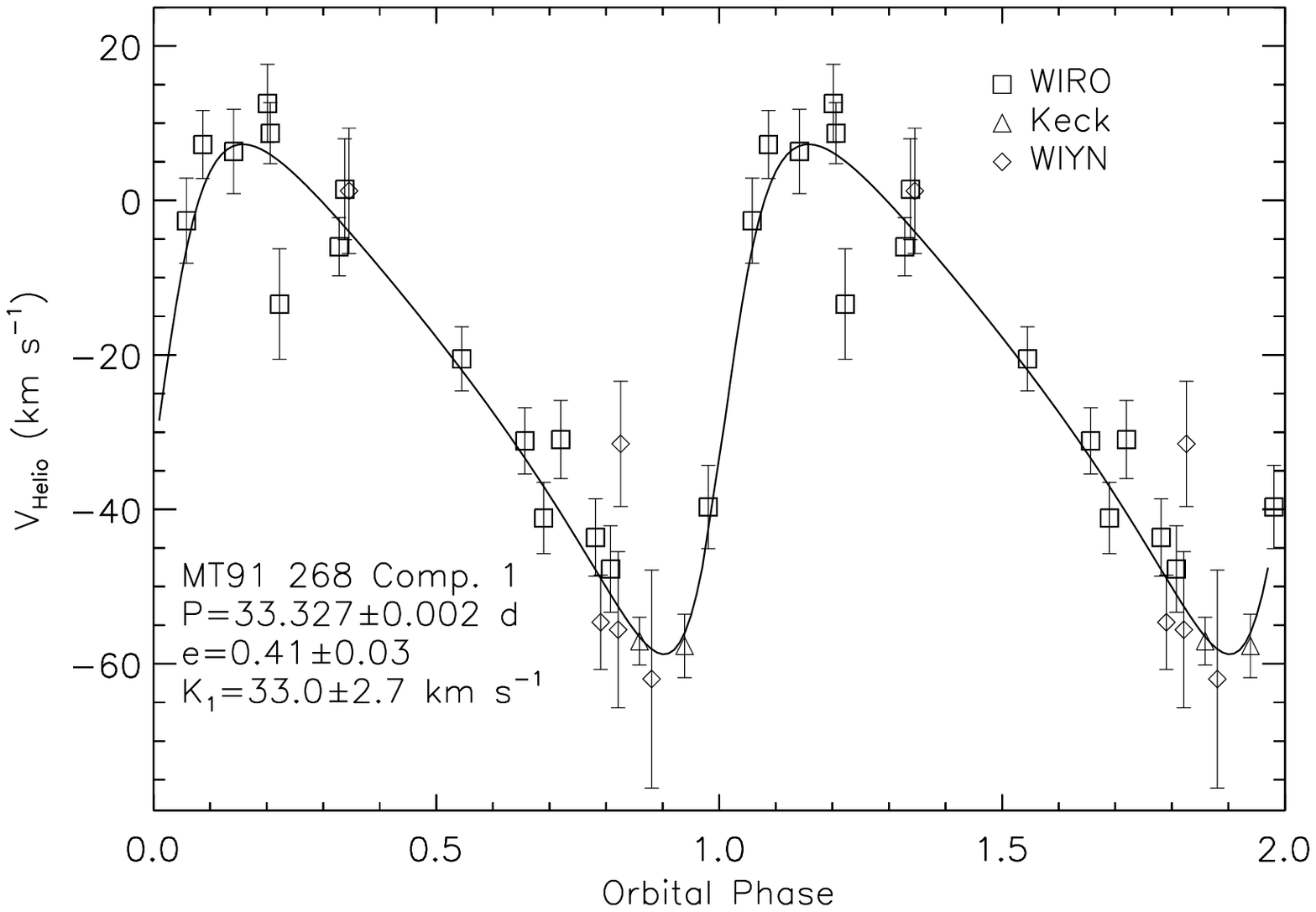}
\caption{Folded radial velocity data and best-fitting solution for the single-lined 
triple system MT91~268, Component 1.  
\label{sol268a}}
\end{figure}

\begin{figure}
\epsscale{1.0}
\centering
\plotone{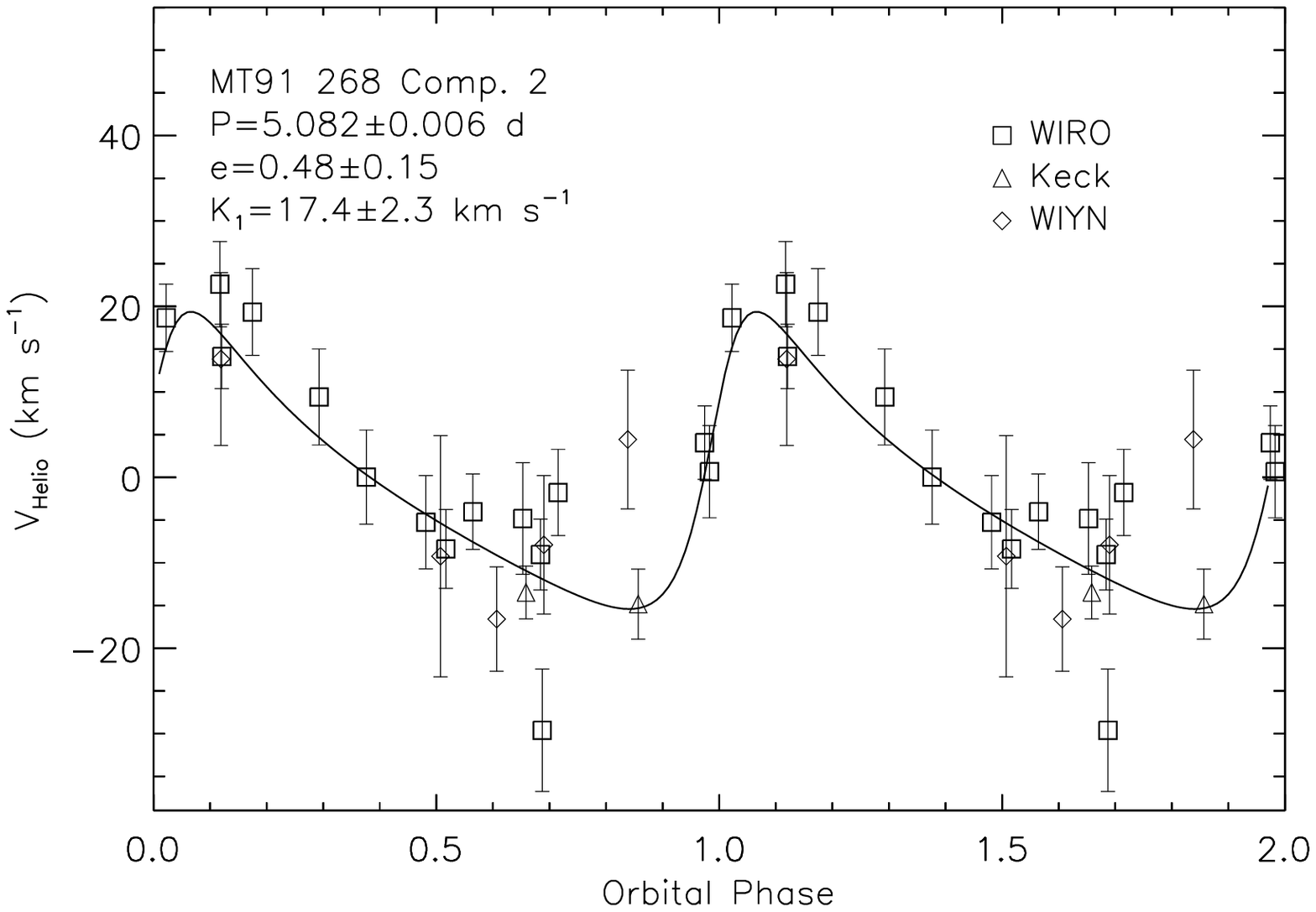}
\caption{Folded radial velocity data and best-fitting solution for the single-lined 
triple system MT91~268, Component 2.   
\label{sol268b}}
\end{figure}

{\it MT91~268---}A B2V (revised from B2.5V in Paper I), 
this single-lined shows
unambiguous evidence for being a triple system.  The initial
single-component solution  at a period near 33.2 d 
displayed systematic residuals with a period  near 5.0
days.  We used these initial guesses to fit a joint orbital
solution for two velocity components using two data from
Keck, five data from WIYN, and 15 data from WIRO between
1999 and 2013.   Our best-fitting solution consists of the
first component having $P_1$=33.327$\pm$0.002 d,
$e_1$=0.41$\pm$0.03, $K_{1-comp1}$=33.0$\pm$2.7 \kms, and
the second component having $P_2$=5.082$\pm$0.006 d,
$e_2$=0.48$\pm$0.15, and $K_{1-comp2}$=17.4$\pm$2.3 \kms.  
Figures~\ref{sol268a} and \ref{sol268b} display the  orbital
solution and folded velocity data for each component.  The
existence of several outlier measurements in both plots is
consistent with the hypothesis of this being a
quadruple system.  

Adopting 11 \msun\ for the mass of the primary star
MT91~268a, the implied minimum masses for the unseen
companions are $M_2\geq$2.9 \msun\ for the 33.2-day component 1
(MT91~268b) and $M_3\geq$0.7 \msun\ for the 5.08-day component
2 (MT91~268c).  This allows that the companions are probable
A--mid-B and G--B dwarfs, respectively.  Given the 
moderate eccentricity, short period, and extreme mass ratio, 
this may be a dynamically unstable system.  

\begin{figure}
\epsscale{1.0}
\centering
\plotone{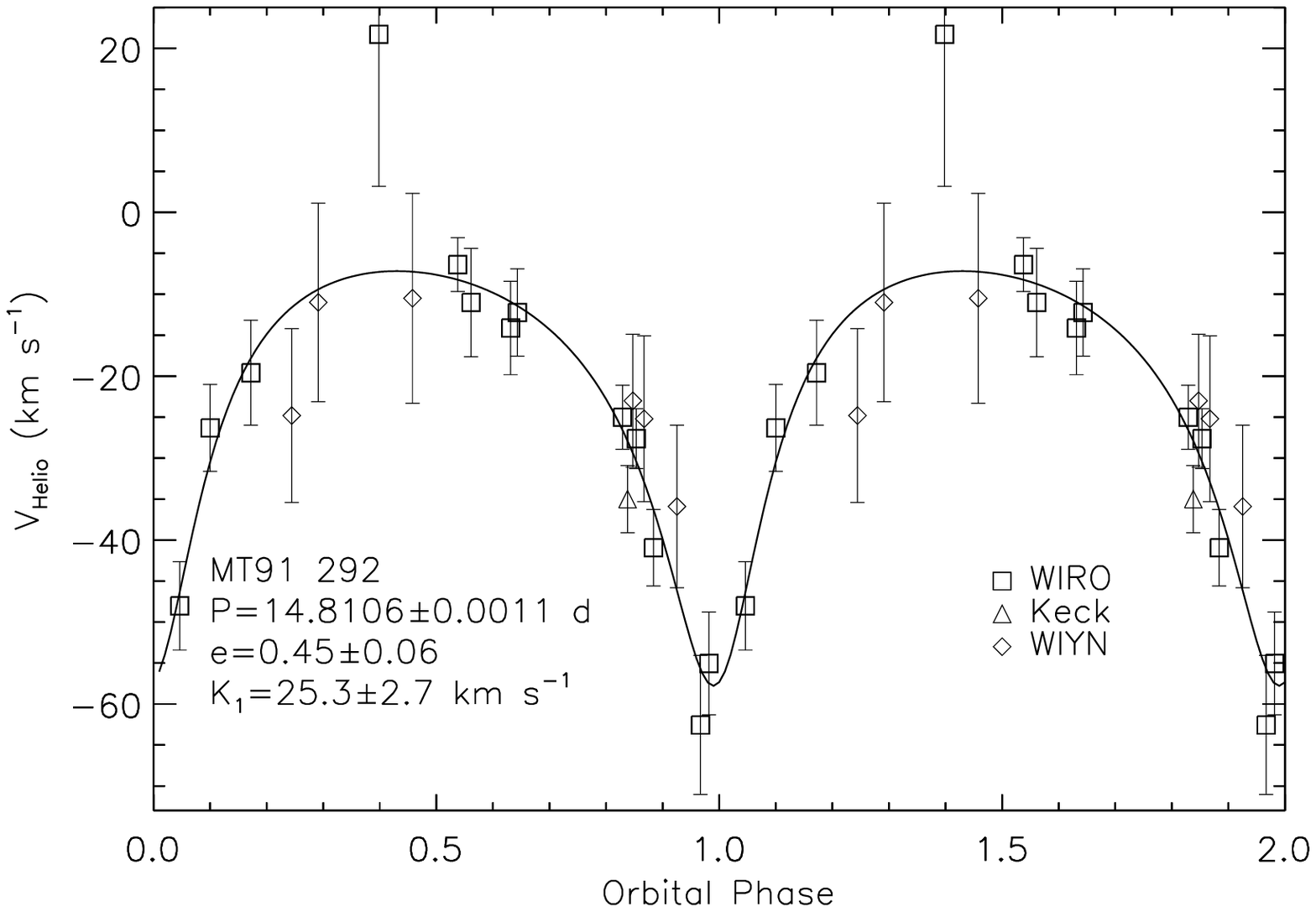}
\caption{Folded radial velocity data and best-fitting solution for MT91~292.   
\label{sol292}}
\end{figure}

{\it MT91~292---}With a period of 14.811$\pm$0.001 d,
this B2V system lies just above the 7--14 day period gap
noticed by \citet{Kiminki2012b}.  The velocity
semi-amplitude of $K_1$= 25.3$\pm$2.6 \kms\ is well-defined by the
combination of 1 Keck data point, six WIYN measurements, 
and 13 WIRO measurements spanning nearly the entire length
of the 2000--2013 survey period.   Adopting a primary mass
of 11 \msun, an inclination of $i$=90\degr\ implies a
minimal secondary mass for MT292b of $M_2$=1.6 \msun.  The
secondary mass approaches that of the primary if
$i\simeq$12\degr. Figure~\ref{sol292} shows the best-fitting 
 orbital solution and folded velocity data.

\begin{figure}
\epsscale{1.0}
\centering
\plotone{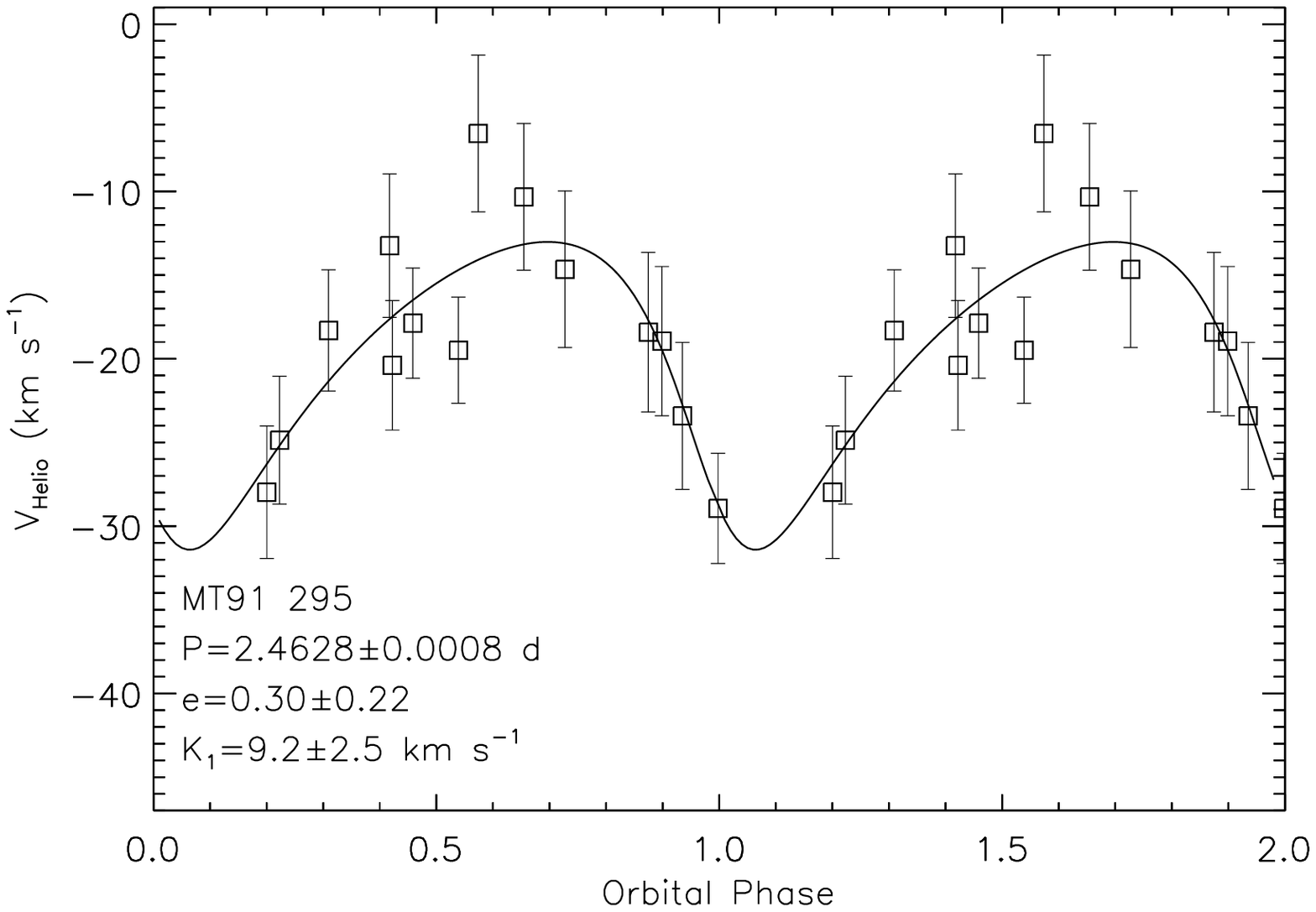}
\caption{Folded radial velocity data and best-fitting solution for MT91~295.   
\label{sol295}}
\end{figure}

{\it MT91~295---}Fourteen measurements from WIRO during 2012
and 2013 indicate that this B1.5V (revised from B2V in Paper
I) has $P$=2.4628$\pm$0.0008 d and $e$=0.30$\pm$0.22.  This
is an unusually large eccentricity for such a short-period
system, but solutions with zero eccentricity yielded
substantially poorer fits.  The semi-amplitude of
$K_1$=9.2$\pm$2.4 \kms\ is small but detectable given the
typical uncertainty of 3.9 \kms.   The power spectrum has
multiple peaks consistent with this being a triple system
having components with periods of 4.67 d and 1.68 d and
velocity semi-amplitudes near 4--6 \kms.     The two-component 
solution provides a somewhat better fit, but the
number of free parameters approaches the number of data
points.  In either case, this is a short-period multiple
system.  Figure~\ref{sol295} shows the best-fitting {\it
single-component} orbital solution and folded velocity data.
Adopting 12 \msun\ for the mass of the primary, the implied
lower limit on the secondary (MT91~295b) mass is $M_2$=0.3
\msun.  The inclination is unconstrained, as $M_2$
approaches $M_1$ for $i$=3\degr.  

\begin{figure}
\epsscale{1.0}
\centering
\plotone{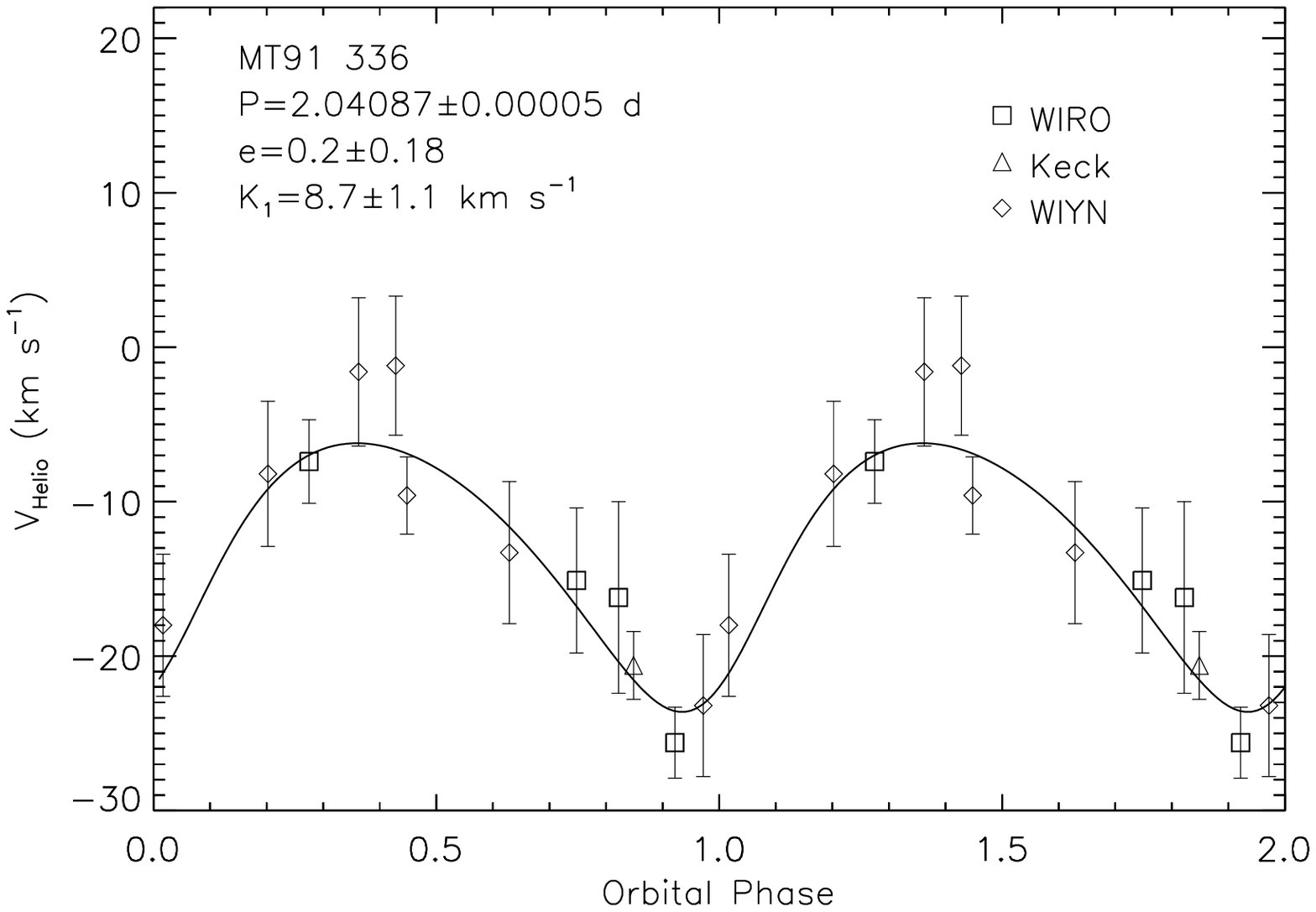}
\caption{Folded radial velocity data and best-fitting solution for MT91~336.   
\label{sol336}}
\end{figure}

{\it MT91~336---}This narrow-lined B2V (revised slightly
from B3III in Paper I)  has one observation from Keck, 
seven from WIYN, and four from WIRO spanning 1999 -- 2014. 
The twelve data points yield a period of 2.04087$\pm$0.00006 with
amplitude $K_1$=8.8$\pm$1.1 \kms\ and 
eccentricity of $e$=0.21$\pm$0.18.  There is a
possible alias near 4.24 days. Figure~\ref{sol336} shows the
best-fitting orbital solution and folded velocity data.
Adopting 11 \msun\ for the mass of the primary, the implied
lower limit on the secondary (MT91~336b) mass is $M_2$=0.3
\msun.  The inclination is unconstrained, as $M_2$
approaches $M_1$ for $i$=3\degr.  

\begin{figure}
\epsscale{1.0}
\centering
\plotone{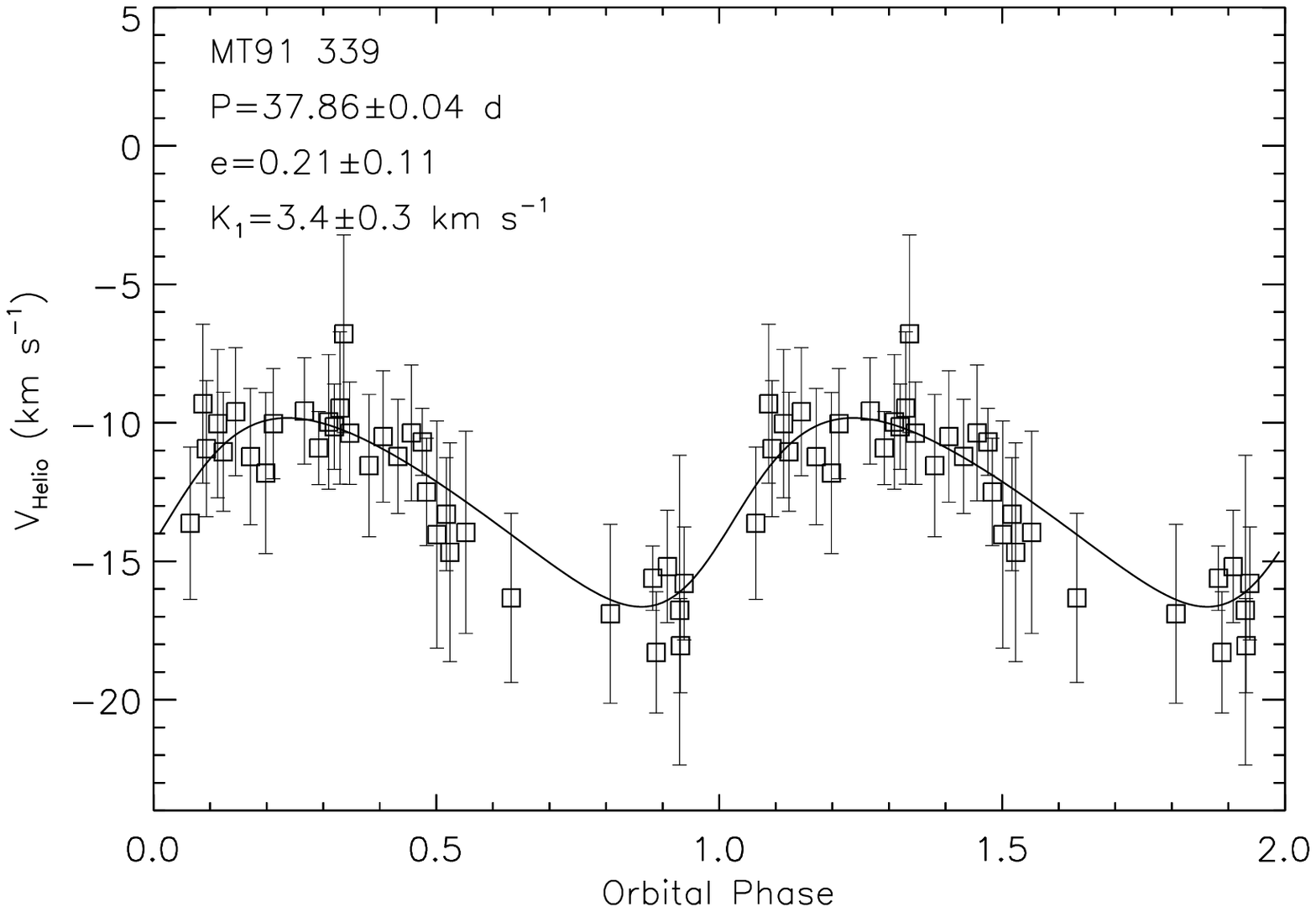}
\caption{Folded radial velocity data and best-fitting solution for MT91~339.   
\label{sol339}}
\end{figure}

{\it MT91~339---}With a velocity amplitude of just
$K_1$=3.4$\pm$0.3 \kms\, this  O8V primary has one of the
smallest amplitudes of any in our survey.   The narrow, deep
He lines allow for a high level of precision, leading to
typical velocity uncertainties of $\sim$2 \kms\ on the 34
WIRO spectra spanning 2008--2013.  The orbital period is
$P$=37.86$\pm$0.04 d with eccentricity 0.21$\pm$0.11.   A
Monte Carlo simulation wherein the  Julian dates are
shuffled randomly among the observed velocities shows that
in only 9\%  of the iterations does the peak in the power
spectrum exceed the observed one. We conclude that the
probability of a real periodicity is high.  
Figure~\ref{sol339} shows the best-fitting orbital solution 
and folded velocity data. The orbital inclination is
unconstrained, allowing for secondary (MT91~339b) masses
from 0.5 \msun\ to 21 \msun.  

\begin{figure}
\epsscale{1.0}
\centering
\plotone{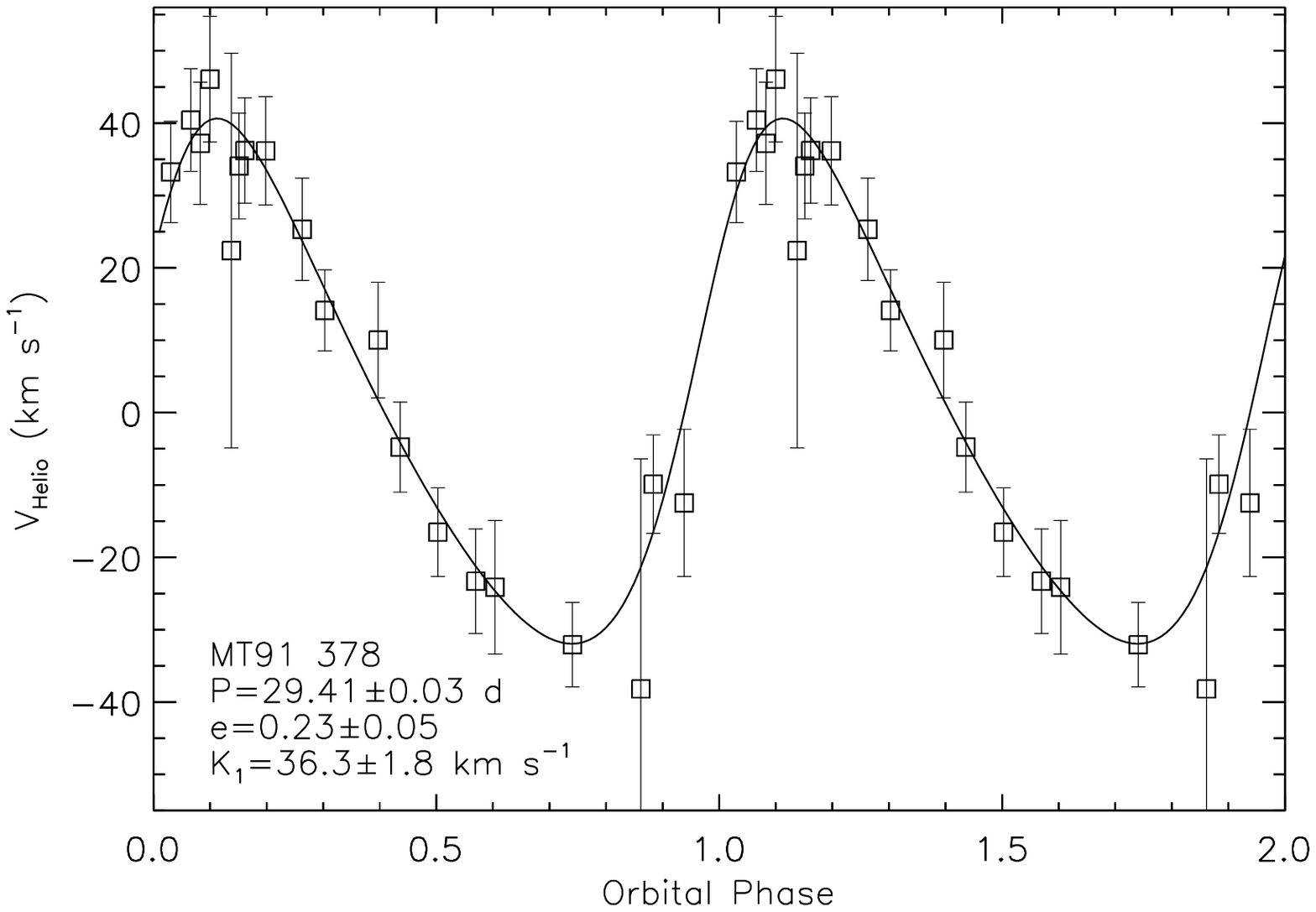}
\caption{Folded radial velocity data and best-fitting solution for MT91~378.   
\label{sol378}}
\end{figure}

{\it MT91~378---}The V=13.49 B0V primary shows a velocity
amplitude of $K_1$=36.3$\pm$1.9 \kms\ and a period of
29.41$\pm$0.03 d.  Figure~\ref{sol378} shows the
best-fitting orbital solution and folded velocity  data.
Although there is some hint of variable line widths
suggestive of a double-lined system, the limited SNR of the
spectra preclude attempts at spectral deblending using the
19 data obtained at WIRO between 2010 and 2013.  For an
adopted primary mass of 18 \msun, an inclination of
$i=$90\degr\  yields $M_2$=4.1 \msun.   The secondary mass
approaches the primary mass if $i$=17\degr.    Hence, the
secondary (MT91~378b)  is limited to be a B spectral type,
assuming it is on the main sequence.  

\begin{figure}
\epsscale{1.0}
\centering
\plotone{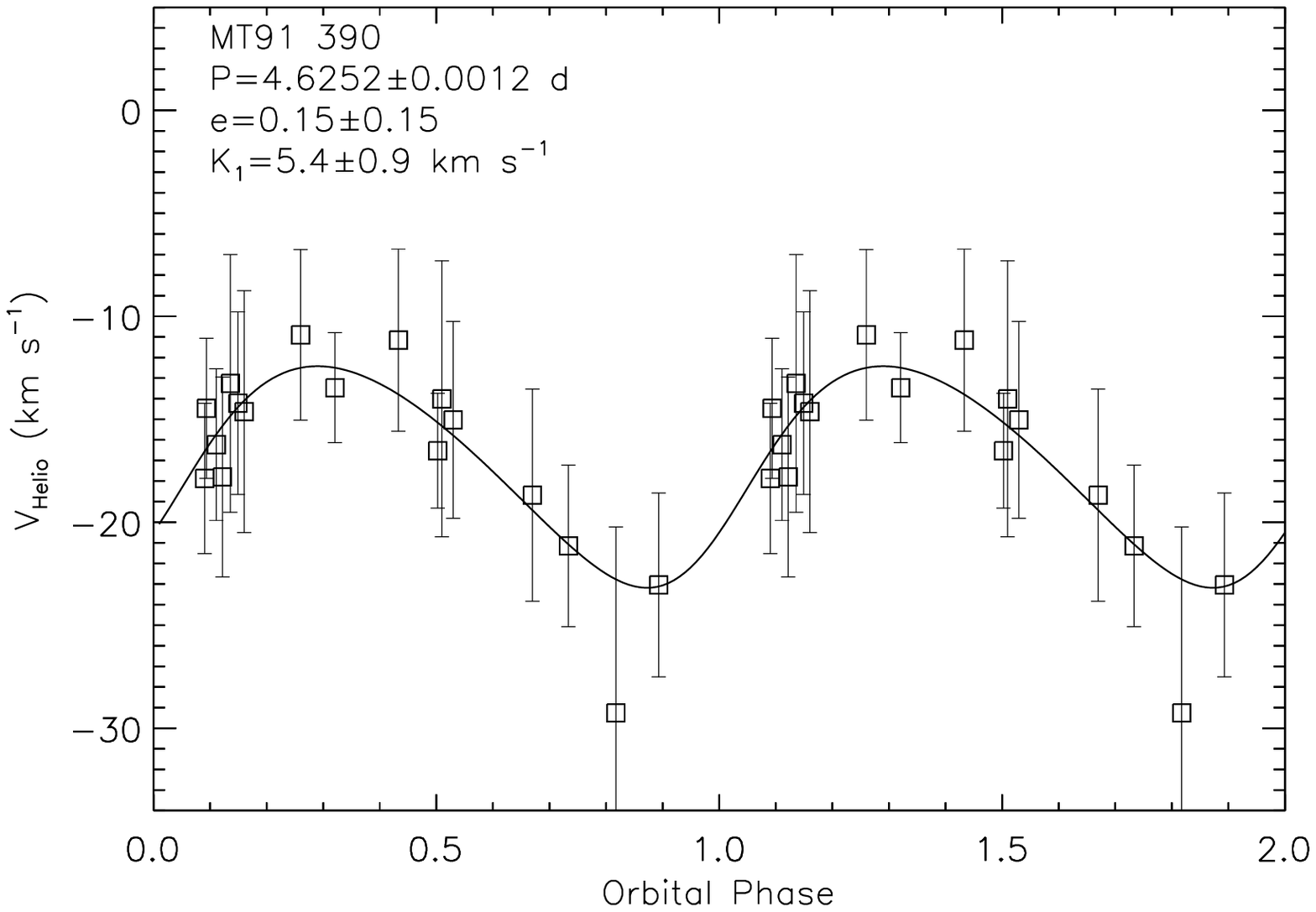}
\caption{Folded radial velocity data and best-fitting solution for MT91~390.   
\label{sol390}}
\end{figure}

{\it MT91~390---}With a velocity amplitude of
$K_1$=5.4$\pm$0.9 \kms, this O8V primary is among the
lowest-amplitude systems observed in the survey.   The
period of $P$=4.625$\pm$0.001 d using 17 radial velocity
measurements from WIRO between 2011 June and 2013 August
places MT91~390 among the conspicuous group of short-period
binaries.   There are no known eclipses, so the inclination
is unconstrained.  For $i$=90\degr\ the implied secondary
mass is $M_2$=0.32 \msun, while for $i$=2\degr\ the
secondary (MT91~390b) mass approaches that of the primary.  
Figure~\ref{sol390} shows the best-fitting orbital solution 
 and folded velocity data.

\begin{figure}
\epsscale{1.0}
\centering
\plotone{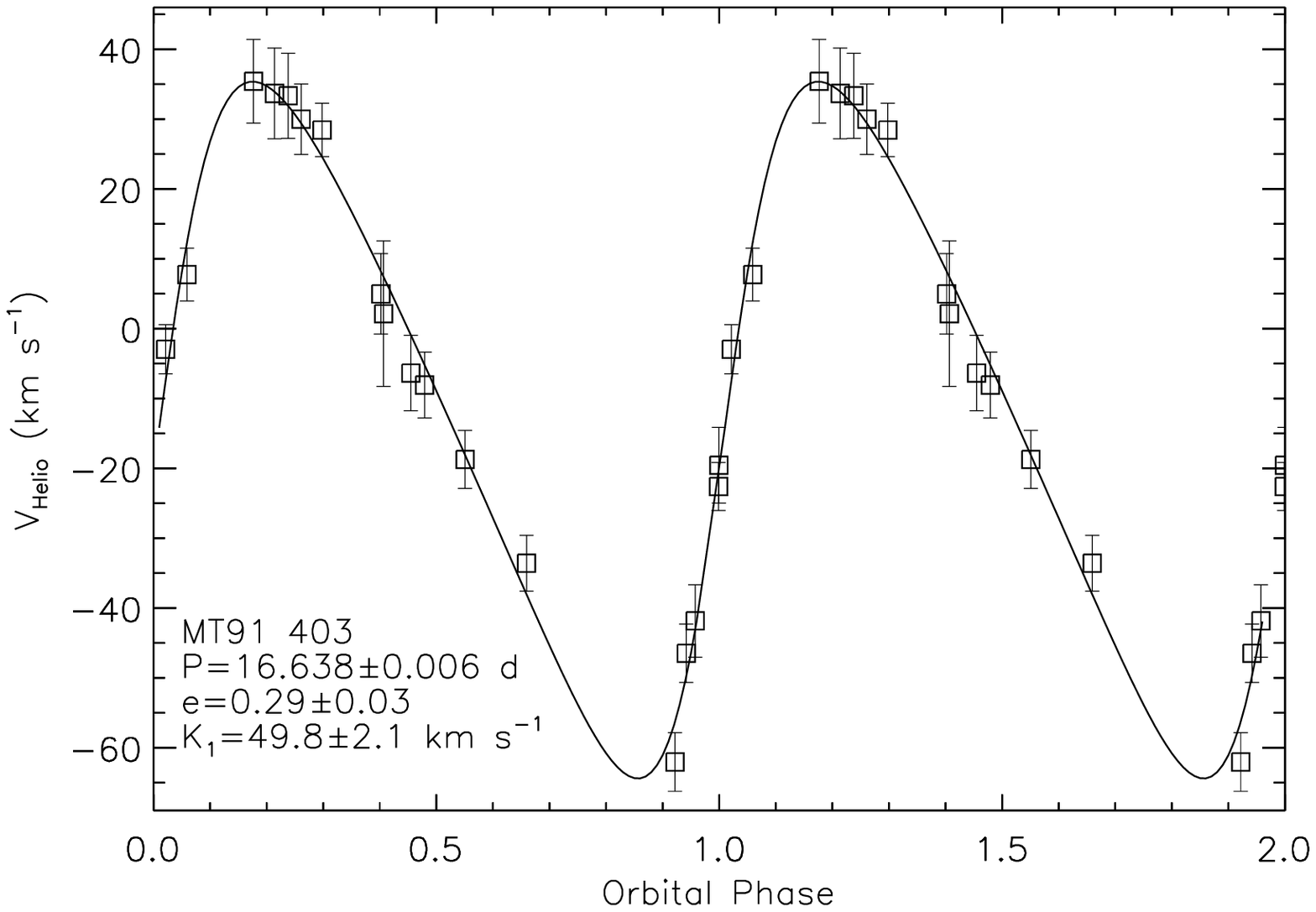}
\caption{Folded radial velocity data and best-fitting solution for MT91~403.   
\label{sol403}}
\end{figure}

{\it MT91~403---}This B1V primary exhibits a velocity
semi-amplitude of $K_1$=49.8$\pm$2.1 \kms\  and a period of
$P$=16.638$\pm$0.006 d using 18 measurements from WIRO. 
With no known eclipses, there are no inclination
constraints.  The adopted primary mass of 14 \msun\ leads to
a secondary mass of M$_2$=4.1 \msun\ for $i$=90\degr\ and
$M_2$=14 \msun\ for $i$=23\degr.  These limits dictate
that the secondary (MT91~403b) is a probable B star.   
Figure~\ref{sol403} shows the best-fitting orbital solution 
and folded velocity curve.

\begin{figure}
\epsscale{1.0}
\centering
\plotone{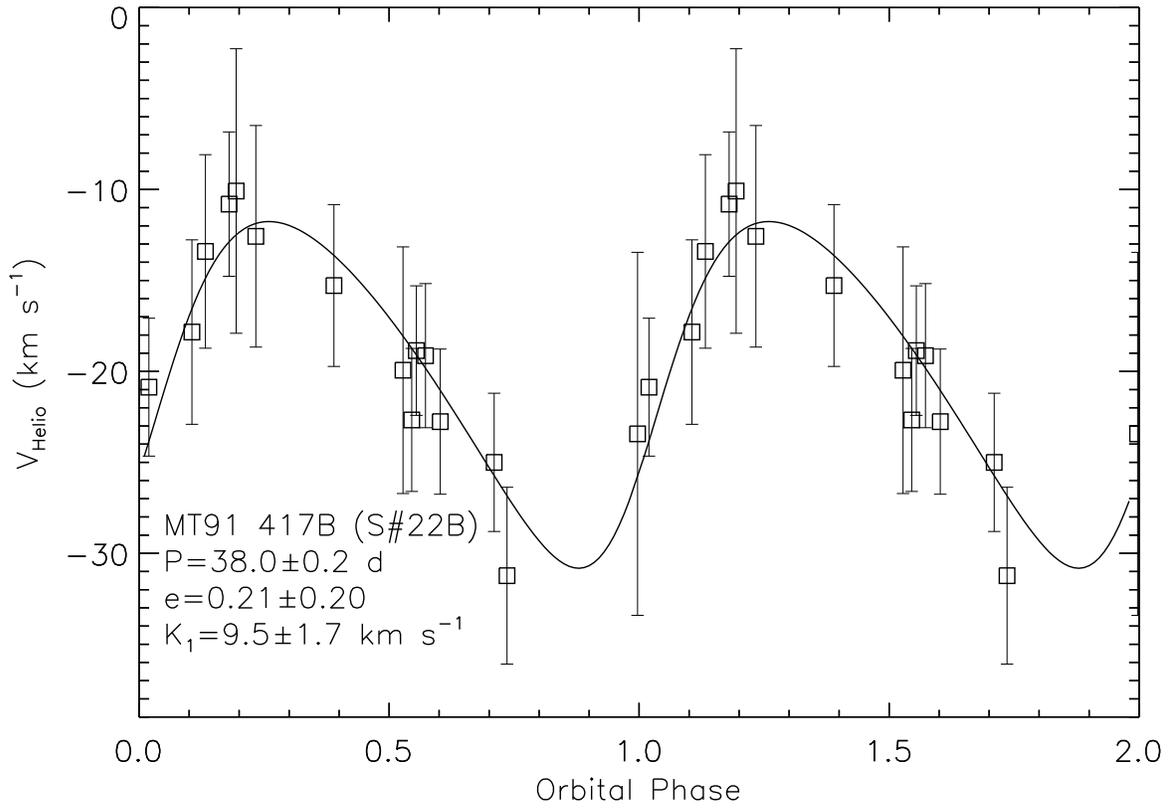}
\caption{Folded radial velocity data and best-fitting solution for MT91~417B (Schulte \#22B).   
\label{sol417b}}
\end{figure}

{\it MT91~417B (Schulte~\#22B)---}MT91~417 is a visual
double consisting of MT91~417A (O3I; the northwest
component) and  MT91~417B (O6V; the southeast component)  at
a separation of 1\farcs5 ($\sim$2100 AU projected separation
at 1.4 kpc) at PA$\simeq$50\degr.    We obtained 14 spectra
of MT91~417B at WIRO in good seeing conditions between 2013
August 6 and 2014 May 26 with the 1\farcs2 slit
oriented perpendicular to the two stars. One additional
measurement from WIRO on 2008 June 26 was used. MT91~417B 
exhibits variability with a period of $P$=38.0$\pm$0.2 d and
$K_1$=9.5$\pm$1.7 \kms.  Figure~\ref{sol417b}  shows the
best-fitting orbital solution  and folded velocity curve. 
The implied secondary (MT91~417B-b) minimum mass is 1.8
\msun\ for $i$=90\degr.   The secondary mass approaches that
of the primary for $i$=5\degr.  The velocity curve is not 
well-sampled at all phases so the derived parameters are
particularly uncertain.  

\begin{figure}
\epsscale{1.0}
\centering
\plotone{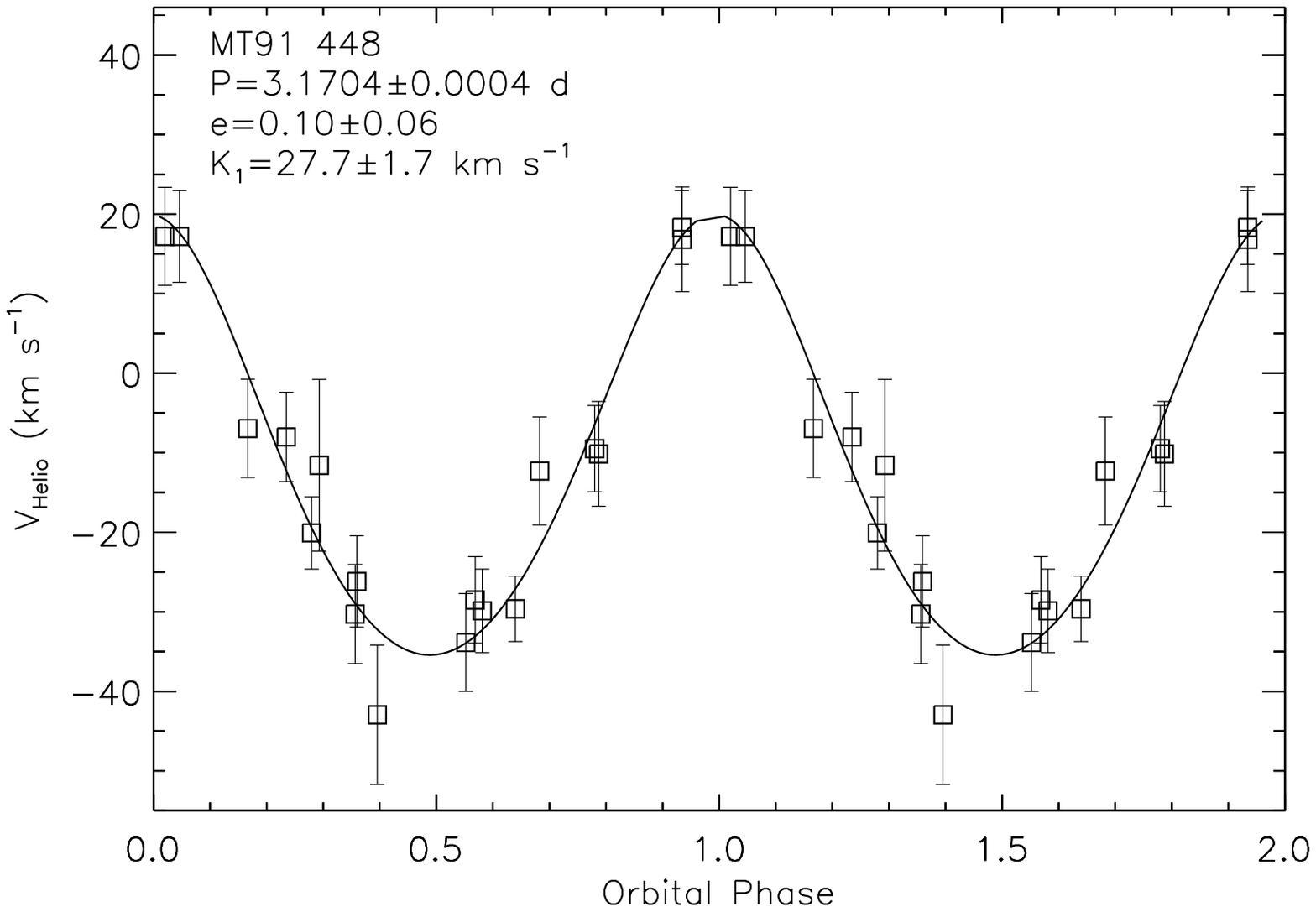}
\caption{Folded radial velocity data and best-fitting solution for MT91~448.   
\label{sol448}}
\end{figure}

{\it MT91~448---}A period of $P$=3.1704$\pm$0.0004 d for
this O6V primary places it among the shortest period systems
in Cyg~OB2.  The 18 WIRO data spanning 2011--2013 yield a
secure velocity semi-amplitude of 27.7$\pm$1.7 \kms\ and a
low eccentricity of $e$=0.10$\pm$0.06.  There are no known
eclipses. An inclination of $i$=90\degr\ yields a minimum
secondary (MT91~448b) mass $M_2$=2.1 \msun, while the
secondary mass approaches that of the primary for
$i$=6\degr.   Figure~\ref{sol448} shows the best-fitting
orbital solution and folded velocity curve data.

\begin{figure}
\epsscale{1.0}
\centering
\plotone{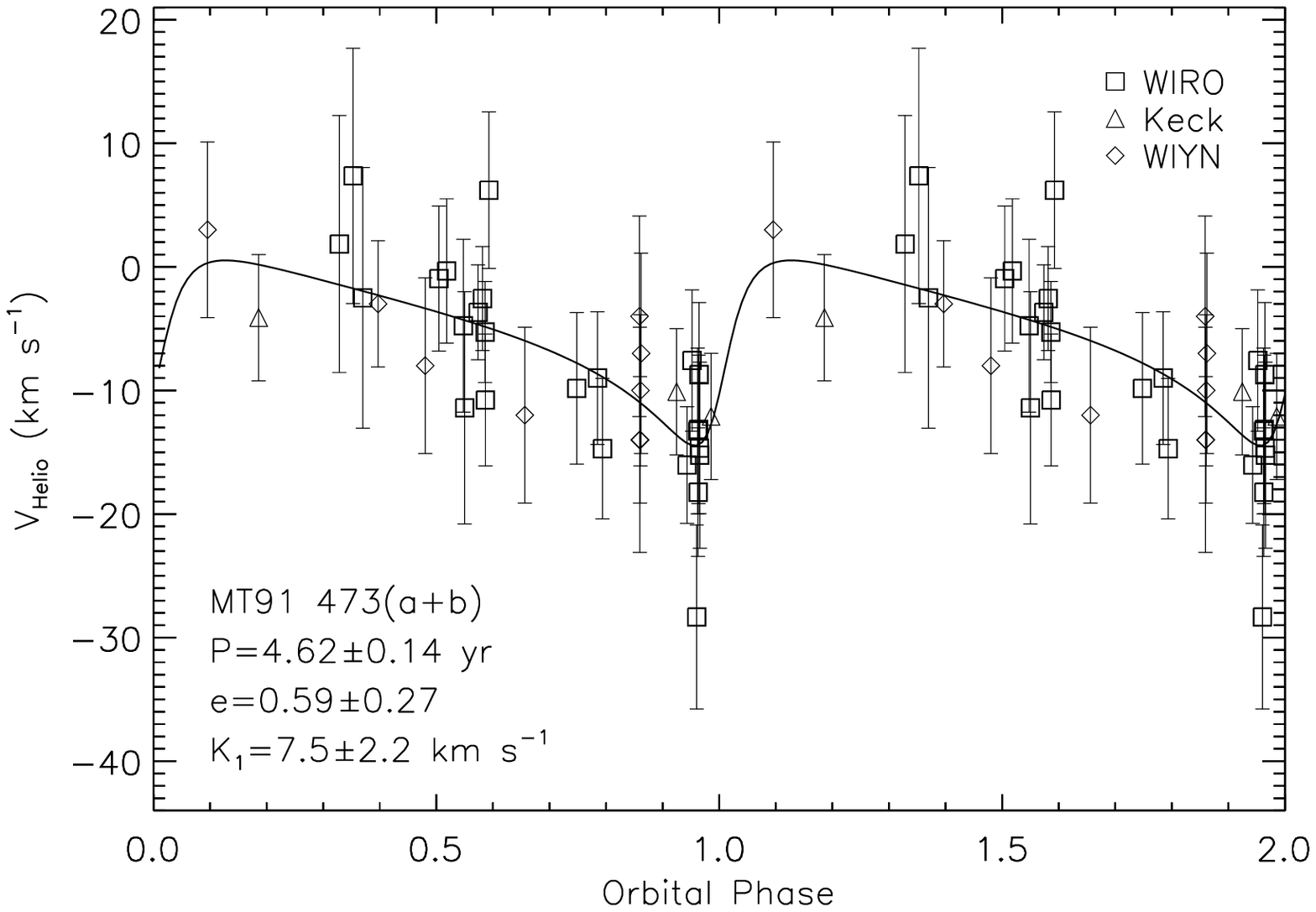}
\caption{Folded radial velocity data and best-fitting solution for MT91~473.   
\label{sol473}}
\end{figure}

{\it MT91~473---}This O8.5V has three measurements from Keck
(1999--2000),  nine from WIYN (2000--2008) and 24 from WIRO
(2010--2013).   The power spectrum is complex, showing
multiple peaks on timescales of 2000 d to 1.9 d.  The
dominant peak at  1700 d yields a folded velocity curve
that appear consistent with  the long-term trends observed
in an unphased velocity curve.   The WIRO data alone suggest
a period in this same vicinity.  A Monte Carlo simulation
wherein the  Julian dates are shuffled randomly among the
observed velocities shows that in only 1.3\%  of the
iterations does the peak in the power spectrum exceed the
observed one. We conclude that the observed periodicity is
highly likely to be real.      Our best-fitting orbital
solution gives a period of $P$=1687$\pm$51 d and amplitude
$K_1$=7.5$\pm$2.2 \kms.  Figure~\ref{sol473} shows the
best-fitting orbital solution and folded velocity 
data.

Smaller peaks in the power spectrum suggest the possibility
of short-period variations on the $\sim$2 d timescale, and a
couple of the KECK HIRES spectra suggest a possibility of a
double-lined system. We fit single Gaussian curves to the
\ion{He}{1} $\lambda$4471 lines in the Keck and WIYN spectra
and the \ion{He}{1} $\lambda$5876 lines in the WIRO
spectra.  The FWHM of the lines varies over the range  3.7
-- 4.4 \AA\ with typical uncertainties of 0.25 \AA.  By
applying a power spectrum analysis to the FWHM values, we
find a peak  at 1.846 d, constituting evidence for a
probable double-lined  system with twice this period, 3.692
d.  The spectra have insufficient SNR and spectral
resolution to attempt a deblending, but we conclude that
MT91~473 likely consists of at least three stars.    In this
scenario, components MT91~473a and MT91~473b would
constitute a tight binary with  a period of about 3.6 d and
similar spectral types near O8.5V.  The inclination  for
this pair must be low, near 10\degr, such  that the line
splitting does not exceed the observed values.    MT91~473c
is the unseen companion responsible for the radial velocity
curve shown in Figure~\ref{sol473}.  Adopting masses of 19 
\msun\ for each of components a and b implies a mass of 5.0
\msun\ for $i$=90\degr, meaning that the secondary must be
at least a mid-B star.   For inclinations as small as
11\degr\ the mass of component c could approach that of
a+b.  

\begin{figure}
\epsscale{1.0}
\centering
\plotone{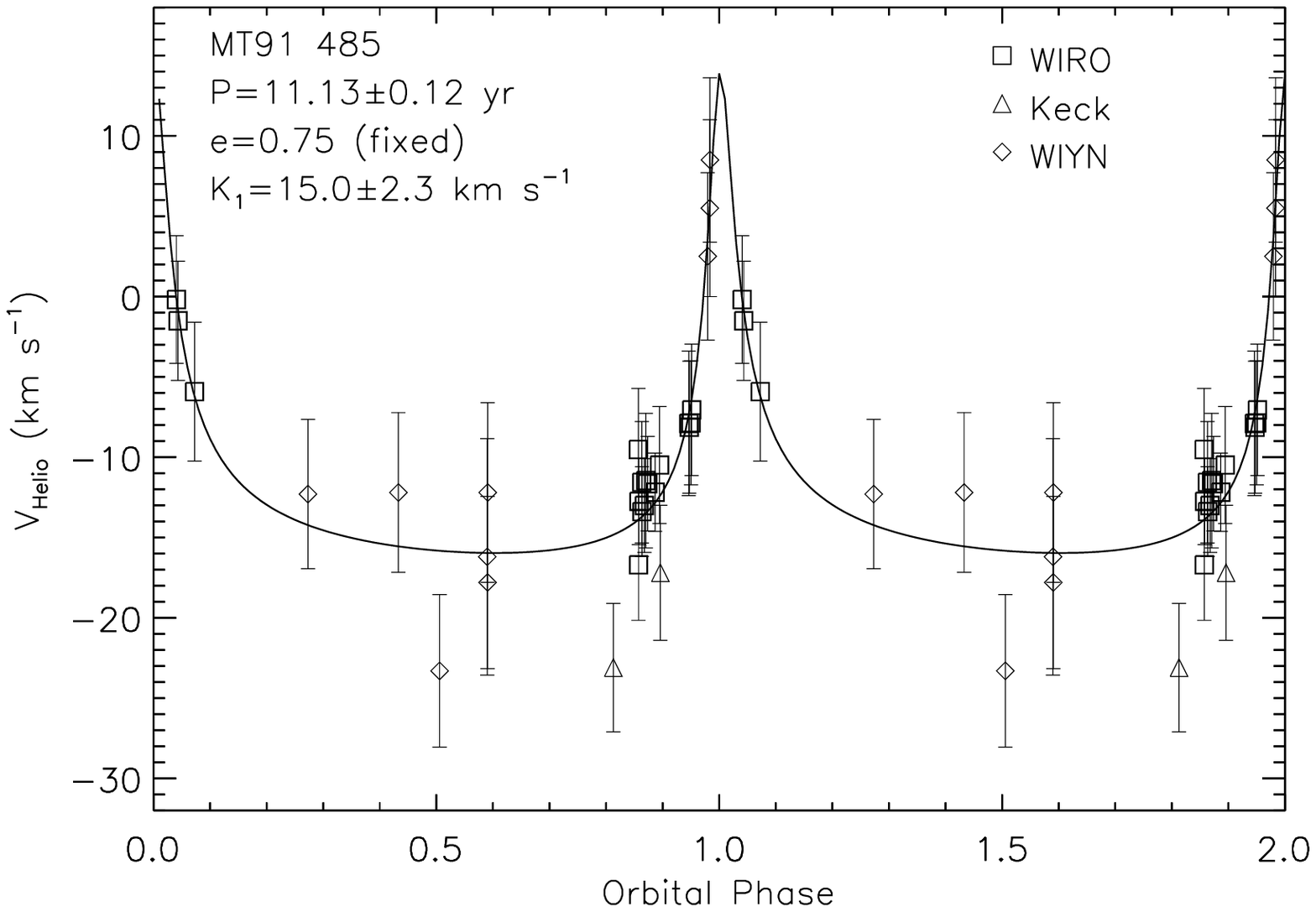}
\caption{Folded radial velocity data and best-fitting solution for MT91~485.   
\label{sol485}}
\end{figure}

{\it MT91~485---} MT91~485 is an O8V showing long-term
radial velocity variations with $P$=4066$\pm$45 d and
semi-amplitude $K_1$=15.0$\pm$2.3 \kms\ based on two Keck
measurements, nine  WIYN measurements, and 17 WIRO
measurements covering the period 2001 August through 2013
October.  We found it necessary to fix the eccentricity 
during fitting to avoid extremely large values.   We estimate
$e$=0.75$\pm$0.20 by manual experimentation.  
Figure~\ref{sol485} shows the best-fitting orbital solution
and folded velocity curve data.  The phase light curve is
essentially the same as the unphased light curve, given that
the period of 11.1 years is nearly as long as the duration
of our survey.  Adopting 21 \msun\ for the primary mass, the
inclination of $i$=90\degr\ leads to a secondary mass of
M$_2$=11.4 \msun. $M_2$ approaches $M_1$ for $i$=40\degr.  
The secondary, MT91~485b, must be more massive than an
early-B dwarf star.  MT91~485 is an example of a highly
eccentric system that could  easily be missed in a
radial velocity survey
if it were not for fortuitous phase coverage.   

\begin{figure}
\epsscale{1.0}
\centering
\plotone{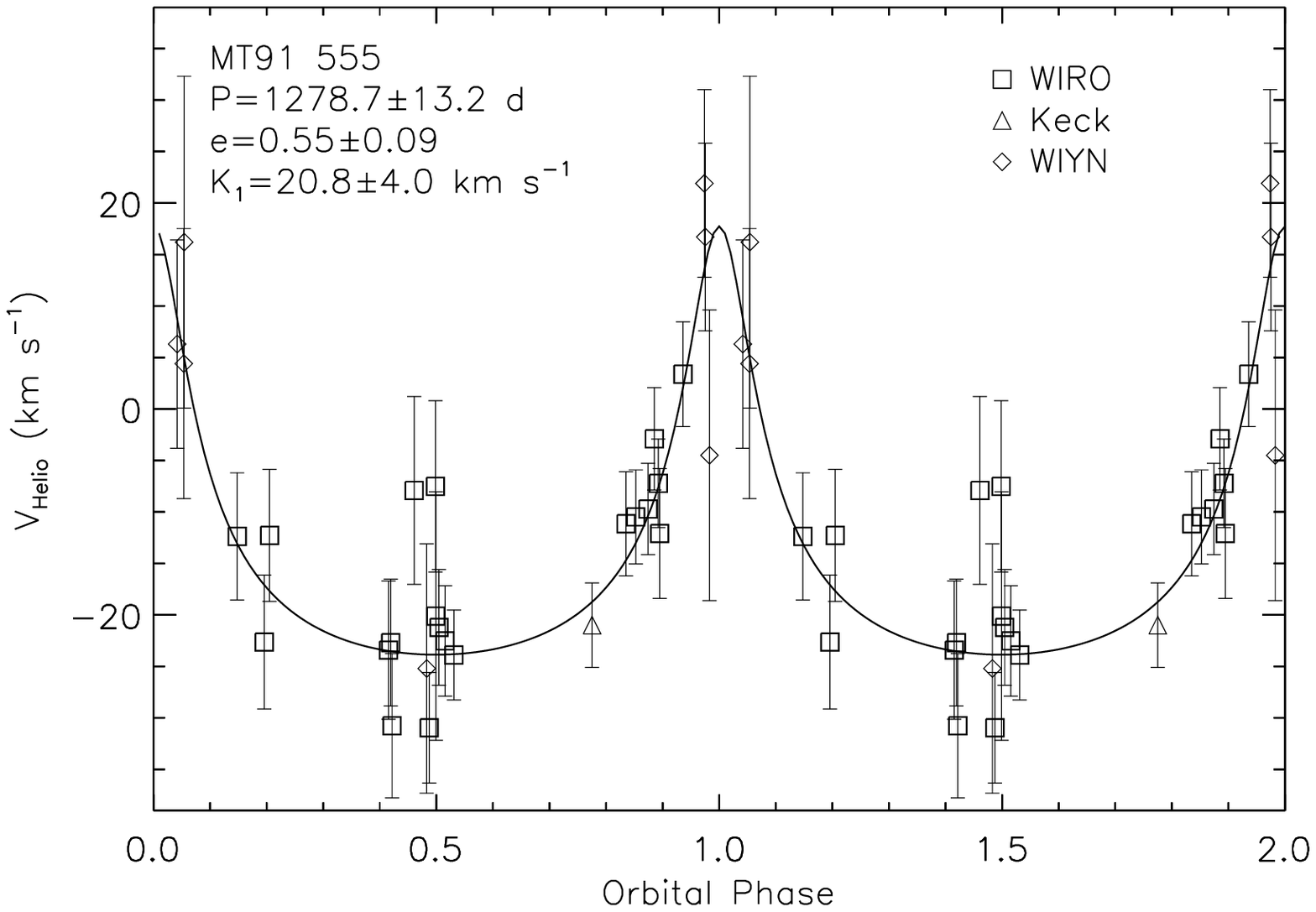}
\caption{Folded radial velocity data and best-fitting solution for MT91~555.   
\label{sol555}}
\end{figure}

{\it MT91~555---}One of the long-term radial velocity
variables in the survey, MT91~555 (O8V) has a period
$P$=1279.5$\pm$13.2 d (3.50 years) and semi-amplitude
$K_1$=20.4$\pm$3.4 \kms. Figure~\ref{sol555}  shows the
best-fitting orbital solution and folded velocity data
using 1 datum from Keck, 7 from WIYN, and 20 from WIRO.  
Adopting a primary mass of 21 \msun\ yields a minimum
secondary mass $M_2$=10.3 \msun, meaning that the secondary
(MT91~555b) is  constrained to be about B2V or earlier.  For
$i$=34\degr\ the secondary mass approaches that of the
primary.

\begin{figure}
\epsscale{1.0}
\centering
\plotone{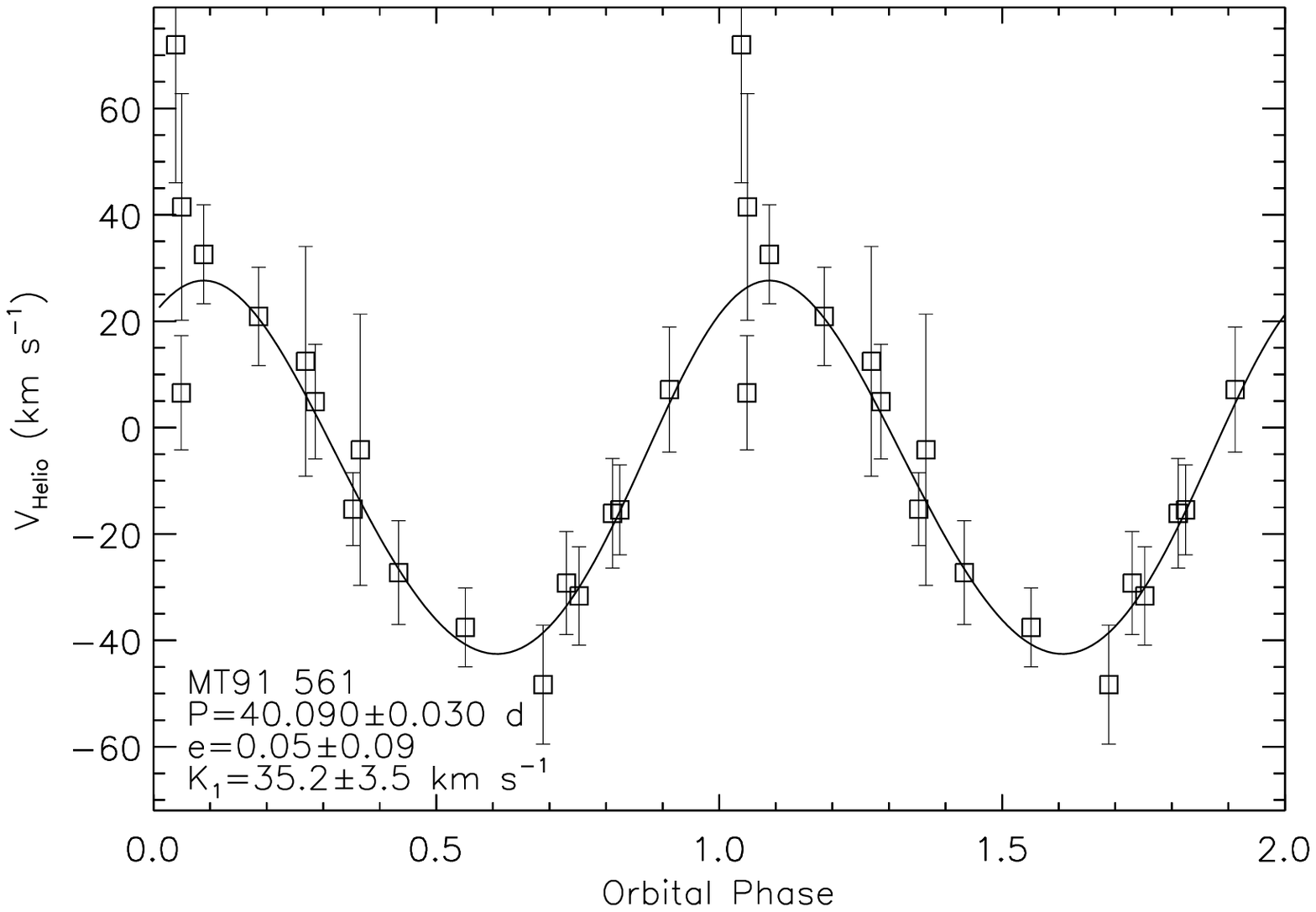}
\caption{Folded radial velocity data and best-fitting solution for MT91~561.   
\label{sol561}}
\end{figure}

{\it MT91~561---}A B2V, the primary star of MT91~561 shows  a
period of $P$=40.09$\pm$0.03 d and a velocity semi-amplitude
of 35.2$\pm$3.6 \kms\ using 17 WIRO data points between 2007 and
2013. Figure~\ref{sol561} shows the best-fitting orbital
solution  and folded velocity curve. For an adopted primary
mass of 11 \msun, the secondary mass is $M_2$=3.3 \msun\ if
$i$=90\degr.  $M_2$ approaches $M_1$ when $i$=23\degr.  This
constrains the  secondary (MT91~561b) to be a least as
massive as a late B main sequence star, and the mass ratio
lies in the range $q$=0.30--1.

\begin{figure}
\epsscale{1.0}
\centering
\plotone{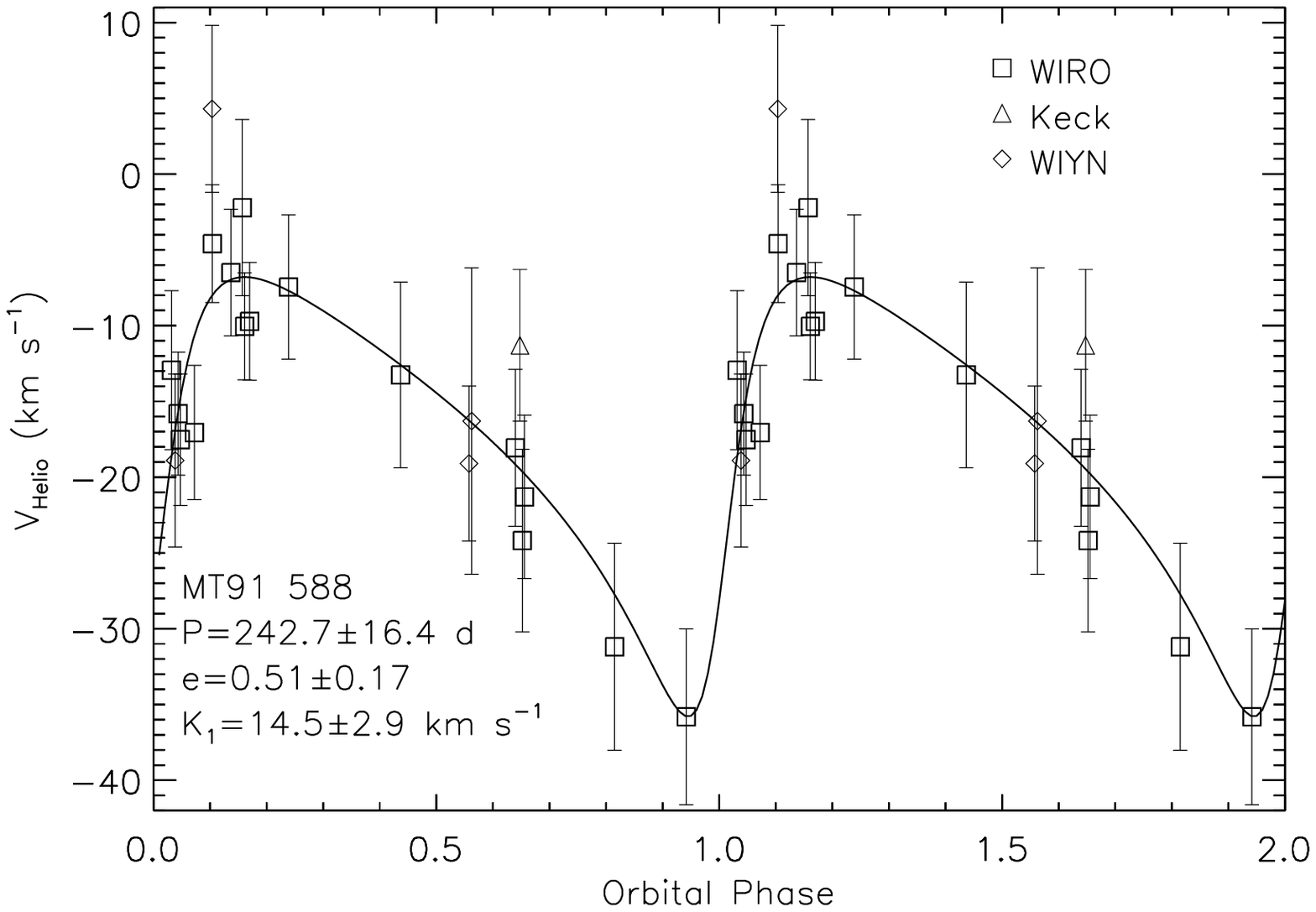}
\caption{Folded radial velocity data and best-fitting solution for MT91~588.   
\label{sol588}}
\end{figure}

{\it MT91~588---}This B0V has a moderately eccentric orbit
with  $P$=245.1$\pm$0.3 d, $e$=0.51$\pm$0.17, and a velocity
semi-amplitude of 14.5$\pm$2.9 \kms\ using one measurement
from Keck, four from WIYN, and 16 from WIRO over the period
1999--2013.  Figure~\ref{sol588} shows the best-fitting
orbital solution  and folded velocity data.  Adopting a
primary mass of 18 \msun\ yields a minimum secondary mass
$M_2$=3.2 \msun, meaning that the secondary (MT91~588b) is 
constrained to be an early A star or earlier.  For
$i$=15\degr\ the secondary mass approaches that of the
primary.  The sampling of the velocity curve is such that
aliases of 34.9 d, 69.8 d, 104 d, 174 d are possible but
less likely.  

\begin{figure}
\epsscale{1.0}
\centering
\plotone{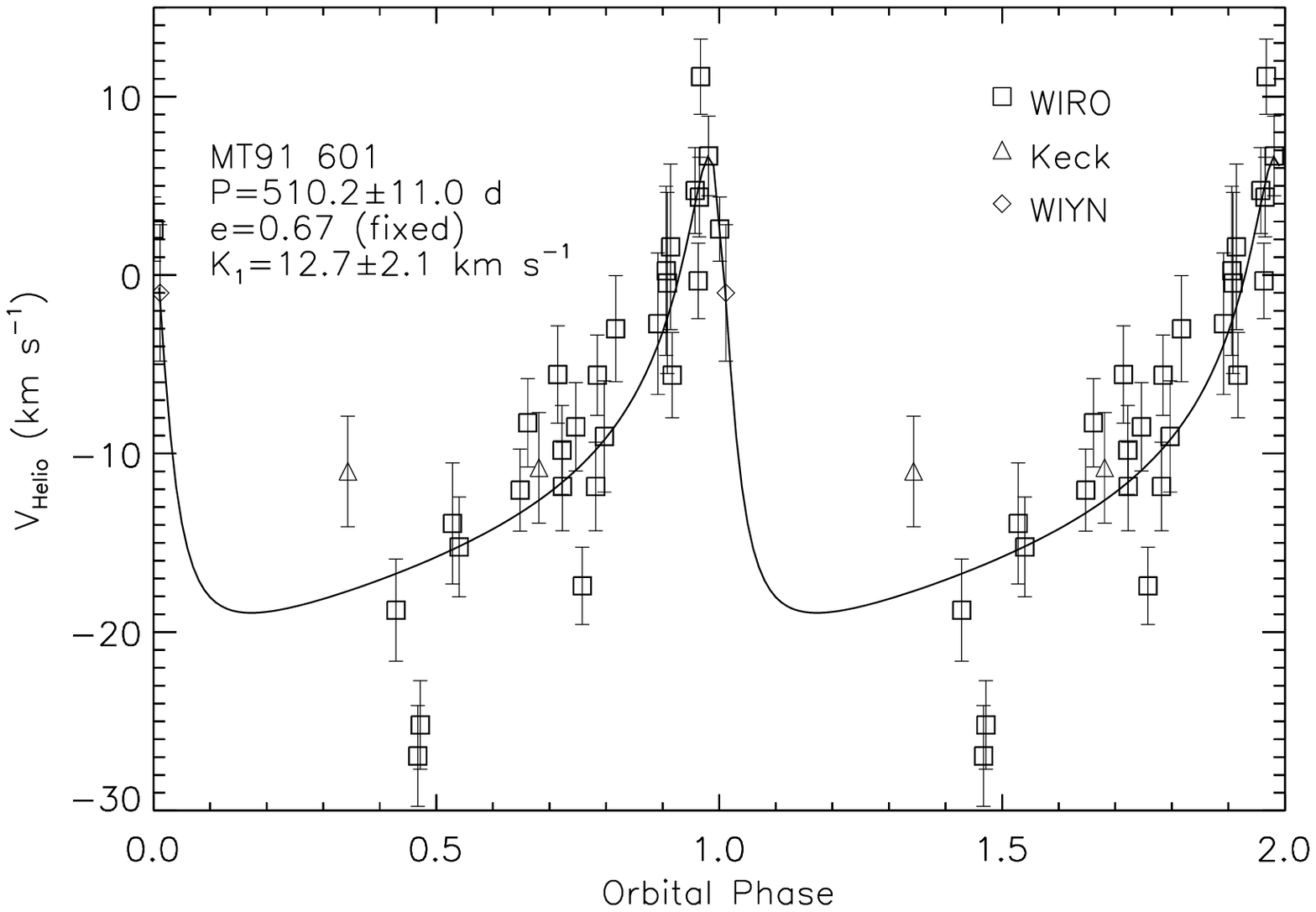}
\caption{Folded radial velocity data and best-fitting solution for MT91~601.   
\label{sol601}}
\end{figure}

{\it MT91~601---}The two Keck, one WIYN and 28 WIRO data of
this O9.5III reveal a period of 510.2$\pm$0.9 d and
$K_1$=12.8$\pm$2.1 \kms.  Owing to the incomplete phase
coverage,  we found it necessary to fix the eccentricity (at
0.67) to avoid extremely large eccentricities that tended to
result during the fitting process. Manual experimentation
suggests an eccentricity uncertainty of $\sim$0.2. 
Figure~\ref{sol601} shows the best-fitting orbital solution 
and folded velocity curve.  Adopting a primary mass of 21
\msun\ yields a minimum secondary mass $M_2$=4.1 \msun,
meaning that the secondary (MT91~601b) is likely a B or O
main sequence star.  The reduced $\chi^2$ of the fit is
unusually large, at 3.4.  We examined the O-C residuals for
signs of periodicity, but no strong peaks were seen.   We
conclude that this evolved primary star exhibits irregular
photospheric variations in addition to the identified
periodic modulation.      

\begin{figure}
\epsscale{1.0}
\centering
\plotone{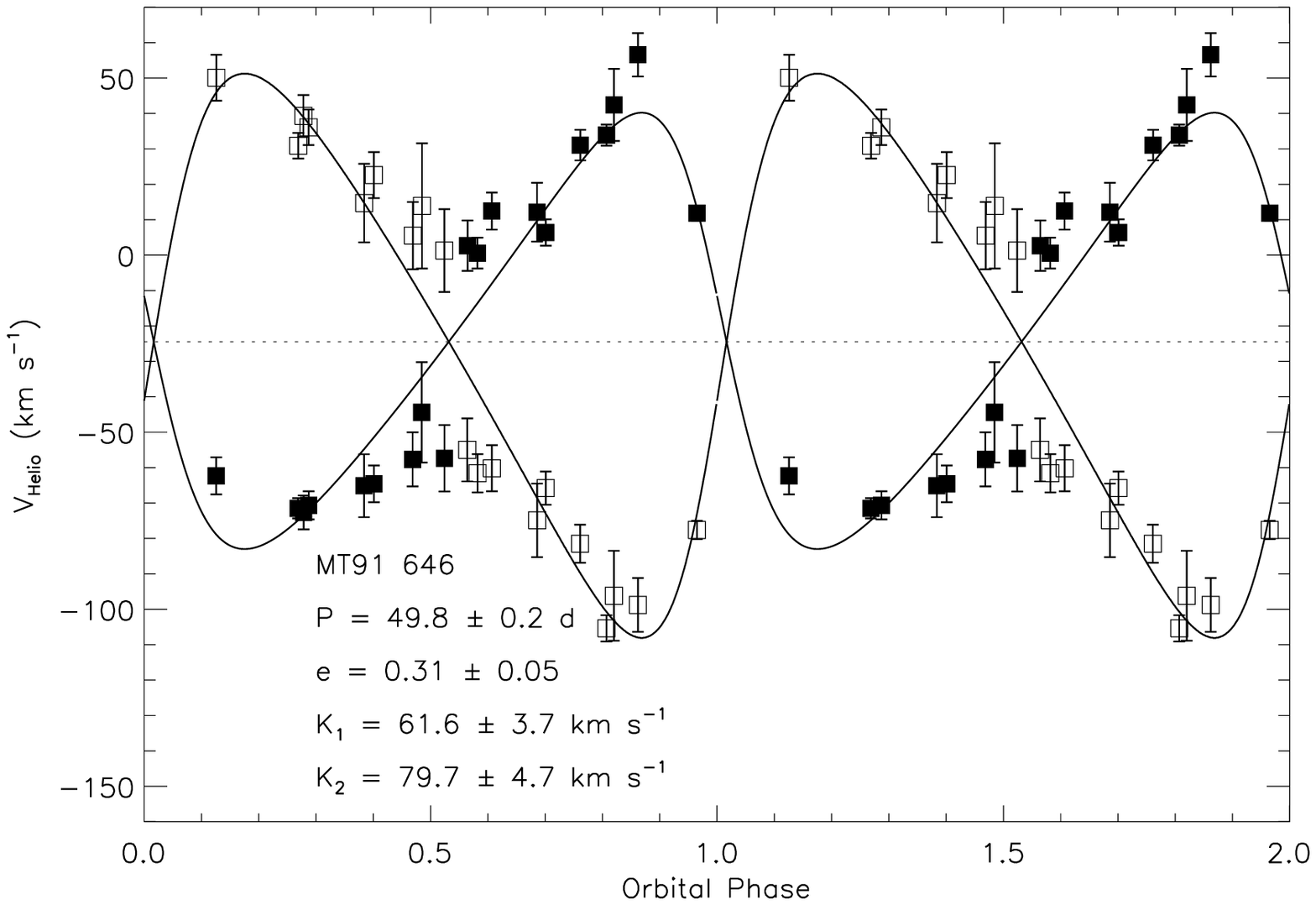}
\caption{Folded radial velocity data and best-fitting solution for MT91~646.   
\label{sol646}}
\end{figure}

{\it MT91~646---}Classified as B1.5V in both \citet{MT91}
and \citet{Kiminki07}, spectra of this system reveal a
variable line width indicating an SB2.  In several of our 19
WIRO spectra  the lines are sufficiently separated to see
that the line depths and widths are similar, having a FWHM
near 2.9 \AA.   We fit the most deblended of the 19 WIRO
spectra with two-component fixed-width Gaussian profiles to
measure velocities for each component.  The solution yields a
period of $P$=49.8$\pm$0.2 d with velocity
semi-amplitudes $K_1$=61.6$\pm$53.7\kms\ and $K_2$=79.6$\pm$4.7
\kms.   This implies a mass ratio of $q$=0.77$\pm$0.05. 
Figure~\ref{sol646}  shows the best-fitting double-lined
orbital solution.  The mass function implies a lower limit
on the system mass of 12.5 \msun, consistent with a B1V and
B1.5V seen at a high inclination.  As such,
MT91~646 is one of the most ``twinlike'' systems among the
Cyg~OB2 sample.   

\begin{figure}
\epsscale{1.0}
\centering
\plotone{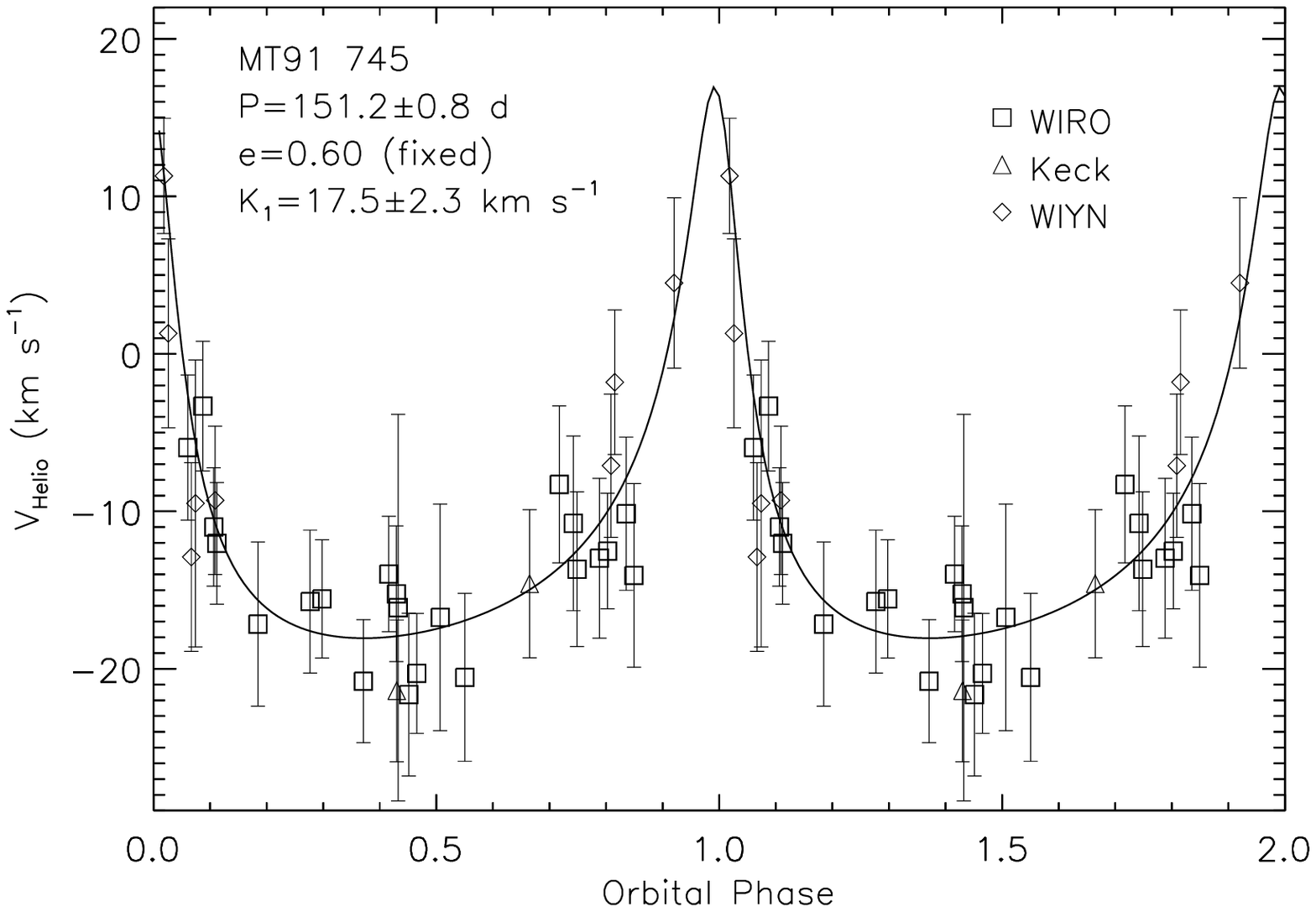}
\caption{Folded radial velocity data and best-fitting solution for MT91~745.   
\label{sol745}}
\end{figure}

{\it MT91~745---}This O7V has two observations from Keck,
eight from WIYN, and 22 from WIRO over 1999--2013, yielding
a period $P$=151.2$\pm$0.8 d, semi-amplitude
$K_1$=17.5$\pm$2.3 \kms, and eccentricity
$e$=0.49$\pm$0.14.  Figure~\ref{sol745}   shows the
best-fitting orbital solution and folded velocity curve. 
Adopting a primary mass of 25 \msun, the implied lower limit
to the secondary mass is $M_2$=4.0 \msun\ for $i$=90\degr. 
$M_2$ approaches $M_1$ for $i$=15\degr.   Thus, the
secondary  (MT91~745b) is at least the mass of a B dwarf
star.     

\subsection{Stars Showing Irregular Variability}

\citet{Garmany1980} noted that some of their O-star sample
exhibited irregular rather than periodic velocity
variations.  This phenomenon is quite common in massive
stars, especially evolved stars, and has been attributed to
line profile variations caused by non-radial pulsations or
clumps in the stellar winds \citep{Vogt1983,
Fullerton1996}.  We note here six objects that have more
than 18 data measurements where significant velocity
variations at the level of 25--30 \kms\ are observed, but no
strong periodicity is evident. Table~\ref{random.tab}  
lists the Heliocentric Julian dates, velocities, and
velocity uncertainties of each measurement.  

{\it MT91~138---}The two Keck data, eight WIYN data, and
22 WIRO data between 1999 and 2008 show that this O8I
displays a low level of irregular variability.   A
long-term, eccentric binary cannot be ruled out. 

{\it MT91~417A (Schulte \#22A---}This O3If shows variations
at the level of about 25 \kms\ on the basis of 20 WIRO data
from 2008 and 2014.   Our lack of long-term data prevents us
from ruling out the possibility of longer term  orbital
variations.  We can only say that the observed variations
appear random on the basis of the data presented in
Table~\ref{random.tab}.  Because of the very weak
\ion{He}{1} lines in this very hot star, velocities are
measured from \ion{He}{2} $\lambda$5411 assuming a rest
wavelength of 5411.45 \AA. \ion{He}{1} should not be present
in such a hot star, but light from the PSF wings of
MT91~417B may contaminate some of our spectra given the
close 2\arcsec\ separation.  

{\it MT91~457---} One Keck, four WIYN, and 11 WIRO data on
this O3If during 1999--2008 show that this star exhibits
non-periodic variations at the level of $\sim$25 \kms.  

{\it MT91~483---}This O5I/III shows low-amplitude (15 \kms)
variations but exhibits power on many scales from days to
months using three data from Keck, three from WIYN, and 25
data from WIRO over 1999 - 2013.   We show, using a Monte
Carlo  analysis which randomly shuffles the dates of 
observation among the observed velocities, that the peaks
near 124 days and five days have at least a  0.32
probability of occurring by chance. Because of the
weak \ion{He}{1} lines in this very hot star, velocities are
measured from \ion{He}{2} $\lambda$5411 assuming a rest
wavelength of 5411.45 \AA. 

{\it MT91~556---}MT91~556 was observed 40 times between
2008 and 2013 at WIRO  with an average velocity uncertainty
of 4 \kms\ and variations of up to  20 \kms.   Although
velocity variations are observed in this B1I,  there is no
dominant periodicity and they are consistent with random
atmospheric fluctuations.  

{\it MT91~632---}MT691~32 was observed 30 times between 2008
and 2011 at WIRO  with an average velocity uncertainty of 3
\kms.   Although velocity variations are observed in this
O9I at the level of 25 \kms,   there is no dominant
periodicity and they are consistent with random atmospheric
fluctuations.  

\subsection{Stars Showing No or Little Variability After
Extensive Observation}

We report here 16 stars having at least 12 velocity
measurements spanning at least five years that show no
evidence for velocity variations.  These stars are
candidates for single stars, stars with very low-mass
companions, or systems seen at  very low inclination
angles.  Among this list there may be undetected binaries
with long periods and/or highly eccentric orbits.  
Additionally, systems with periods near multiples of one
year undergoing periastron during December--April when
Cygnus is minimally observable have a low probability of
being detected in this survey.  For each non-variable system
we report the mean velocity, the rms velocity dispersion,
and the mean velocity uncertainty.  

{\it MT91~005---}This O6V was observed twice at Keck,
three times at WIYN, and ten times with WIRO over the period
1999--2013.  The radial velocities have a mean systemic
velocity of $-$5.5 \kms, an  rms of 2.1 \kms\ and mean uncertainties
of 2.5 \kms.   The very narrow lines (1.9 \AA\ FWHM) suggest
that the rotational axis may be nearly parallel to the line
of sight.  

{\it MT91~020---}Twelve measurements at WIRO, four from
Keck, and two from WIYN spanning 1999--2011 show no
evidence for variability, with a mean velocity of $-$3.2 \kms,
an rms of 4.3 \kms,  and mean uncertainties of 4.2 \kms. 

{\it MT91~083---}With 22 measurements at three
observatories spanning 1999--2008, this B1I star shows no
evidence for variability.  The mean systemic velocity is
$-$1.3 \kms, with an rms of 1.7 \kms\ and mean uncertainties of 2.4 \kms
in the nine WIRO data, one Keck datum, and 7 WIYN data.  

{\it MT91~217---}Having two measurements from Keck, seven
from WIYN, and 17 from WIRO spanning 1999--2011, this O9V
star shows no evidence for variability.  The mean systemic
velocity is $-$11.1 \kms, with an rms of 2.4 \kms\ and mean
uncertainties of 3.0 \kms.  

{\it MT91~227---}Two data from Keck,  six from WIYN, and
26 from WIRO over the period 1999--2011, are remarkably
constant for this O9V.  The mean  velocity is $-$3.8
\kms\ with an rms of 5.5 \kms\ and mean uncertainties of 5.1
\kms.  

{\it MT91~259---}One Keck, six WIYN, and seven WIRO data
 over the period 1999--2011 show a mean velocity of $-$14.0
\kms\ with an rms of 1.3 \kms\ and mean uncertainties of 2.1 \kms.  

{\it MT91~299---}This O7V has two data from Keck, six from
WIYN, and 37 from WIRO  between 1999 and 2013. The mean 
velocity is $-$14.4 \kms\ with an rms of 6.0 \kms\ and a
typical velocity uncertainty of 5.5 \kms. This object shows
some evidence for a periodicity near 405 days, but the
amplitude is low compared to the observational
uncertainties, and a convincing orbital solution was not
obtained.  

{\it MT91~317---}This O8V has two data from Keck and 11 from
WIYN between 1999 and 2007. The mean  velocity is $-$4.2 \kms\ 
with an rms of 4.4 \kms\ and a typical velocity uncertainty
of 5.5 \kms. This object shows some evidence for a
periodicity near 405 days, but the amplitude is low compared
to the observational uncertainties,  and a convincing
orbital solution was not obtained.  

{\it MT91~376---}On the basis of two Keck and 10 WIYN data
over the period 1999--2006 this O8V shows a mean velocity of
$-$23.7 \kms\ and rms of 6.2 \kms\ for a mean velocity
uncertainty of 7.5 \kms.  

{\it MT91~455---}  This O8V has two data from Keck, 3 from
WIYN, and 20 from WIRO between 1999 and 2013. The mean 
velocity is -12.2 \kms\ with and rms of 3.3 \kms\ and a
mean velocity uncertainty of 3.6 \kms. 

{\it MT91~462---}  This O7III has 12 data from from WIRO
between 2008 and 2014. The mean  velocity is -11.9 \kms\
with and rms of 2.8 \kms\ and a mean velocity uncertainty of
4.1 \kms. 

{\it MT91~470---}Two measurements from Keck, five from
WIYN, and  nine from WIRO over the the period 1999--2013
show no evidence for velocity variations beyond the uncertainties. 
The mean velocity is $-$21.3 \kms\ with an rms of 6.9 \kms\
and a mean velocity uncertainty of 5.2 \kms. 

{\it MT91~480---}We measured both \ion{He}{1} and
\ion{He}{2}  velocities for this broad-lined O7V using 4
WIYN and 14 WIRO data between 2001 and 2013, finding no
compelling evidence for velocity variations.   The line
profiles are broad and appear to vary 
without evidence of periodicity.  The mean velocity is
$-$16.2 \kms\ with an rms of 13.3 \kms\ and a mean velocity
uncertainty of 11.7 \kms. 

{\it MT91~507---}Two Keck, two WIYN, and 14 WIRO data
between 1999 and 2013  yield a mean  velocity of $-$8.9 \kms
with an rms of 5.1 \kms\ and a mean velocity uncertainty of
6.5 \kms\ for this O9V star. 

{\it MT91~534---}Three Keck, two WIYN, and 14 WIRO data
between 1999 and 2013  yield a mean  velocity of $-$5.0 \kms
with an rms of 3.8 \kms\ and a mean velocity uncertainty of
3.1 \kms\ for this O8.5V star. 

{\it MT91~611---}Two Keck and 14 WIYN data
between 1999 and 2008  yield a mean  velocity of $-$23.1 \kms\ 
with an rms of 3.7 \kms\ and a mean velocity uncertainty of
3.9 \kms\ for this O7V.

\subsection{Survey Summary To Date} 

The Cygnus OB2 Radial Velocity Survey was designed to
produce a complete census of massive binaries drawn from a
photometrically selected sample, namely that of
\citet{MT91}.  The parent sample contained 150 stars---the
146 listed in Table~1 of Paper~I plus MT91~267
(inadvertently excluded initially but reported
as a binary in Paper VI), MT91~417B (not initially
recognized as a close companion to MT91~417A), Schulte \#3
and Schulte~\#73.  Of these 150, six were found to be
probable emission-line objects (MT91~186, 213, 488, 522,
650, 793)  and excluded from further observation because
radial velocities were considered unreliable.  Another 16
objects were determined to have spectral types later than B3
and excluded from further observation (MT91~170, 189, 196,
222, 239, 271, 273, 427, 444, 453, 459, 493, 539, 554, 576,
641).  This leaves 128 objects in the ``unbiased'' survey
sample that we deem to be the complete photometrically
selected targets.  

In the above sections we identified 6 systems (4.6\%), all
supergiants, as exhibiting irregular velocity variation and
16 objects (12.5\%) as consistent with having no intrinsic 
velocity variability.  Forty-five objects of the 128 in the
complete unbiased sample (35.1\%) have orbital solutions
presented either in this work or previous papers in this
series.   This leaves 63 systems (about half) that have
insufficient data to make a reliable determination.  The
vast majority of these are B1--B2.5 primaries, although a
few O stars remain in this category because there are fewer
than 12 spectra of suitable quality.  The average number of
spectra among the 62  ``indeterminate'' systems is five.  None
have evidence for significant velocity variations.  We
expect that most of these will turn out to  be
constant-velocity objects, but we also anticipate discovery
of a handful of new binaries in future observing seasons. 
We expect most of these to be long-period and/or
high-eccentricity systems that have eluded detection because
of limited data.   Nevertheless, the completeness analysis
in Section 3.5 leads us to conclude that the vast
majority of the detectable  binaries in Cyg~OB2 now have
complete orbital solutions, as published herein.  

\subsection{Modeling the Survey Completeness}  

We estimated the completeness of the Survey as a function of
orbital period  using a Monte Carlo analysis identical to
that described in \citet{Kiminki2012b}. Our code generates
populations of binary systems with periods 1 -- 5000 days
having mass ratios and eccentricities described by power
laws: Prob($q$) $\propto$ $q$$^\alpha$,  and  Prob($e$)
$\propto$ $e$$^\gamma$.   The nominal power-law exponents
are  $\alpha\simeq 0$ and  $\gamma\simeq 0$, as inferred, in
part, by \citet{Kiminki2012b} and strengthened by the
additional data in this work.  A random inclination is
assigned to each system before being sampled at the actual
dates and times of the Survey observations.  A nominal
detection threshold of 7 \kms\  is used to determine whether
the system would be observed  as a binary in the Survey. 
This is lower than the more conservative value of 15 \kms\
used in \citet{Kiminki2012b}, but it better represents the
sensitivity of the data to radial velocity variability given
the typical resolutions of $R\simeq$4500 (60 \kms\ FWHM)
achieved  in the majority of WIRO spectra.  Velocity
precisions of better than 0.1 of the spectral FWHM are
typically attained by centroiding on the strong \ion{He}{1}
$\lambda$5876 feature.  Precisions are 8--15 \kms\ in some
objects where rotation broadens the line profile.    Upon
careful examination,  the detection threshold is really a
complex function of  the signal-to-noise ratio of the
spectra, the spectral type of the star, and the breadth of
the spectral features (i.e., rotational velocity);
nevertheless, the adopted value of 7 \kms\ is generally
applicable as a Survey average, while 15 \kms\ is a
conservative value.  

\begin{figure}
\epsscale{1.0}
\centering
\plotone{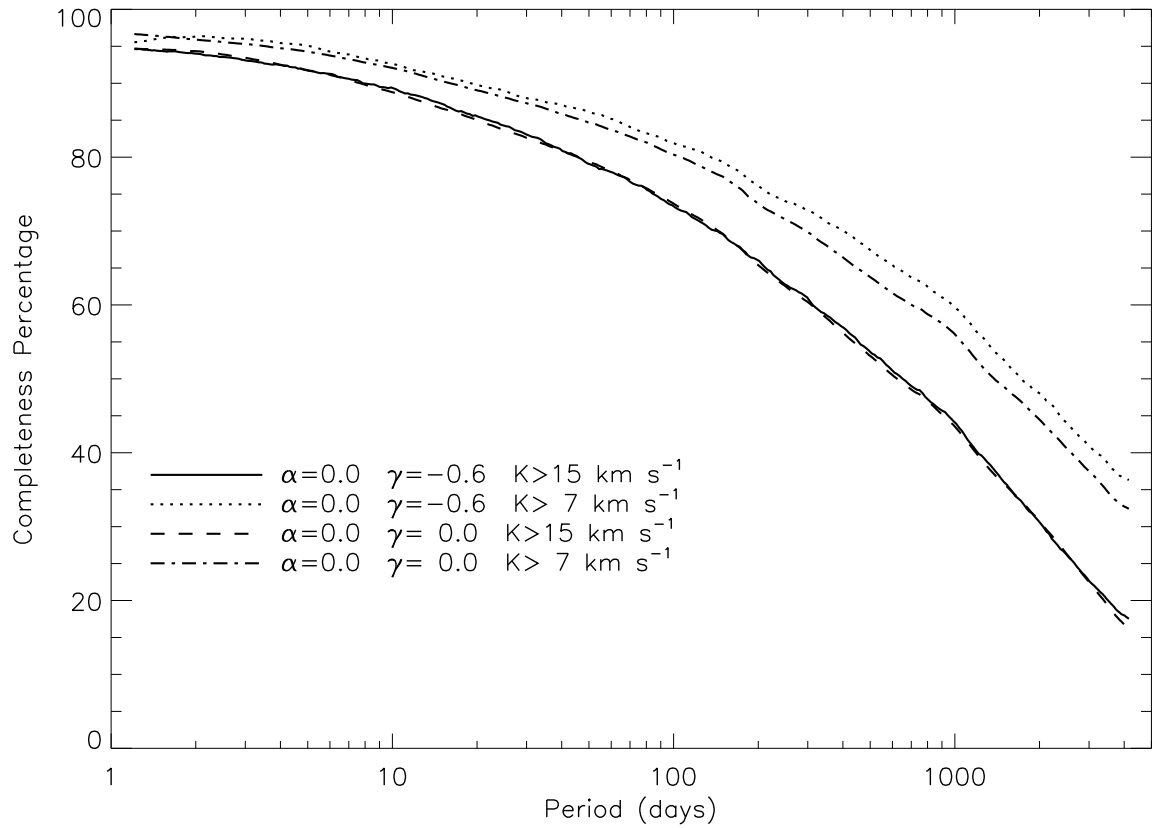}
\caption{Survey completeness as a function of orbital period,
based on the Monte Carlo simulations discussed in the text.
Line styles denote different combinations of 
power law indices adopted to describe the distribution
of mass ratios and eccentricities.  
\label{complete}}
\end{figure}

Figure~\ref{complete} displays the completeness as a
function of orbital period, where completeness is defined as
the  ratio of detected binaries to all binaries of the same
period in the synthesized population.  Different line styles
depict different values for the power law exponent $\gamma$
and different choices for the  detection velocity
threshold.  Completeness exceeds 80\% for systems with
$P<$100 days in the case of a 7 \kms\ detection threshold
and is only $\sim$10\% lower  for the conservative threshold
of 15 \kms.  At $P$=1000 d we find that  40--60\% of 
binaries are still detected.  The smooth decline in
completeness with orbital period reflects the difficulty in
detecting  wide, long-period systems.   While the
completeness estimates are only as  valid as the input
power-law assumptions for the mass ratio and period
distributions, the conclusion that the vast majority of
systems with periods less than a few hundred days have been
detected is a robust one.  Completeness estimates beyond
about 2000 d should be regarded as highly uncertain.

\section{Discussion and Analysis of the Orbital Parameters from Cyg~OB2}
\subsection{Summary of Binaries}

Table~\ref{bigtable.tab} summarizes the spectroscopic binary
type (SB1/SB2), spectral types for the components, orbital
period, primary semi-amplitude, eccentricity,  and mass
ratio constraints for all 48 massive multiple systems now
known in Cygnus OB2.  Twenty three are new results from this
work,  twenty are reported in previous papers in this
series, and five stem from other published works.  The final
two columns provide  references to published orbital
solutions and additional notes on each system.  The table
contains six stars not initially included in Table~1 of
\citet{Kiminki07} among the unbiased list survey targets.  
MT91~267 and MT91~417B, explained in Section 3.4 constitute
two of these.  The double-lined system CPR2002~B17 was
reported and analyzed by \citet{Stroud10}.  Schulte \#5 has
been known as a binary  since \citet{Wilson48}.  CPR2002~A36
and A45 were added later after being reported as suspected
binaries.  In summary, Schulte \#5, CPR2002~A36,
CPR2002~A45, and CPR2002~B17 should not be included  in the
``unbiased'' sample used to estimate binary fractions 
because they were targeted only after evidence of their
binary nature became known.  

The compilation of 48 targets in Table~\ref{bigtable.tab} 
includes 16 SB2 and 32 SB1 systems.   Fourteen contain at
least one evolved component.  Eight are probable triple
systems based on either radial velocity data or
high-angular-resolution imaging (see discussion in Section 3).
Eight are known to show eclipses.  

\begin{figure}
\epsscale{1.0}
\centering
\plotone{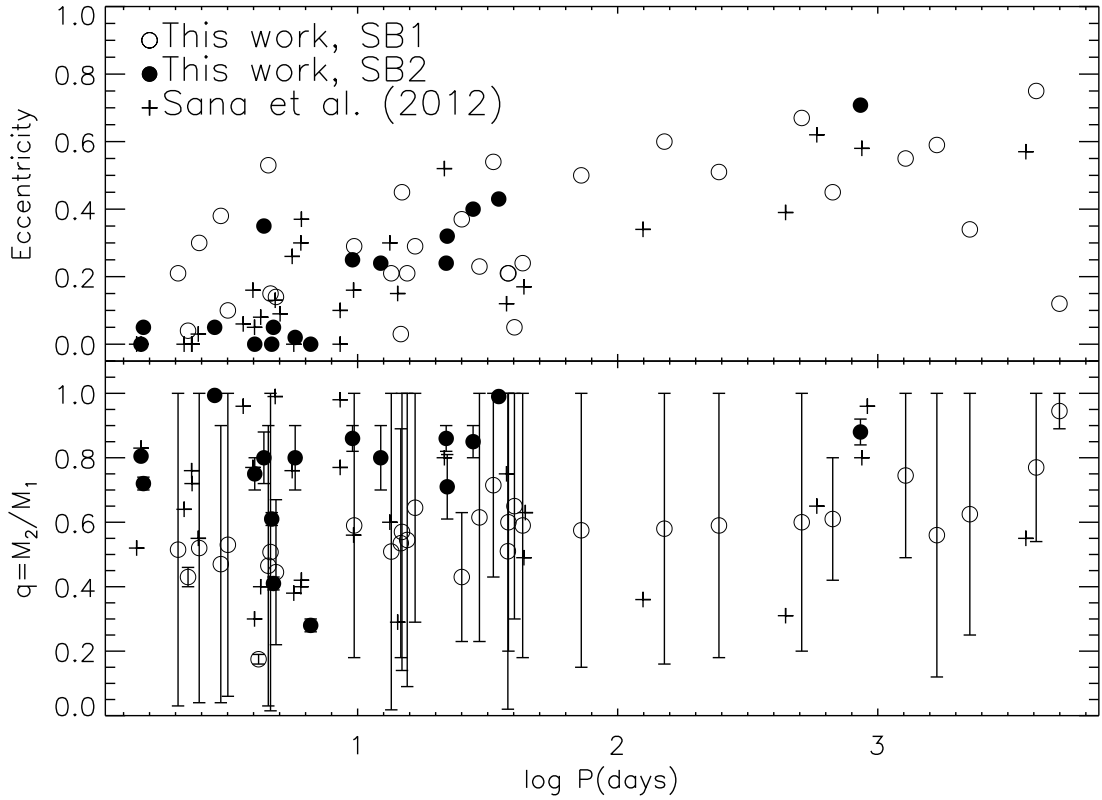}
\caption{Eccentricity (top panel) and mass ratio (lower panel) 
versus orbital period in $\log$(days) for SB1 (open circles)
and SB2 (filled circles) systems from Table~\ref{bigtable.tab}.  
Pluses denote binary systems from \citet{Sana2012}.  Error
bars in the lower panel designate the range of allowed mass ratios,
given the extent of possible inclination angles.   
\label{peq}}
\end{figure}

Figure~\ref{peq} plots  eccentricity (top panel) and mass
ratio (lower panel) versus orbital period  for all 48 massive
binaries.    Filled circles mark SB2 systems and open
circles  denote SB1 systems from Cyg~OB2.  Pluses show the
O-star binaries from \citet{Sana2012}.  Error bars in the
lower panel illustrate the range of mass ratios allowed by
the available constraints on orbital inclination. 
Statistically, systems will preferentially lie toward the
lower end of this range since it is more likely to observe
the orbital plane edge-on than face-on.  The overall trend
in the top panel has been noted elsewhere.   The shortest-period systems
have low eccentricities while longer-period
systems have larger eccentricities.  Any trend
in the lower panel is less obvious.  Mass ratios span the
the range from  near 1.0 for some of the shortest-period
systems to about 0.1, the  approximate lower limit
detectable by the velocity precision of the Survey. 
Long-period systems generally have weak constraints on the
eccentricity and mass ratio as a result of incomplete phase
coverage.  The two samples plotted here are affected by
different target selection, sampling, and sensitivity
biases, so appropriate caution should be exercised in
attempts to statistically  compare them.   Nevertheless, the
distribution of points among the two samples is similar.  

\subsection{The Eccentricity Distribution} 

\begin{figure}
\epsscale{1.0}
\centering
\plotone{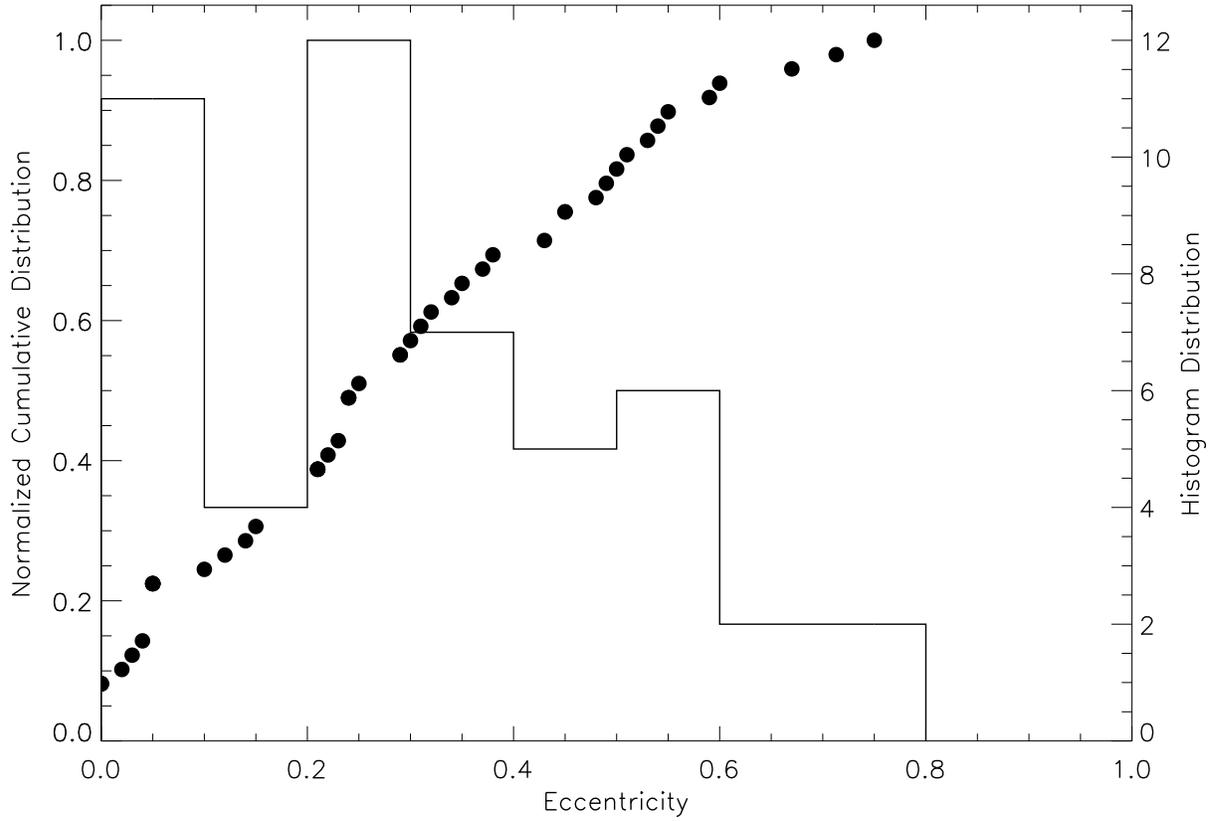}
\caption{Observed distribution of eccentricity for the sample of
Cyg~OB2 binaries.  The left-hand y-axis shows the cumulative probability, 
corresponding to the plotted points.  
The right-hand y-axis provides the number of objects
in each histogram bin.  The distribution is approximately uniform 
between $e=0$ and $e\simeq$0.6.    
\label{edist}}
\end{figure}

Figure~\ref{edist} shows the observed distribution of
orbital eccentricities for the full sample of 48 Cyg~OB2
binaries.   The left-hand y-axis labels and the plotted
points shows the cumulative fraction.    The right-hand
y-axis and plotted histogram shows the number of objects in
each eccentricity bin.  The distribution is approximately
uniform (Prob($e$)$\propto e^0$) between $e=0$ and
$e\simeq$0.6.  A two-sided Kolmogorov-Smirnov (K-S) test
indicates  that the eccentricities have a 75\% chance of
being drawn from the a uniform distribution between the
limits $0.1<e<0.6$. We also employed the Anderson-Darling
(A-D) two-sample statistic\footnote{ The Anderson-Darling
statistic and its advantages over the K-S statistic are
described in more detail  at \citet{FeigelsonBabu2012}.  We
use the code in the Python scipy.stats package to compute
the A-D probabilities.} which is more sensitive to
differences between two populations when  differences in the
the cumulative distributions are both positive and negative
across the sample range, or when these differences occur
near the ends of the distributions.  The A-D statistic
concurs that the probability of the eccentricities being
drawn from a uniform distribution between  $0.1<e<0.6$ is
75\%.   However, over the full eccentricity range there   is
an overabundance of $e<0.1$ systems compared to a  uniform
distribution. The paucity of points at high eccentricity may
partly be an observational bias against detecting  highly
eccentric systems, and it may partly reflect a real dearth
of such loosely bound, easily disrupted systems.  Our
completeness analysis above  assumes two different values, 0
and $-$0.6 \citep[the nominal value from][]{Kiminki2012b}, 
for $\gamma$, the power law exponent describing the
eccentricity distribution.  Completeness is shown to be
insensitive to $\gamma$.  Nevertheless,  a few long-period,
high-eccentricity systems have likely evaded detection with
the current dataset.  Unless there are an disproportionate
number of high-$e$ systems in Cyg~OB2,  the $e$ distribution
can be described as approximately uniform out to periods of
$\sim$1000 days.    

\subsection{The Mass Ratio Distribution} 

\begin{figure}
\epsscale{1.0}
\centering
\plotone{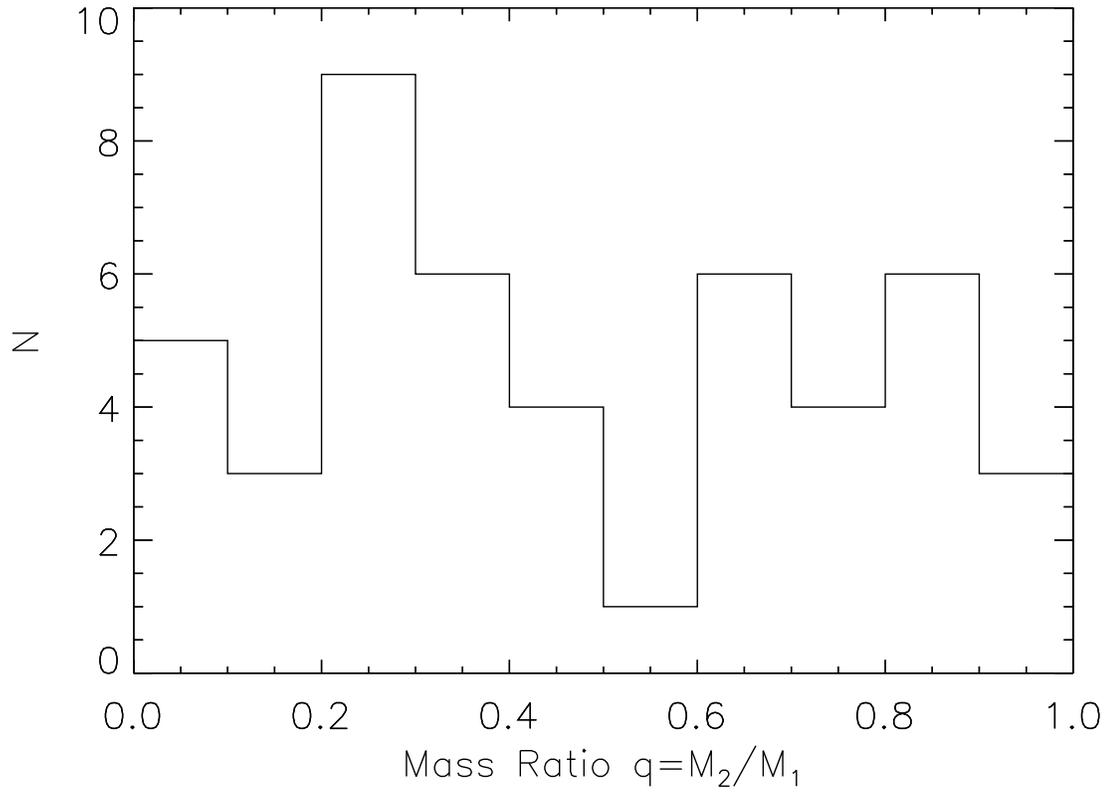}
\caption{A statistical distribution of mass ratios for
48 known massive binaries in Cygnus OB2, based on 
Monte Carlo realizations over allowed inclination angles.   
\label{qdist}}
\end{figure}

The distribution of mass ratios among our sample can be
recovered, statistically, using Monte Carlo methods, by
assigning randomly chosen inclinations on the unit sphere. 
In most cases an upper limit on the inclination is estimated
by either the observed eclipse profile or the lack of
observed eclipses;  for long-period systems the upper limit
on the inclination is effectively 90\degr.   A lower limit
on the inclination is obtained by letting the mass of the
secondary approach that  of the primary; the lack of
observed spectral features from a secondary star is a weak
constraint on this lower bound. For each system we use the
measured orbital period, adopted primary mass, observed
velocity amplitude, and a random inclination between the 
observationally imposed limits to solve for the mass of the
secondary star.   The mean value of $q$ is then computed
after 1000 such iterations for each system. 
Figure~\ref{qdist} shows the resulting histogram of mass
ratios.  The overall distribution is approximately uniform.
However, incompleteness appreciably affects the bins at
$q\lesssim0.2$, so the true number of binaries in these bins
is highly uncertain.  Given the large uncertainties on any
individual mass ratio value, we elect not to pursue a
statistical analysis of these data. In principle we could
estimate the completeness factor at $q<0.2$ given
assumptions about the functional form of the underlying
secondary mass distribution and period distribution (e.g.,
by adopting a power law distribution, following \citet{kf07,
Kiminki2012b, Sana2012}).  However, as we will show below, 
the  true distribution of orbital parameters, in particular
the periods, may not be a single power law.  

\subsection{The Orbital Period Distribution}
\subsubsection{General Characteristics}

\begin{figure}
\epsscale{1.0}
\centering
\plotone{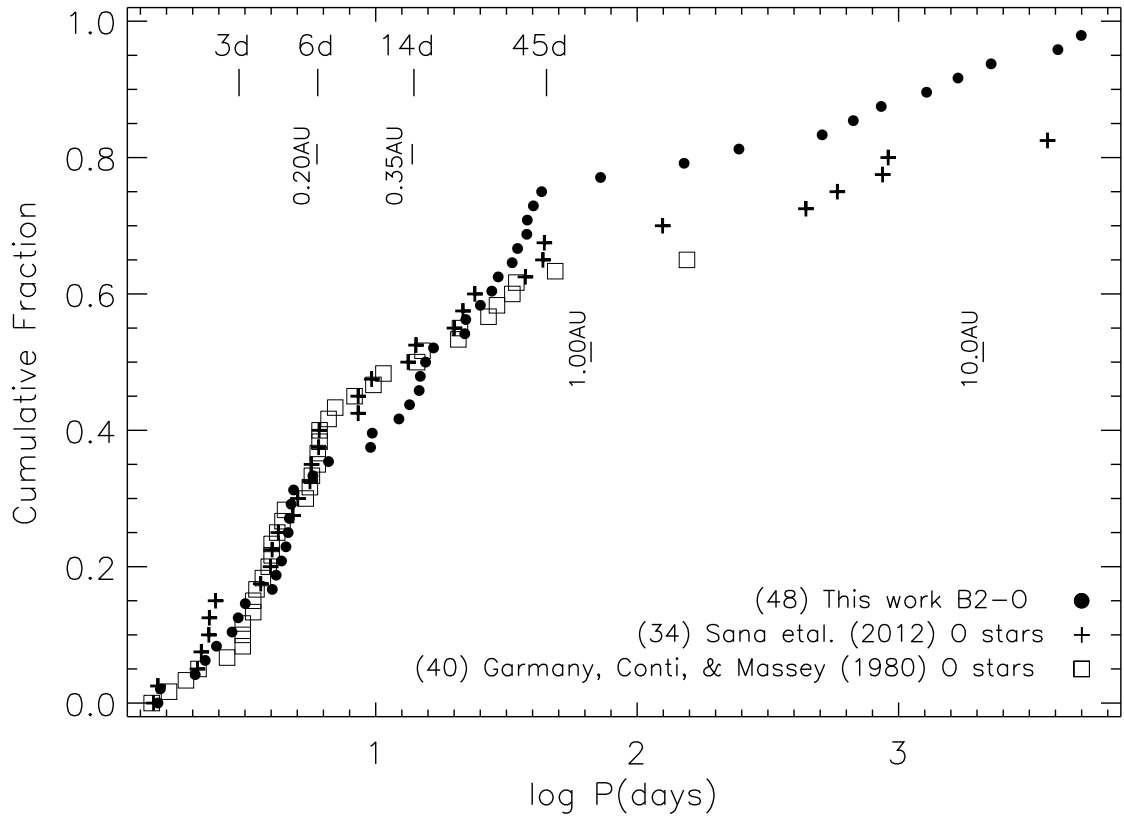}
\caption{Observed cumulative distribution of orbital periods for
the Cyg~OB2 sample ({\it filled circles}), the \citet{Sana2012}
sample of O stars in Galactic open clusters ({\it pluses}),
and O stars from \citet{Garmany1980} ({\it open squares}).    
\label{pcum1}}
\end{figure}

Figure~\ref{pcum1} displays the observed cumulative
distribution of orbital periods (in $\log$ $P$ ) for  the
full sample of 48 massive (B3 and earlier) systems in Cyg~OB2
(filled circles), O stars from six Galactic open clusters
\citep[][plusses]{Sana2012}, and  bright O stars drawn from
the general Galactic population from \citet[][open
squares]{Garmany1980}.  Fiducial marks near the top of the
plot indicate the period in days. Other annotations mark the
corresponding semi-major axis for the given period, assuming
a total system mass of 30 \msun.   The \citet{Garmany1980} 
sample has been normalized by a factor of 1.5 to allow
better comparison with the other two surveys in the
short-period limit where all three are highly complete.  
Both the \citet{Sana2012} sample and the Cyg~OB2 sample are
reasonably complete to periods of several hundred days,
while the \citet{Garmany1980} sample  lacks such
longer-period systems---a consequence of the limited
observing campaign.   A constant slope in this plot 
corresponds to a uniform distribution in $\log$ $P$.  A 
steeper slope means a larger number of detected binaries
per $\log$ $P$ interval than a shallower slope. 

The observed cumulative period distributions for the three
samples appear remarkably similar at short periods, rising
rapidly  with similar slope from a short-period limit of
about 1.4 days to near 7 days.   An A-D test shows that the
Cyg~OB2 and \citet{Sana2012} observed distributions have a
91\% chance of being drawn from the same parent populations
in this range, increasing to 97\% if the complete instead of the unbiased 
Cyg~OB2 sample is used.  All three samples show an abundance of
short-period binaries, noted previously by
\citet{Kiminki2012b, Sana2012, Zinnecker07}. All three
samples also show a flattening of the slope beginning at
about 6 days.   The Cyg~OB2 sample exhibits the most
pronounced flattening between 6 and 14 days, indicating a
paucity of systems in this range, as noted previously by
\citet{Kiminki2012b}.   There is then an upturn between 14
and $\sim$45 days, with a slope approximately matching that
of the very short-period systems.  The \citet{Sana2012} and
\citet{Garmany1980} samples, by contrast,  show lesser
degrees of flattening longward of six days, with the former
being steeper than the latter.  The limited number of data
points precludes any strong statement regarding the
similarity of the three samples in this restricted range, or
the reality of the change in slope near six days.

Both the  Cyg~OB2 and the \citet{Sana2012} samples flatten
or exhibit a break near 45 d.   The \citet{Garmany1980} sample
has become considerably incomplete at these periods and is
not considered further.  Both samples exhibit similar slopes
out to periods of several thousand days where both become
significantly incomplete.   \citet{SanaEvans2011} and
\citet{Sana2012} considered the possibility that period
distribution could be characterized by a double-\"Opik
\citep[i.e., uniform in $\log{P}$;][]{Opik24} distribution with
a break near 10 days, however they found that such a
description did not provide a better fit to the data. 
Corrections for observational bias (i.e., incompleteness)
become important at periods greater than several hundred
days, and we address these in the context of the Cyg~OB2
sample below.  

\subsubsection{Statistical Analysis in the Short-Period and Long-Period Regimes}

A two-sample Anderson-Darling\footnote{We use the
Anderson-Darling statistic exclusively hereafter in lieu of
the more popular  but more problematic Kolmogorov-Smirnov
statistic.  We find that the K-S statistic yields similar
probabilities to the A-D statistic when when the cumulative
distributions are smooth; the A-D statistic yields lower
probabilities of the null hypothesis that the two
distributions are drawn from the same parent population when
the cumulative distributions show multiple points of
inflection. This is consistent with the discussion in
\citet{FeigelsonBabu2012}.}  test shows that the Cyg~OB2 
and the \citet{Sana2012} observed cumulative distributions are 
individually consistent with uniform (in $\log$ P given by
$\beta=0$)  between 1.4 and 45 days at the 42\% and 45\%
levels, respectively; the Cyg~OB2 and \citet{Sana2012}
samples becomes consistent with uniform at the 86\% and 62\%
level for periods $<$25 days.   The Cyg~OB2 and 
\citet{Sana2012} samples are consistent with each other at
the 91\% levels for $P<$ 25 d  but only 19\% for $P<$ 45
d.   The Cyg~OB2 and the \citet{Garmany1980} samples are
consistent with each other at the 40\% and 5\% levels for
upper period limits of 25 d and 45 d, respectively. Taken
together, these statistics indicate that the period
distributions in the Cyg~OB2 and \citet{Sana2012} surveys
are  probably consistent with each other $P<25$ d, but not
at $P>25$ d  and  are not consistent with a uniform distribution
even at the shortest periods.

When all three surveys are combined, the probability that
the combined sample is consistent with uniform drops to 14\%
and 21\%  for upper  period limits of 25 d and 45 d,
respectively.  We interpret this to be evidence for 
structure in the period distribution that becomes apparent
with sufficiently large numbers.  Thus, the period
distribution does not appear to be scale-free; it is
suggestive of features imposed by physical phenomena
occurring during the formation and/or evolution of massive
systems.  

The most obvious feature in the period distribution at
$P<45$ d is the flattening of the slope near 6 days seen in
all three datasets.   \citet{Kiminki2012b}  characterized
this structure as a  possible excess of  3--6
d systems accompanied by a deficit of 7--14 d systems,  observed
here as a flattening of the slope of the cumulative
distribution over that range in all three samples.   This
possible feature seen in the Cyg~OB2 distribution is not as
obvious in the other two data sets.  Incompleteness in
\citet{Garmany1980} sample, though not explicitly quantified
in that work, probably begins in the 15--30 d range and could easily
mask such a signal if it were present.  By contrast,  the
completeness of the Cyg~OB2 survey is $>$90\% for $P$$<$10
d.     The reality of such a signal in the \citet{Sana2012} 
sample  is uncertain owing to the small number of points
between 6 d  and 45 d (10 systems), even though the sample
is likely to be similarly complete.  We conclude that the
Cyg~OB2 period data is not convincingly consistent with
uniform (a single power law of slope zero) at
$P$$<$45 d.  There is evidence for structure in the form of
a 3--6 d period ''excess''  and the 6--14 d period
``deficit'' that becomes more pronounced when the three
samples are combined.

In the period range 60 -- 4000 d  the Cyg~OB2 and
\citet{Sana2012} samples, as observed,  are consistent with
uniform at the 94\% and 55\% levels.  The two samples are
consistent with each other at the 99\% level.  By
contrast, over the full period range 1.4 -- 5000 days, both
samples are inconsistent with uniform, having a $<$0.01\%
chance of being drawn from such a distribution.  The single
$\beta=0$ power law is a poor description of the observed
period distribution over the full range, but observational
biases against detection of long-period systems must be
considered before  strong conclusions may be drawn.  

\subsubsection{The Period Distribution Corrected for Survey Completeness}

\begin{figure}
\epsscale{1.0}
\centering
\plotone{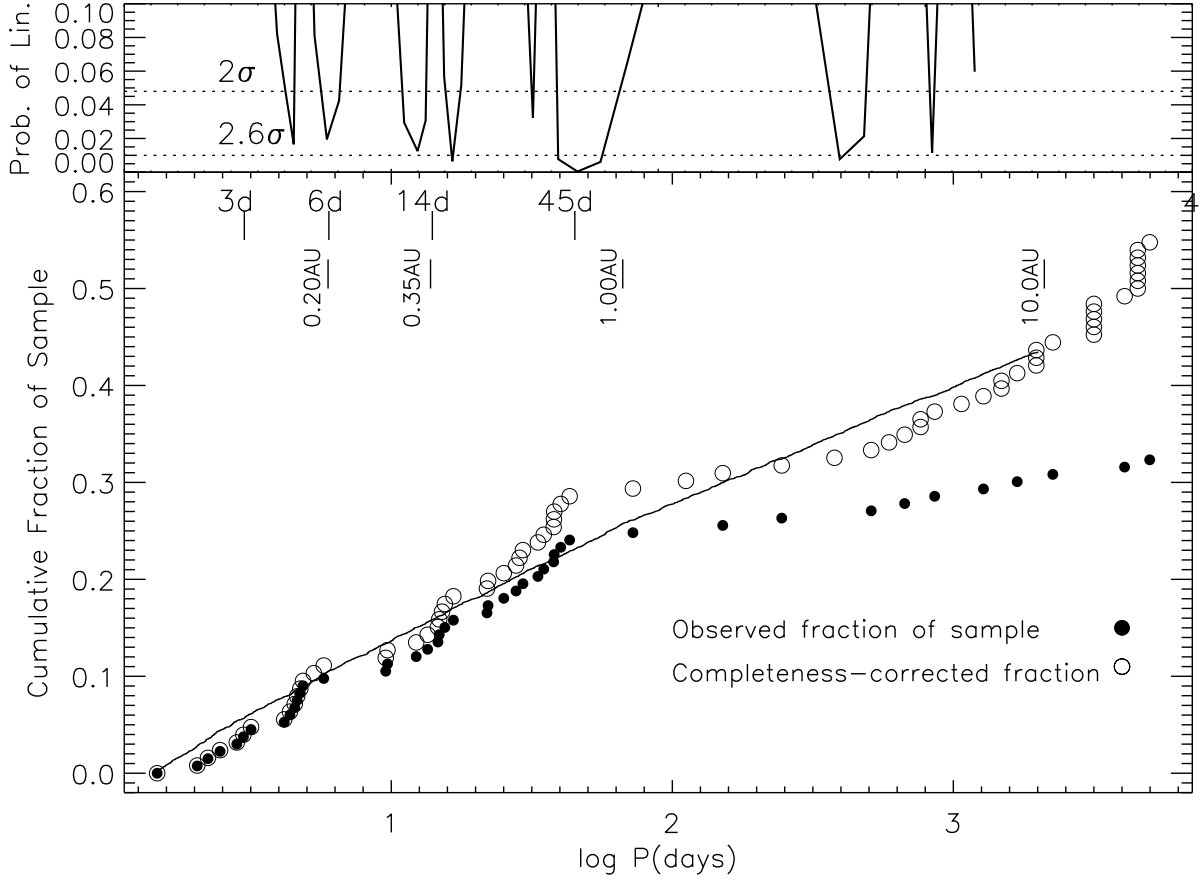}
\caption{(Lower panel) Observed cumulative distribution of orbital periods 
in the unbiased Cyg~OB2 sample as a fraction of the
entire sample ({\it filled circles}) and the
the inferred true distribution of orbital periods ({\it open circles}) using the completeness
corrections shown in Figure~\ref{complete} and assumptions for the
underlying distribution of eccentricities and mass ratios, as described in the text.
The solid curve is a power law with slope $\beta=-0.22$.  (Upper panel) 
Probability that the observed distribution is consistent with
uniform in a moving window of width seven observed systems.  Probable
changes in slope near 6 days, 14 days, and 45 days appear as dips
in the probability curve.  
\label{pcum2}}
\end{figure}

Figure~\ref{pcum2} (lower panel) shows the observed
cumulative period distribution for the unbiased sample of 45
systems as observed (filled symbols)  and after correction
for completeness (open symbols) as described by
completeness curve  in Figure~\ref{complete} using the solid
line representing the 15 \kms\ (most conservative) 
detection threshold.  The  y-axis is now scaled to indicate
the fraction of the total number of stars in the unbiased
Cyg~OB2 sample. The correction from observed to underlying
period distribution is performed by  tabulating the
cumulative incompleteness (1-completeness) at the location
of each observed system, N,  starting from the shortest
periods, and inserting an additional ``system'' when the
cumulative  incompleteness reaches an integer value; systems
are added at periods halfway between the period of system N
and N-1.   This can result in the appearance of two or more
systems plotted at the same orbital period when the
completeness is low (i.e.,  for long periods).      

Open symbols in Figure~\ref{pcum2}  trace a curve similar to
the filled symbols until the correction for completeness
becomes a large factor beyond about 2000 days.  The
difference between the filled and open symbols is small for
periods less than a few hundred days but approaches 20\% at
the 5000-day limit of the Survey.  The open symbols
display an apparent change in slope near 45 days, indicating
that the putative break in the cumulative distribution is 
not likely to be caused by observational biases.  Under the
adopted prescriptions for the mass ratio distribution (i.e.,
uniform) and the eccentricity distribution (i.e., uniform) 
adopted in the Monte Carlo modeling, this figure shows that
the binary fraction reaches 30\% by $P$=45 d. Fifty-five
percent of the systems are binaries with periods less than
5000 days.  This is nearly identical to the binary fraction
of 51\% computed for O stars in the Large Magellanic Cloud
\citep{Sana2013} and may indicate that the binarity characteristics
of massive stars 
are insensitive to metallicity.  
 An extrapolation of the quasi-linear trend
defined by systems with $P>$100 days would suggest a binary
fraction near 70\% for systems with periods less than 
$10^5$ days.  On the other hand, a naive extrapolation of
the much steeper slope at $P<$45 days would reach a binary
fraction of 100\% near $10^5$ days.

We used the A-D test to quantify the probability of a change
in slope of the completeness-corrected cumulative period
distribution by comparing it to a hypothetical uniform
(i.e., linear) distribution in a moving window of
encompassing seven observed systems.  Window widths of five
to nine do not appreciably change the results.  
Figure~\ref{pcum2} (upper panel) plots the probability that
the completeness-corrected period distribution is consistent
with uniform.  The probability drops below 1\% (2.6$\sigma$,
shown  by the dotted horizontal line) in the vicinity of the
hypothesized 45-day break.   The low probability near six
days and 14 days supports the  hypothesis of a change slope
near these locations as well.   The similarity of the
Cyg~OB2 and the  \citet{Sana2012} observed cumulative
distribution slopes above 45 days in Figure~\ref{pcum1} is
noteworthy as is the apparent flattening near 45 days
($\simeq$0.9 A.U).   The shallower slope above 45 d may
indicate either a decreasing binary fraction for systems
with semi major axes of $>$1 A.U.  Alternatively,  it may
indicate a change in the distribution of secondary star
masses toward a preference for lower masses.

We interpret the evidence for structure in the cumulative
period distribution as an indication that the processes
responsible for setting massive binary separations are not
scale-free. Nevertheless, we investigated whether a single
power law might at least approximate the data.  The solid
curve in the lower panel of Figure~\ref{pcum2} is a best-fit
single $\beta=-0.22$ power law to the completeness-corrected
data over the range 1.4--2000 d.  The completeness
correction is highly uncertain beyond 2000 d, so we do not
consider this regime in the fit.   An A-D test indicates a
24\% probability that the data is drawn from such a single
power-law distribution.  This probability does not
constitute strong evidence in favor of  this particular
power law, especially given that the results are sensitive
to the magnitude of the  completeness corrections required
for long-period systems. Nevertheless, the
$\beta\simeq-0.22$ power law is not an unreasonable {\it
approximation} over 3.3 orders of magnitude in period (1.4
days to 2000 days), even if it does not describe the fine
details of the distribution at a statistically compelling
level of significance.     

We conclude by noting that the agreement between the
observed orbital period distributions of the Cyg~OB2 sample,
where OB stars formed across an extent of nearly 50 pc,
compared to the \citet{Sana2012} sample from several young
Galactic clusters, suggests that the large-scale physical
environment does not determine the orbital parameters of
massive binaries formed therein.  This stands in contrast
to  suggestions that the binary fraction depends on local
stellar density, being lower in the densest regions
\citep{Sana08, Zinnecker07}. Rather, the small-scale physics
of cloud collapse and fragmentation on the sub-A.U. scale
appear more important in establishing the high fraction of
binaries among massive stars and their mass ratio
distribution which clearly shows that massive stars
preferentially have massive companions.  The results of the
massive star surveys stand in stark contrast to those from
radial velocity surveys of solar-type stars where the
distribution of periods is log-normal with a mean of 180
years \citep{DM91}.   The inflections in the cumulative
period distribution near  0.2 A.U. and near 0.9 A.U. are
suggestive of key physical size  scales beyond  which the
formation of close massive binaries becomes less
efficient.  

\section{Conclusions}

The addition of 23 new massive binary orbital solutions
presented herein raises the total number known  in the
Cygnus OB2 Association to 48.  This is the largest
collection of complete solutions obtained within a single
environment and constitutes a substantial fraction of all
early type  orbits published to date.  At least 35\% of
massive systems in Cyg~OB2 are binary or higher-order
multiples, with several more long-period or highly eccentric
systems expected to be discovered.  The number of 
unobserved high-inclination or extreme-mass-ratio systems
has been  modeled statistically.   These computations
suggests that the true binary fractions are near 55\% for
periods $P\leq$5000 d, nearly identical to the results from
O stars in the LMC, suggesting a degree of metallicity
independence.  Extrapolation to longer periods
suggests binary fractions in the range 65--80\% for periods
$P<10^4$ d.  

The distribution of orbital periods is described as
approximately uniform in $\log~P$ between the short-period
limit of 1.2 d and $\simeq$45 d, but there is evidence for
structure in this distribution, with an excess of 3--6 d
systems and a possible dearth of 7--14 d systems, along with
a break near 45 days.  Between 45 d and the long-period
survey limit of $\sim$5000 d the distribution is also
approximately uniform, but with either a lower binary
fraction or a population of secondary masses favoring
lower-mass companions compared to the 1--45~d  subsample. 
No single power-law distribution describes the period data
at a statistically significant level.   A single power law
of slope $\beta=-0.22$ also provides a rough description of
the cumulative period distribution between 1.4 and
$\sim$2000 d if structure in the distribution is ignored. 
The statistical power of even the present
completeness-corrected sample has not yielded a convincing
power-law or other analytic description of the present-day
binary period distribution.   It seems likely that 
evolutionary effects such as period migration and stellar
mergers have altered the primordial distributions to the
ones observed as the present-day distributions in the 3--4
Myr old Cygnus OB2 Association and other young O-star
cluster. Such evolutionary effects may occur during the
pre-main-sequences phases, rendering any simple primordial
distribution unobservable even in principle.

One goal of this work is to provide a robust  measure of
massive binary parameters to help inform stellar population
synthesis models, including those used to predict cosmic
rates of high-energy phenomena such as supernova and
$\gamma$-ray bursts.  It is our hope that such models may
now be grounded more solidly in the data.   This compilation
may also prove helpful in guiding models of massive star
formation and evolution by providing observational constraints on the
binary frequency and distribution of periods and mass ratios
among early-type systems. We expect that there are several
more discoverable binary (and multiple) systems among the
Survey targets, but this present compilation likely contains
the vast majority of systems discoverable using radial
velocity techniques.

\acknowledgements  Insightful comments from an expert
anonymous referee greatly improved this manuscript.   We
thank our long-term collaborator Chris Fryer for continued
encouragement, in particular for a timely conversation after
a colloquium at UC Santa Cruz in 1998 that launched the
Cygnus OB2 Radial Velocity Survey.  We acknowledge continued
support from the National Science Foundation through
Research Experience for Undergraduates (REU) program grant
AST 10-63146, through grant AST 03-07778, and grant AST
09-08239.   The Wyoming NASA Space Grant Consortium provided
student support through grant \#NNX10A095H. We thank Tyler
Ellis, Sarah Eftekharzadeh, Ian Ewing, and Garrett Long for
assistance with observing at WIRO.  We are grateful to WIRO
staff James Weger and Jerry Bucher whose diligent support
work enabled this program to obtain spectra on more than 290
nights at the Wyoming Infrared Observatory 2.3 m telescope. 
Generous allocations of telescope time at the Lick and Keck
telescopes through the University of California and at the
WIYN 3.5 m telescope through NOAO and the University of
Wisconsin made possible such a large observational
project.   Some of the data presented herein were obtained
at the W.M. Keck Observatory, which is operated as a
scientific partnership among the California Institute of
Technology, the University of California and the National
Aeronautics and Space Administration. The Observatory was
made possible by the generous financial support of the W.M.
Keck Foundation.  The WIYN Observatory is a joint facility
of the University of Wisconsin-Madison, Indiana University,
Yale University, and the National Optical Astronomy
Observatory.

\textit{Facilities:} \facility{WIRO ()}, \facility{WIYN ()}, \facility{Keck:I ()}

\appendix

\section{Re-analysis of Orbital Solutions for Systems from Papers II and III}

In this Appendix we briefly re-analyze systems originally
presented in Papers II and III for purposes of
self-consistency, using the methods employed in this work
and incorporating any new spectra, when available.  Most of
the solutions are essentially unchanged, but results are
presented in a manner consistent with the solutions reported
in Papers IV, VI, and in this contribution.   In three cases
(MT91~252, S~73, CPR2002~A45), additional data allowed us
to discover that the original solutions were aliases and to 
present a better solution.

In this work, we disentangled the component spectra of double-lined
binaries using the method of \citet{GL2006}. One of the
strengths of this method is that the radial velocities can be refined
via cross correlation after each iteration (i.e, cross-correlating the
the residual spectra with the resultant component spectrum as the
template). The cross-correlated velocities generally have smaller
uncertainties and utilize more lines, leading to a better measurement
of the true stellar velocity and the binary systemic velocity.

\begin{figure}
\epsscale{1.0}
\centering
\plotone{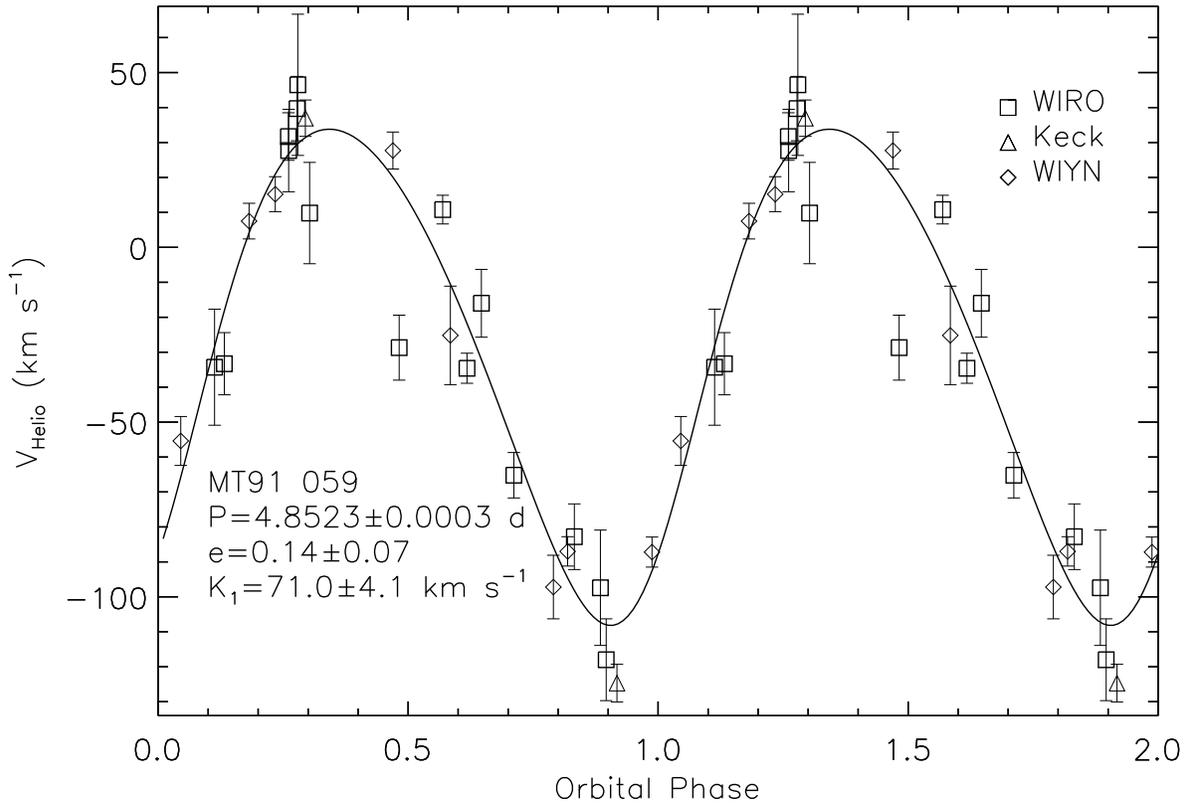}
\caption{Folded radial velocity data and best-fitting solution for MT91~059,
re-analyzed and updated slightly from Paper II.  The large reduced $\chi^2$ of
4.3 suggests either the presence of an additional velocity component or 
additional photospheric variability.   
\label{sol059}}
\end{figure}

{\it MT91~059---}Table~\ref{velupdated1.tab} lists the
updated ephemeris and Table~\ref{solutionsupdated1.tab}
provides refined orbital elements for the single-lined
binary MT91~059 originally presented in Paper II. 
Figure~\ref{sol059} displays the best-fitting solution and
folded velocity data.  The original period of 4.85 d is 
changed only slightly, but we note here the large reduced
$chi^2$ of 4.3 that indicates either an additional velocity
component or photosspheric variability.  Attempts to identify
periodicities in the observed-minus-computed  velocities in
Table~\ref{velupdated1.tab} did not yield convincing
evidence for an additional velocity component.  Hints of
emission in the cores of the \ion{He}{1} $\lambda$4471 lines
seeing in our two  high-resolution spectra from Keck support
the possibility of extra-photospheric  variability, perhaps
as part of an accretion stream in this close ($\sim$0.18
A.U.) binary.

\begin{figure}
\epsscale{1.0}
\centering
\plotone{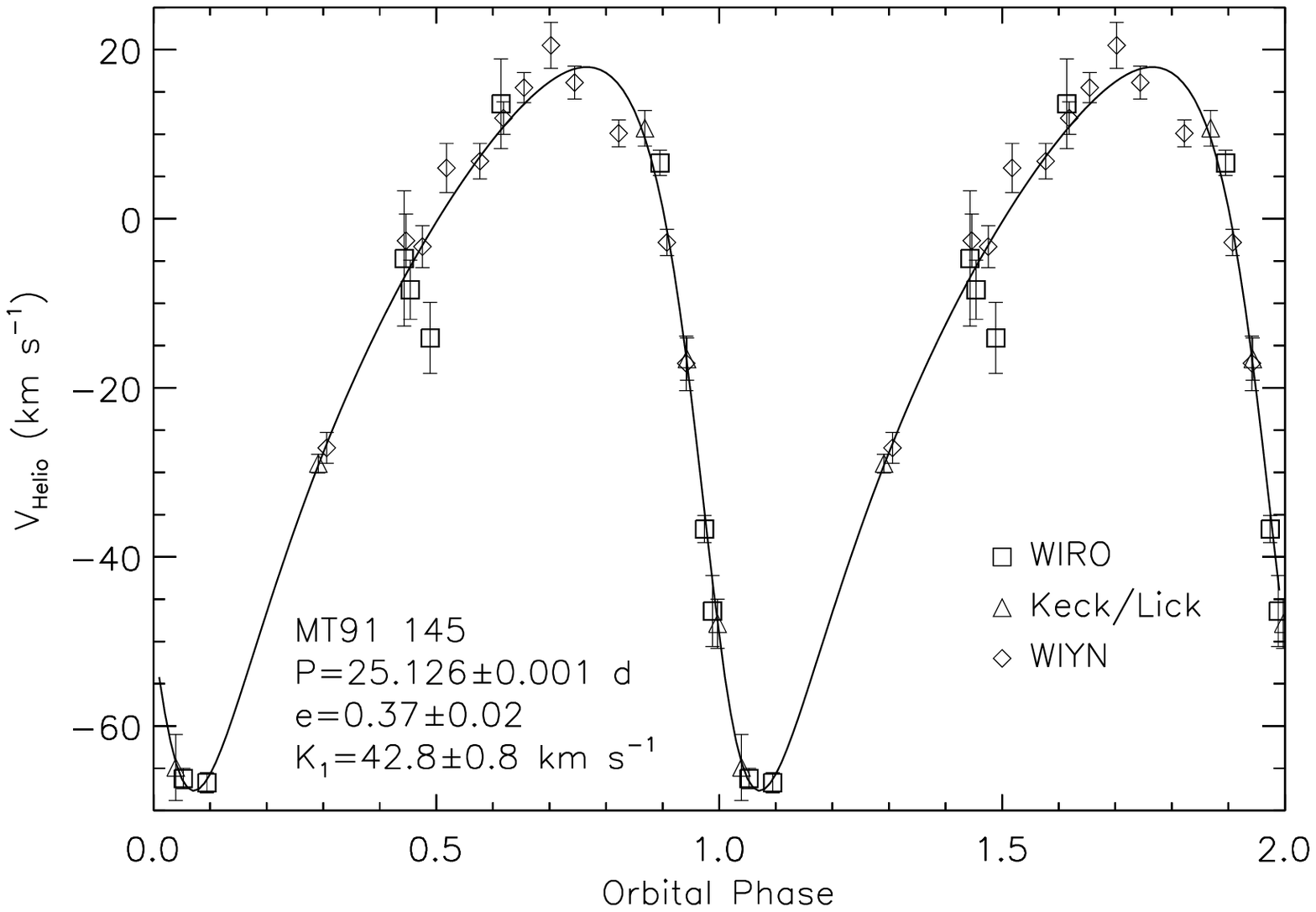}
\caption{Folded radial velocity data and best-fitting solution for MT91~145,
re-analyzed and updated slightly from Paper III.  
\label{sol145}}
\end{figure}

{\it MT91~145---}Table~\ref{velupdated1.tab} lists the
updated ephemeris and  Table~\ref{solutionsupdated1.tab}
provides refined orbital elements   for this single-lined
O9II star originally presented in Paper III. The orbital
elements are only slightly revised from the original
published estimate.   Figure~\ref{sol145} displays the best-fitting 
solution and folded velocity data.

\begin{figure}
\epsscale{1.0}
\centering
\plotone{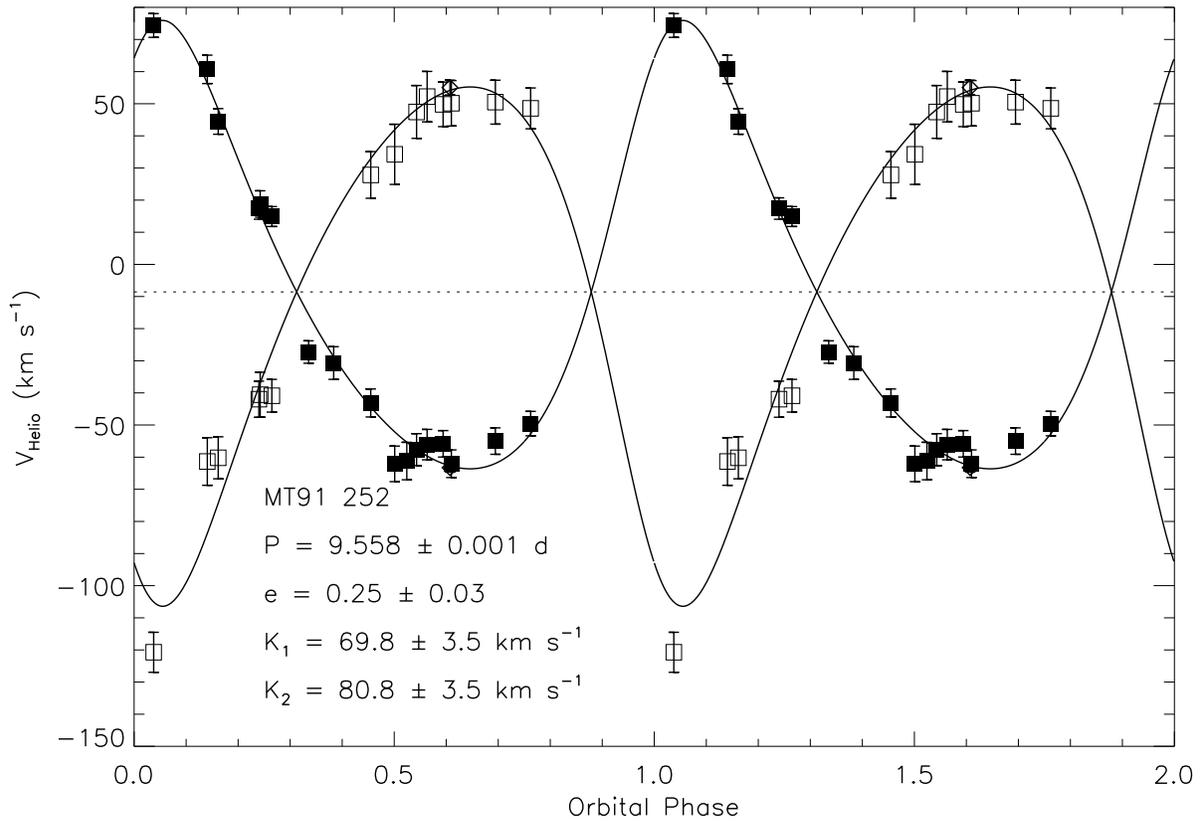}
\caption{Folded radial velocity data and best-fitting solution for MT91~252,
re-analyzed and updated from Paper II.  
\label{sol252}}
\end{figure}

{\it MT91~252---}We identified this system as an SB2 in
Paper~II but could only perform a limited analysis owing to
the small number of spectra and low SNR of the data. Using
the technique described in this paper, we included 17
epochs of WIRO spectra and a KECK+HIRES spectrum to compute
the solution listed in Table~\ref{solutionsupdated2.tab}.
The previously estimated period of 18--19~days has been
revised to $9.558\pm 0.001$~days. The power spectra for both
components showed an alias at $\sim$11.4~days. However, the
folded velocity curve showed significantly higher scatter and was
ruled out. Figure~\ref{sol252} shows the computed solution.
No revisions to the estimated spectral classifications are
provided. By adopting a mass of 11--14~\msun\ for
the B1--2V components, the period and velocity amplitudes 
imply an inclination between 30 and 40~degrees.

\begin{figure}
\epsscale{1.0}
\centering
;\plotone{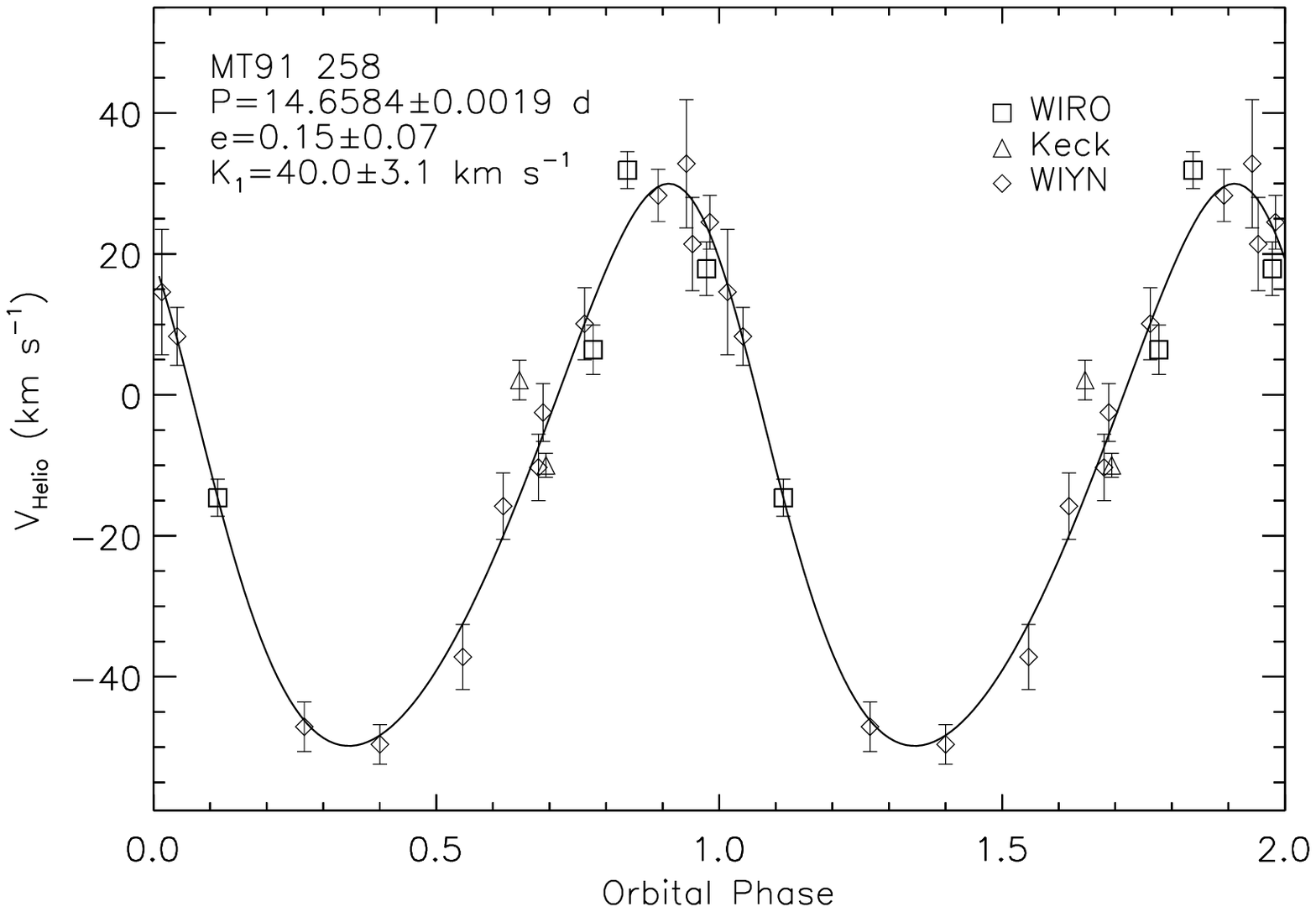}
\caption{Folded radial velocity data and best-fitting solution for MT91~258,
re-analyzed and updated slightly from Paper II.  
\label{sol258}}
\end{figure}

{\it MT91~258---}Originally presented as an SB1 in Paper
II, the updated ephemeris for this 14.658 d O8V appears in
Table~\ref{velupdated1.tab} and the orbital parameters
appear in Table~\ref{solutionsupdated1.tab}.
Figure~\ref{sol258} displays the best-fitting solution and
folded velocity curve. We attribute the unusually high reduced
$\chi^2$ value to velocity zero point differences between
data from different observatories used in the solution.

{\it MT91~372---}Identified as a 2.2 d binary in Paper
III, this eclipsing double-lined system is analyzed in
greater detail in conjunction
with the eclipsing binary distance to Cygnus OB2 in a
forthcoming work \citep{Kobulnicky2014}. 

{\it MT91~696---}An updated ephemeris for this eclipsing
O9.5+B0V+B? triple system from Paper IV will be presented in
\citep{Kobulnicky2014}.

{\it Schulte \#3---}An updated ephemeris for this eclipsing
O6IV: + O9III system from Paper II will be presented in
\citep{Kobulnicky2014}. 

\begin{figure}
\epsscale{1.0}
\centering
\plotone{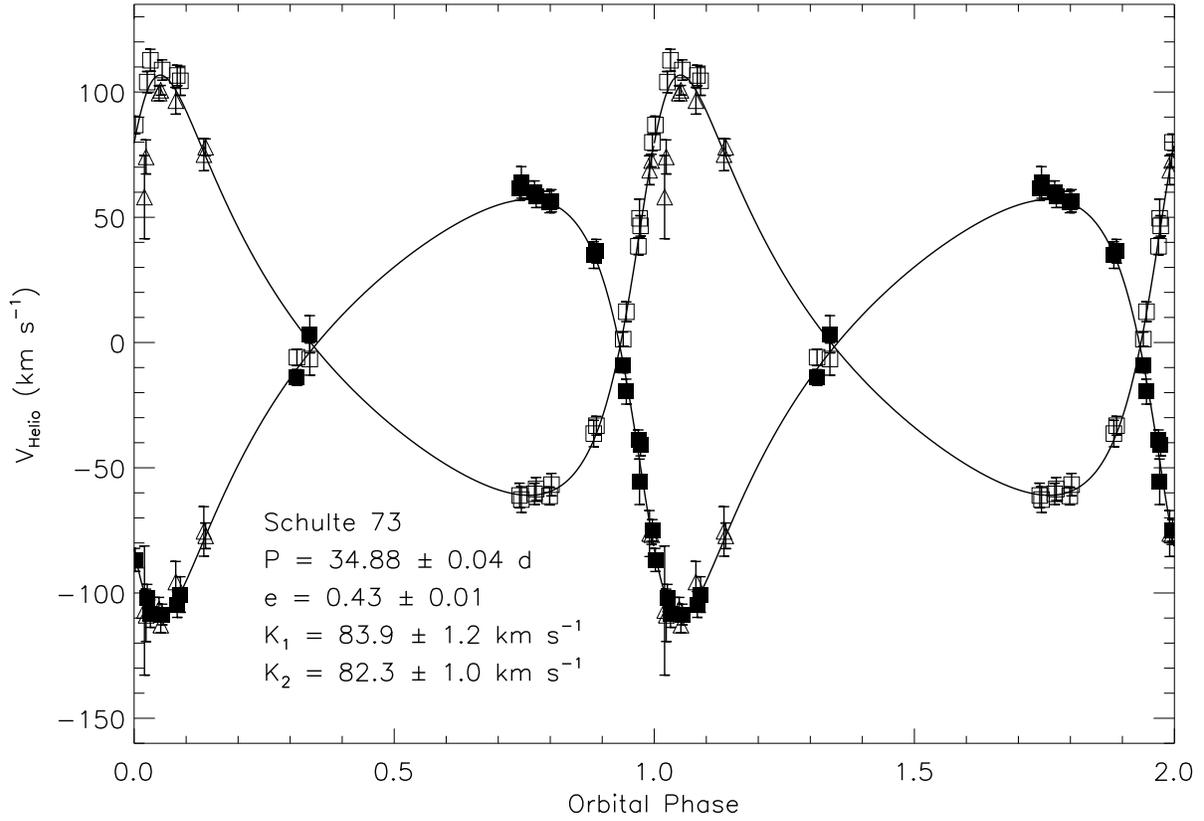}
\caption{Folded radial velocity data and best-fitting solution for Schulte \#73,
re-analyzed and updated from Paper III.  
\label{solS73}}
\end{figure}

{\it Schulte \#73---}This system was identified as a
17.28~day SB2 in Paper~III. Velocities remeasured with the
technique described in Papers IV, VI, and this work were
used as the initial guesses for the \citet{GL2006} method
of spectral deconvolution and cross correlation
for all WIRO data. Owing to the limited phase coverage of
the WIYN+Hydra spectra, we did not apply the \citet{GL2006}
method to these data.  However, since the
\ion{He}{1}~5876\AA\ velocities show a strong linear correlation
with the final cross-correlation velocities, we used this
correlation to correct the WIYN \ion{He}{1}~4471\AA\ velocities. The
final WIRO+WIYN data yielded a period 34.88~days. The power
spectra showed additional strong signals at $\sim$10~days,
$\sim$17~days, $\sim$57~days, and $\sim$67~days. However,
these were ruled out based on the large scatter in the
corresponding folded velocity curves. The updated solution is
provided in Table~\ref{solutionsupdated2.tab} and shown in
Figure~\ref{solS73}.  We slightly revise the spectral
classifications, based on equivalent width ratios of
\ion{He}{2}~5411\AA\ to \ion{He}{1}~5876\AA, as O8.5III: and
O9.0III:. Using the theoretical masses from
\citet{Martins2005}, these classifications indicate that the
system has an inclination near $\sim$40\deg.

{\it CPR2002 A36---}An updated ephemeris for this eclipsing 
B0Ib + B0III system from Paper III will be presented in
\citep{Kobulnicky2014}. 

\begin{figure}
\epsscale{1.0}
\centering
\plotone{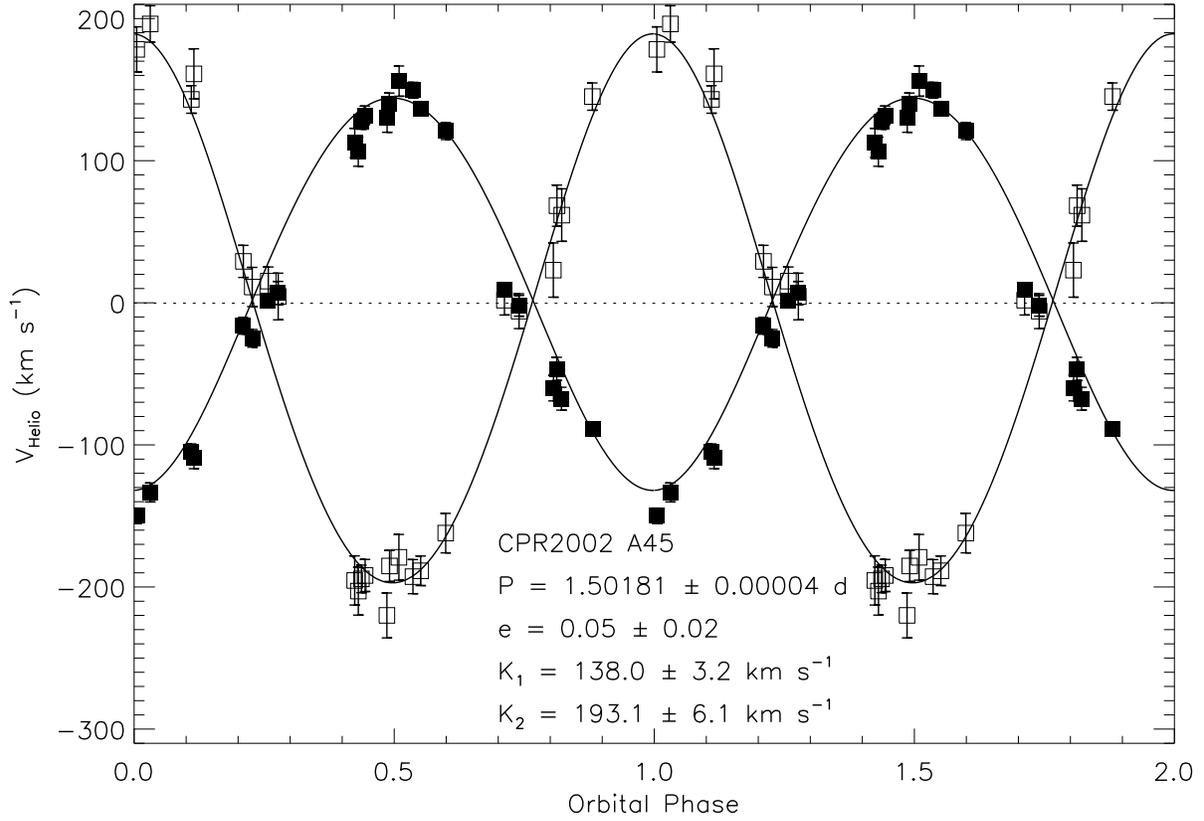}
\caption{Folded radial velocity data and best-fitting solution for CPR2002~A45,
re-analyzed and updated from Paper III.  
\label{solA45}}
\end{figure}

{\it CPR2002 A45---}Also identified as an SB2 in Paper~III,
this system was previously listed with a period of
2.884~days and an uncharacteristically high eccentricity of
0.273. Using the \citet{GL2006} spectral deconvolution
method and some additional observations from WIRO, the
revised radial velocities indicate a period near half the
original solution. Both components' power spectra show a
singular strong signal at 1.5020~days. The final combined
solution yields a period of $1.50181\pm0.00004$~d and a
considerably smaller eccentricity of $0.05\pm0.02$. The
revised solution is listed in
Table~\ref{solutionsupdated2.tab} and shown in
Figure~\ref{solA45}. We retain the primary spectral
classification of B0.5V, but owing to the new, higher mass
ratio of 0.72$\sim$0.02, we revise the secondary spectral
classification to B1V--B2V. Based on the theoretical masses
for stars of this type, we estimate an  inclination  between
40 and 45\deg.

\clearpage

\begin{deluxetable}{lcrrr}
\centering
\tabletypesize{\tiny}
\tabletypesize{\scriptsize}
\tablewidth{0pc}
\tablecaption{Ephemerides for New Binary Systems \label{vel.tab}}
\tablehead{
\colhead{} & 
\colhead{} & 
\colhead{$V_{r1}$} &
\colhead{$\sigma_{V}$} &
\colhead{$O_1-C_1$} \\
\colhead{Date (HJD-2,400,000)} &
\colhead{$\phi$} &
\colhead{(\kms)} &
\colhead{(\kms)} &
\colhead{(\kms)}}
\tablecolumns{5}
\startdata  
\cutinheadc{MT91~021}
 55418.648 &  0.367 &     -37.9 &18.0  &  -0.20 \\
 55489.731 &  0.695 &       1.9 &17.9  &  -4.77 \\
 55780.694 &  0.691 &      11.6 & 6.4  &   5.47 \\
 55791.720 &  0.828 &      20.8 & 4.2  &  -1.34 \\
 55805.681 &  0.267 &     -41.1 & 6.1  &   7.09 \\
 56209.582 &  0.906 &      18.7 & 5.3  &  -1.94 \\
 56215.552 &  0.522 &     -27.4 & 5.4  &  -9.83 \\
 56446.902 &  0.372 &     -38.0 & 8.7  &  -1.02 \\
\enddata 
\end{deluxetable}

\clearpage
 
\begin{deluxetable}{lrrrrr}
\centering
%\rotate
\tabletypesize{\scriptsize}
\tablewidth{0pt}
\tablecaption{Orbital Elements of New Binaries \label{solutions1.tab}}
\tablehead{
\colhead{Element} &
\colhead{MT91~021} &
\colhead{MT91~187} &
\colhead{MT91~202} &  
\colhead{MT91~234} &
\colhead{MT91~241}  }
\startdata            
                          % 021              % 187              %202               % 234	   % 241(done)     
$P$ (days)               & 9.70$\pm$0.01     & 13.531$\pm$0.002 & 43.05$\pm$0.05   & 4996$\pm$330  &671$\pm$2	    \\
$e$                      & 0.29$\pm$0.13     & 0.51$\pm$0.12	& 0.24$\pm$0.11    & 0.12$\pm$0.16 &0.45$\pm$0.05   \\
$\omega$ (deg)           & 81$\pm$31	     & 15$\pm$19	& 160$\pm$40	   & 328$\pm$54    &338$\pm$6	    \\
$\gamma$ (\kms)          & -16.2$\pm$4.5     & -9.6$\pm$0.6	& -10.0$\pm$2.0    & -17.6$\pm$1.5 &-16.7$\pm$1.5   \\
$T_0$ (HJD-2,400,000)    & 56336.58$\pm$0.79 & 56282.7$\pm$0.5  & 56468.29$\pm$4.3 & 55024$\pm$722 &125161$\pm$189  \\
$K_{1}$ (\kms)           & 37.8$\pm$5.5      & 6.5$\pm$1.1	& 19.7$\pm$2.6     & 17.1$\pm$2.2  &18.7$\pm$1.2    \\
$K_{2}$ (\kms)           & \nodata	     & \nodata  	& \nodata          & \nodata	   &\nodata	    \\
S.~C.$_1$                & B1.5V  	     & B1V		& B2V              & B1.5V	   &B1.5V 	    \\
\enddata
\end{deluxetable} 
 
\clearpage

\begin{deluxetable}{lrrrrr}
\centering
%\rotate
\tabletypesize{\scriptsize}
\tablewidth{0pt}
\tablecaption{Orbital Elements of New Binaries  \label{solutions2.tab}}
\tablehead{
\colhead{Element} &
\colhead{MT91~268-c1} &
\colhead{MT91~268-c2} &
\colhead{MT91~292}  &  
\colhead{MT91~295}  &
\colhead{MT91~236} 
}
\startdata            
                          % 268-c1(done)     % 268-c2(done)    %292		   % 295	      %336	     
$P$ (days)                &33.327$\pm$0.002  &5.082$\pm$0.006  &14.8106$\pm$0.0011 &2.4628$\pm$0.0008 &2.04087$\pm$0.00006 \\
$e$                       &0.41$\pm$0.03     &0.48$\pm$0.15    &0.45$\pm$0.06	   &0.30$\pm$0.22     &0.21$\pm$0.18     \\
$\omega$ (deg)            &256$\pm$13	     &294$\pm$28       &191$\pm$13	   &137$\pm$46        &216$\pm$51         \\
$\gamma$ (\kms)           &-22.5$\pm$1.7     &-1.5$\pm$2.5     &-21.4$\pm$1.6	   &-20.2$\pm$1.4     &-13.4$\pm$1.1     \\
$T_0$ (HJD-2,400,000)     &56667.7$\pm$1.1   &56528.8$\pm$0.7  &56058.94$\pm$0.38  &56567.1$\pm$0.3   &2456793.1$\pm$0.3   \\
$K_{1}$ (\kms)            &33.0$\pm$2.7      &17.4$\pm$2.3     &25.3$\pm$2.6	   &9.2$\pm$2.4       &8.8$\pm$1.1      \\
$K_{2}$ (\kms)            &\nodata	     &\nodata	       &\nodata 	   &\nodata	      &\nodata           \\
S.~C.$_1$                 &B2V	             &B2V	       &B2V		   &B2V 	      &B2V               \\
\enddata
\end{deluxetable}

\clearpage

\begin{deluxetable}{lrrrrr}
\centering
%\rotate
\tabletypesize{\scriptsize}
\tablewidth{0pt}
\tablecaption{Orbital Elements of New Binaries  \label{solutions3.tab}}
\tablehead{
\colhead{Element} &
\colhead{MT91~339} &
\colhead{MT91~378} &
\colhead{MT91~390} & 
\colhead{MT91~403} & 
\colhead{MT91~417B} 
}
\startdata            
                          %339  	   %378 	      %390		 %403		      %417B		  
$P$ (days)                &44.63$\pm$0.03  &29.41$\pm$0.03    &4.6252$\pm$0.0012 &16.6379$\pm$0.0062  &38.0$\pm$0.2	  \\
$e$                       &0.21$\pm$0.11   &0.23$\pm$0.05     &0.15$\pm$0.15	 &0.29$\pm$0.03       &0.21$\pm$0.20	  \\
$\omega$ (deg)            &251$\pm$32	   &299$\pm$14        &240$\pm$52	 &260$\pm$9	      &244$\pm$31	  \\
$\gamma$ (\kms)           &-13.0$\pm$0.3   &0.4$\pm$1.4       &-17.4$\pm$0.6	 &-12.9$\pm$1.5       &-20.4$\pm$1.2	  \\
$T_0$ (HJD-2,400,000)     &55229.7$\pm$3.9 &56310.98$\pm$1.16 &55971.80$\pm$0.67 &55973.13$\pm$0.33   &56719.9$\pm$1.6    \\
$K_{1}$ (\kms)            &3.4$\pm$0.3     &36.3$\pm$1.9      &5.4$\pm$0.9	 &49.8$\pm$2.1        &9.5$\pm$1.7	  \\
$K_{2}$ (\kms)            &\nodata	   &\nodata	      &\nodata  	 &\nodata	      &\nodata  	  \\
S.~C.$_1$                	  \\	   &B0V 	      &O8V		 &B1V		      &O6V		  \\
S.~C.$_2$                 &O--K 	   &B:  	      &O--K		 &B:		      &O6V-A		  \\
$i$  (degrees)            &2--90	   &19--90	      &2--90		 &23--90	      &5--90		  \\
$M_{1}$ (adopted; \msun)  &21		   &18	              &21		 &14		      &31		  \\ 
$M_{2}$ (\msun)           &21--0.5	   &17--4.1	      &21--0.32 	 &14 --4.0	      &31--1.8  	  \\ 
$q$                       &1--0.02	   &1--0.23	      &1-0.015  	 &1--0.29	      &1--0.06  	  \\
$a$ (AU)                  &$\sim$0.7	   &$\sim$0.55        &$\sim$0.15	 &$\sim$0.35	      &$\sim$0.7	  \\
$\chi_{red}^2$            &0.32 	   &0.48	      &0.17		 &0.85  	      &0.25		  \\
\enddata
\end{deluxetable} 

\clearpage

\begin{deluxetable}{lrrrrr}
\centering
%\rotate
\tabletypesize{\scriptsize}
\tablewidth{0pt}
\tablecaption{Orbital Elements of New Binaries  \label{solutions4.tab}}
\tablehead{
\colhead{Element} &
\colhead{MT91~448} &
\colhead{MT91~473} &
\colhead{MT91~485} &  
\colhead{MT91~555} &  
\colhead{MT91~561} }
\startdata            
                         %448		    %473			  %485  	 %555		  %561  	 
$P$ (days)               &3.1704$\pm$0.0004 &1687$\pm$51\tablenotemark{a} &4066$\pm$45   &1278.7$\pm$13.2 &40.09$\pm$0.04\\
$e$                      &0.10$\pm$0.06     &0.59$\pm$0.26		  &0.75 (Fixed)  &0.55$\pm$0.09   &0.05$\pm$0.09 \\
$\omega$ (deg)           &3$\pm$34	    &255$\pm$28 		  &352$\pm$9	 &2$\pm$13	  &325$\pm$118   \\
$\gamma$ (\kms)          &-10.6$\pm$1.2     &-5.0$\pm$1.6		  &-12.1$\pm$1.1 &-14.4$\pm$1.3   &-8.9$\pm$2.5  \\
$T_0$ (HJD-2,400,000)    &55971.2$\pm$0.3   &64990$\pm$438		  &52229$\pm$29  &57208$\pm$49    &55977$\pm$13.3\\
$K_{1}$ (\kms)           &27.7$\pm$1.7      &7.5$\pm$2.3		  &15.0$\pm$2.3  &20.8$\pm$4.0    &35.2$\pm$3.6  \\
$K_{2}$ (\kms)           &\nodata	    &\nodata			  &\nodata	 &\nodata	  &\nodata	 \\
S.~C.$_1$                &O6V		    &O8.5V+O9V: 		  &O8V  	 &O8V		  &B2V  	 \\
\enddata
\tablenotetext{a}{Probable triple system; described here are the parameters for the 
combined $\sim$3.6 d binary a+b components (O8.5V+O9V?) treated as a single-lined
system under the influence of the unseen c component, a probable early A -- O dwarf star.  }
\end{deluxetable}

\begin{deluxetable}{lrrrr}
\centering
%\rotate
\tabletypesize{\scriptsize}
\tablewidth{0pt}
\tablecaption{Orbital Elements of New Binaries  \label{solutions5.tab}}
\tablehead{
\colhead{Element} &
\colhead{MT91~588} &
\colhead{MT91~601} &
\colhead{MT91~646} & 
\colhead{MT91~745} }  
\startdata            
                          %588  	    %601	     %646	       %745
$P$ (days)                &245.1$\pm$0.3    &510.2$\pm$0.9   &49.8$\pm$0.2     &151.2$\pm$0.8	  \\
$e$                       &0.51$\pm$0.17    &0.67 (fixed)    &0.31$\pm$0.05    &0.60$\pm$0.2	  \\
$\omega$ (deg)            &242$\pm$16	    &41$\pm$20       &260$\pm$8        &15$\pm$10	  \\
$\gamma$ (\kms)           &-17.8$\pm$1.4    &-12.6$\pm$1.6   &-24.4$\pm$1.9    &-10.7$\pm$0.8	  \\
$T_0$ (HJD-2,400,000)     &56549.1$\pm$8.4  &56222.3$\pm$7.4 &56476.7$\pm$1.1  &56543.7$\pm$3.8   \\
$K_{1}$ (\kms)            &14.5$\pm$2.9     &12.8$\pm$2.1    &61.6$\pm$3.7     &17.5$\pm$2.3	  \\
$K_{2}$ (\kms)            &\nodata	    &\nodata	     &79.6$\pm$4.7     &\nodata 	  \\
S.~C.$_1$                 &B0V  	    &O9.5III	     &B1V	       &O7V		  \\
S.~C.$_2$                 &B--A 	    &O--B	     &B1.5V	       &A--B		  \\
$i$  (degrees)            &15--90	    &18--90	     &75--90	       &15--90  	  \\
$M_{1}$ (adopted; \msun)  &18		    &21 	     &14	       &25		  \\ 
$M_{2}$ (\msun)           &18--3.2	    &21--4.1	     &$\sim$12         &25--4.0 	  \\ 
$q$                       &1--0.18	    &1--0.20	     &0.77$\pm$0.05    &1--0.16 	  \\
$a$ (AU)                  &$\sim$2.2	    &$\sim$3.7       &$\sim$0.6       &$\sim$1.8  \\
$\chi_{red}^2$            &1.0  	    &3.4	     &4.5	       &0.60		  \\
\enddata
\end{deluxetable}

\begin{deluxetable}{lrr}
\centering
\tabletypesize{\tiny}
\tabletypesize{\scriptsize}
\tablewidth{0pc}
\tablecaption{Velocities of Stars Showing Aperiodic Variations\label{random.tab}}
\tablehead{
\colhead{} & 
\colhead{$V_{Helio}$} &
\colhead{$\sigma_V$}\\
\colhead{Date (HJD-2,400,000)} &
\colhead{(\kms)} &
\colhead{(\kms)}}
\tablecolumns{3}
\startdata  
\cutinheadc{MT91~138}
52146.686 & -14.1 & 5.2 \\
52161.793 &  -8.8 & 3.5 \\
52162.651 &  -5.2 & 6.6 \\
53338.617 & -31.6 & 9.1 \\
53989.652 &  -6.6 & 5.3 \\
53990.848 &  -1.6 & 4.1 \\
54285.925 & -23.4 & 4.8 \\
54286.904 & -19.6 & 3.9 \\
\enddata
\tablenotetext{a}{Velocities measured from \ion{He}{2}
$\lambda$5411 assuming a rest wavelength of 5411.45 \AA. }
\end{deluxetable}

\begin{deluxetable}{lcccccclc}
\centering
\tabletypesize{\scriptsize}
%\rotate
\setlength{\tabcolsep}{0.06in}
\tablewidth{0pc}
\tablecaption{OB Binaries in Cyg OB2 \label{bigtable.tab}}
\tablehead{
\colhead{} & 
\colhead{} & 
\colhead{} & 
\colhead{$P$} &
\colhead{$K_1$} &
\colhead{} &
\colhead{} &
\colhead{} \\
\colhead{Star} &
\colhead{Type} &
\colhead{S.C.} &
\colhead{(days)} &
\colhead{(km~s$^{-1}$)} &
\colhead{e} &
\colhead{q} &
\colhead{Ref.} &
\colhead{Notes}}
\startdata  
MT91~021      & SB1     & B1.5V + B + ?           & 9.70$\pm$0.01       & 37.8$\pm$5.5 & 0.29$\pm$0.13	 & 0.18--1	    & 0    & a  \\  %done
MT91~059      & SB1     & O8V + B                 & 4.8523$\pm$0.0003   & 71.0$\pm$4.1 & 0.14$\pm$0.07	 & 0.23--0.67	    & 2,0  & b  \\  %done
MT91~070      & SB1     & O9III + B               & 6.19~yr	        & 9$\pm$1      & 0.34$\pm$0.11	 & 0.25--1	    & 1    &	\\  % leave as in Paper IV
MT91~103      & SB2     & B1V + B2V:              & 22.104$\pm$0.002    & 72$\pm$4.0   & 0.32$\pm$0.05	 & 0.71$\pm$0.10    & 1    &	\\  % leave as in Paper IV
MT91~145      & SB1     & O9III + mid B           & 25.1261$\pm$0.0012  & 42.8$\pm$0.8 & 0.37$\pm$0.02	 & 0.23--0.63	    & 3,0  & c  \\  % done
MT91~174      & SB1     & B2III  + G--B           & 4.536$\pm$0.020     &  9$\pm$1     & 0.53$\pm$0.13	 & 0.03--0.9	    & 1    &	\\  % leave as in Paper IV
MT91~187      & SB1     & B1V + G--B              & 13.472$\pm$0.013    & 6.5$\pm$1.1  & 0.21$\pm$0.24	 & 0.018--1	    & 0    & d  \\  %done
MT91~202      & SB1     & B2V + A--B              & 43.07$\pm$0.05      & 19.7$\pm$2.6 & 0.24$\pm$0.11	 & 0.18--1	    & 0    &	\\
\enddata
\tablecomments{Photometric types EW/KE, EA, and EB stand for eclipsing
system of the W UMa type (ellipsoidal; $P<1$ day), Algol type (near
spherical), and $\beta$ Lyr type (ellipsoidal; $P>1$ day)
respectively. The mass ratio for MT91~421 is calculated using the O star
masses of \citet{Martins2005} and interpolated AB masses of
\citet{drilling}. \\
a -- Probable triple system.\\
b -- Re-analyzed and updated from Paper II.\\
c -- Re-analyzed and updated from Paper III.\\
d -- An alternative period of 5.998 d is also possible, but less likely.\\
e -- Preliminary updates on published orbital parameters presented here.  A more complete analysis will
     be presented in Paper 8.  \\
f -- Exhibits irregular velocity variations in addition to the listed orbital motion.\\
}

\tablerefs{
(0)This work; 
(1) Paper VI; %MT070,MT103,MT174,MT267,MT734
(2) Paper~II; %MT059,MT252,MT258,MT720,MT771,S3
(3) Paper~III;%MT372,MT145,S73,A36,A45
(4) Paper~IV;%MT311,MT605,MT696,MT720,MT771,(Schulte 3?)
%MT372
(5) \citet{Wozniak04};
%MT421 and MT429
(6) \citet{PJ98}; 
%MT696
(7) \citet{Rios04};
%Schulte 3
(8) \citet{Kiminki2012b}; 
%Schulte 5
(9) \citet{Wilson48}; 
(10) \citet{Wilson51}; 
(11) \citet{Mics53};
(12) \citet{Wal73}; 
(13) \citet{Contreras97}; 
(14) \citet{Rauw99}; 
(15) \citet{Hall74};
%Schulte 8a
(16) \citet{Romano69}; 
(17) \citet{Debeck04};
%Shulte 9
(18) \citet{Naze08};
(19) \citet{Naze2012};
%A36
(20) \citet{NSVSa};
(21) \citet{Hanson03};
%B17
(22) \citet{Stroud10};
(23) \citet{NSVSb}}

\end{deluxetable}

\clearpage

\begin{deluxetable}{lcrrr}
\centering
\tabletypesize{\tiny}
\tabletypesize{\scriptsize}
\tablewidth{0pc}
\tablecaption{Reanalyzed Ephemerides for Systems from Paper II and Paper III\label{velupdated1.tab}}
\tablehead{
\colhead{} & 
\colhead{} & 
\colhead{$V_{r1}$} &
\colhead{$\sigma_{V}$} &
\colhead{$O_1-C_1$} \\
\colhead{Date (HJD-2,400,000)} &
\colhead{$\phi$} &
\colhead{(\kms)} &
\colhead{(\kms)} &
\colhead{(\kms)}}
\tablecolumns{5}
\startdata  
\cutinheadc{MT91~059}
 51467.913 &  0.294 &      37.0 & 5.2  &   5.52 \\
 51805.750 &  0.918 &    -124.7 & 5.4  & -16.85 \\
 52146.697 &  0.182 &       7.5 & 5.1  &   3.36 \\
 52162.650 &  0.470 &      27.7 & 5.3  &   7.63 \\
 53338.617 &  0.819 &     -87.0 & 4.2  &   7.59 \\
 53570.693 &  0.646 &     -16.0 & 9.7  &  16.08 \\
 53573.675 &  0.261 &      27.7 &11.8  &   0.98 \\
 53633.634 &  0.618 &     -34.6 & 4.3  & -12.84 \\
\enddata 
\end{deluxetable}

\clearpage

\begin{deluxetable}{lrrr}
\centering
%\rotate
\tabletypesize{\scriptsize}
\tablewidth{0pt}
\tablecaption{Updated Orbital Elements for SB1 Systems from Prior Papers\label{solutionsupdated1.tab}}
\tablehead{
\colhead{Element} &
\colhead{MT91~059} &
\colhead{MT91~145} &
\colhead{MT91~258} }
\startdata            
                         %059                %145                %258		     
$P$ (Days)               &4.8523$\pm$0.0003  &25.126$\pm$0.001   &14.6584$\pm$0.0019   \\
$e$                      &0.14$\pm$0.07      &0.37$\pm$0.02      &0.15$\pm$0.07        \\
$\omega$ (deg)           &225$\pm$22         &125$\pm$3          &43$\pm$28	       \\
$\gamma$ (\kms)          &-30.1$\pm$2.9      &-15.8$\pm$0.6      &-14.4$\pm$2.1        \\
$T_0$ (HJD-2,400,000)    &56338.2$\pm$0.3    &53922.94$\pm$0.17  &2454287.5$\pm$1.0    \\
$K_{1}$ (\kms)           &71.0$\pm$4.1       &42.8$\pm$0.8       &40.0$\pm$3.2         \\
$K_{2}$ (\kms)           &\nodata            &\nodata            &\nodata	       \\
S.~C.$_1$                &08V                &O9III              &O8V		       \\
S.~C.$_2$                &O--midB            &O--midB            &O--B  	       \\
$i$  (degrees)           &19--90             &19--90             &15--90	       \\
$M_{1}$ (adopted; \msun) &21                 &23                 &21		       \\ 
$M_{2}$ (\msun)          &21--5              &23--5.3            &21--3.9	       \\ 
$q$                      &1--0.23            &1--0.23            &1--0.19	       \\
$a$ (AU)                 &$\sim$0.18         &$\sim$0.59         &$\sim$0.3	       \\
$\chi_{red}^2$           &4.3                &2.1                &4.9		       \\
\enddata
\tablecomments{Solutions for MT91~059, MT91~145, and MT91~258 make use of the \ion{He}{1} $\lambda$4471 line velocities 
instead of \ion{He}{1} $\lambda$5876, so the systemic velocity should be regarded as particularly uncertain
relative to the rest of the Cyg~OB2 sample.  }
\end{deluxetable}

\begin{deluxetable}{lrrr}
\centering
%\rotate
\tabletypesize{\scriptsize}
\tablewidth{0pt}
\tablecaption{Updated Orbital Elements for SB2 Systems from Prior Papers\label{solutionsupdated2.tab}}
\tablehead{
\colhead{Element} &
\colhead{MT91~252} &
\colhead{Schulte \#73} &
\colhead{CPR2002~A45}}  
\startdata            
                                   %MT252              %S73                   %A45         
$P$ (Days)                         & 9.558$\pm$0.001   & 34.88$\pm$0.04       &1.50181$\pm$0.00004   \\
$e$                                & 0.25$\pm$0.03     & 0.43$\pm$0.01        &0.05$\pm$0.02   \\
$\omega$ (deg)                     & 326$\pm$8         & 133$\pm$2            &181$\pm$25      \\
$\gamma$ (\kms)                    & -9$\pm$1          & -1.7$\pm$0.7         &-1$\pm$2        \\
$T_0$ (HJD-2,400,000)              & 54704.7$\pm$0.2   & 54698.83$\pm$0.09    &72825$\pm$6     \\
$K_{1}$ (\kms)                     & 70$\pm$2          & 84$\pm$1             &138$\pm$3        \\
$K_{2}$ (\kms)                     & 81$\pm$4          & 83$\pm$1             &193$\pm$6        \\
S.~C.$_1$                          & B1--2V	       & O8.5III:	      &B0.5V		\\
S.~C.$_2$                          & B1--2V	       & O9.0III:	      &B1--2V		\\
$i$  (degrees)                     & 30--40	       & $\sim$40	      &40--45		\\
$M_{1}/mathrm{sin}^{3}i$ (\msun)   & 1.6$\pm$0.2       & 6.2$\pm$0.2          &3.3$\pm$0.2      \\ 
$M_{2}/mathrm{sin}^{3}i$ (\msun)   & 1.4$\pm$0.2       & 6.1$\pm$0.3          &2.3$\pm$0.2      \\ 
$q$                                & 0.86$\pm$0.04     & 0.99$\pm$0.02        &0.72$\pm$0.02    \\
$a$ (/rsun)                        & 0.127$\sim$0.005  & 0.242$\sim$0.004     &0.046$\pm$0.001  \\
$\chi_{red}^2$                     & 2.3               & 1.3                  &2.7               \\
\enddata
\end{deluxetable}

\clearpage

\clearpage

\clearpage

\clearpage

\clearpage

\clearpage

\clearpage

\clearpage

\clearpage

\clearpage

\clearpage

\clearpage
\clearpage

\clearpage

\clearpage

\clearpage

\clearpage

\clearpage

\clearpage

\clearpage

\clearpage

\clearpage

\clearpage

\clearpage

\clearpage

\clearpage

\clearpage

\clearpage

\clearpage

\clearpage

\clearpage

\clearpage

\clearpage

\clearpage

\clearpage

\clearpage

\clearpage

\end{document}